# Ultrafast demagnetization in ferromagnetic materials: origins and progress

Xiaowen Chen[1,2,3], Roman Adam[1,4], Daniel E. Bürgler[4], Fangzhou Wang[1,4], Zhenyan Lu[5], Lining Pan[6], Sarah Heidtfeld[4], Christian Greb[4], Meihong Liu[7], Qingfang Liu[7], Jianbo Wang[7,8], Claus M. Schneider[4,9,10], and Derang Cao[1,2,4,8*]

[1]College of Physics, Qingdao University, 266071 Qingdao, China

[2]National Demonstration Center for Experimental Applied Physics Education, Qingdao University, 266071 Qingdao, China

[3]Spin-X Institute, South China University of Technology, 511442 Guangzhou, China

[4]Peter Grünberg Institute, Forschungszentrum Jülich, 52425 Jülich, Germany

[5]Hunan Provincial Key Laboratory of Intelligent Sensors and Advanced Sensor Materials, School of Physics and Electronics, Hunan University of Science and Technology, 411201 Xiangtan, China

[6]CAS Key Laboratory of Magnetic Materials and Devices, Ningbo Institute of Materials Technology and Engineering, Chinese Academy of Sciences, 315201 Ningbo, China

[7]Key Laboratory for Magnetism and Magnetic Materials of Ministry of Education, Lanzhou University, 730000 Lanzhou, China

[8]Key Laboratory of Special Functional Materials and Structural Design, Ministry of Education, Lanzhou University, 730000 Lanzhou, China

[9]Faculty of Physics, University of Duisburg-Essen, 47057 Duisburg, Germany

[10]Department of Physics, University of California Davis, Davis, California 95616-5270, USA

Email: caodr@qdu.edu.cn

Since the discovery of ultrafast demagnetization in Ni thin films in 1996, laser-induced ultrafast spin dynamics have become a prominent research topic in the field of magnetism and spintronics. This development offers new possibilities for the advancement of spintronics and magnetic storage technology. The subject has drawn a substantial number of researchers, leading to a series of research endeavors. Various models have been proposed to elucidate the physical processes underlying laser-induced ultrafast spin dynamics in ferromagnetic materials. However, the potential origins of these processes across different material systems and the true contributions of these different origins remain challenging in the realm of ultrafast spin dynamics. This predicament also hinders the development of spintronic terahertz emitters.

In this review, we initially introduce the different experimental methods used in laser-induced ultrafast spin dynamics. We then systematically explore the magnetization precession process and present seven models of ultrafast demagnetization in ferromagnetic materials. Subsequently, we discuss the physical processes and research status of four ultrafast demagnetization origins (including spin-flipping, spin transport, non-thermal electronic distribution, and laser-induced lattice strain). Since attosecond laser technique and antiferromagnetic materials exhibit promising applications in ultrahigh-frequency spintronics, we acknowledge the emerging studies used by attosecond pules and studies on ultrafast spin dynamics in antiferromagnets, noting the significant challenges that need to be addressed in these burgeoning field.

# Contents

# 1. Introduction

## 1.1 Methods for manipulating the magnetization

Magnetic storage plays a crucial role in the rapid advancement of the information society, driving the need for substantial improvements in both memory storage density and data read/write speeds. However, magnetic storage technologies are heavily dependent on the accurate manipulation of spin dynamics. Manipulating electron spin and elucidating the principles governing spin-related charge transport have consistently attracted significant attention in the field of spintronics. The primary methods of manipulating spin or magnetization include giant magnetoresistance (GMR) [1, 2] and tunneling magnetoresistance (TMR) [3], both of which are associated with magnetic field-related effects [4, 5]. The subsequent advanced techniques such as spin-transfer torque (STT) [6, 7], spin-orbit torques (SOT) [8, 9], heat-assisted magnetic recording (HAMR) [10, 11], microwave-assisted magnetic recording (MAMR) [12, 13], and ultrafast spin dynamics triggered by femtosecond lasers [14, 15] have significantly accelerated the development of spin dynamics manipulation.

The application of a magnetic field is the most common and direct method for switching magnetization in ferromagnetic materials. Due to the dissipation of energy and angular momentum, the magnetization initially oriented opposite to the effective magnetic field undergoes precession toward the field's direction, ultimately aligning with the effective magnetic field. The entire magnetization reversal process occurs on a nanosecond timescale. Although reducing the magnetization reversal time to the picosecond scale is achievable by increasing the applied magnetic field, experimental evidence has shown that excessively high magnetic fields can lead to chaotic magnetization switching [16, 17] and ultimately prolong the magnetization reversal time.

Compared to the traditional method of applying a magnetic field, current-induced STT and SOT effects offer significant advantages in terms of power consumption, speed, and the integration of spintronic devices. The spin valve structure, which consists of a ferromagnetic 1 /nonmagnetic /ferromagnetic 2 (FM1/NM/FM2) system, serves as a typical model for studying STT [15, 18]. When an electric current flows through the pinned layer (FM1), the spin-filtering effect results in an unequal quantity of spin-up and spin-down electrons, generating a spin-polarized current within FM1. This spin-polarized current subsequently tunnels through the non-magnetic layer (NM) and influences the free layer (FM2). In this process, according to the conservation of spin angular momentum, a transverse component of the spin current is generated, exerting torque on FM2. This torque drives the magnetization in the free layer to align with the magnetization direction of the pinned layer. Nevertheless, magnetization reversal via STT requires a relatively high current density. Another challenge is overcoming the speed and barrier reliability bottlenecks of the spin current during the STT process.

The SOT exhibits a lower write current compared to STT and is expected to overcome the limitations of STT. In the SOT process, an electric current flows into the NM layer of the heterostructure, resulting in the generation of a transverse spin current through the spin Hall effect (SHE) [19, 20] or the interfacial Rashba effect [21]. This spin current is subsequently injected into the adjacent FM layer. In the case of the SHE, the magnetization reversal of the sample is achieved through the SOT interaction with the magnetization of the FM layer. Regarding the Rashba effect, the direction of equivalent magnetic field generated in this process depends on the momentum direction of the electrons. When electron flow occurs, the momentum of electrons at the heterostructure interface becomes imbalanced in a specific direction. Since the momentum direction of the electrons is closely correlated with the spin polarization direction, this imbalance in electron momentum distribution leads to a corresponding imbalance in spin polarization at the interface between the NM and FM layers. The spin-polarized electrons induce an SOT that interacts with the magnetization of the FM layer, resulting in magnetization reversal. Nevertheless, it presents a challenge to achieve magnetization reversal of the FM layer solely through the pure SOT interaction in a magnetic tunnel junction film with perpendicular magnetic anisotropy.

HAMR technology is also a highly promising approach for manipulating the magnetic moments of ferromagnetic materials [10, 11]. This method works by reducing the coercivity of the magnetic recording medium through the application of laser heating. Magnetization reversal occurs when the coercive field of the medium becomes lower than the magnetic field induced by the recording head. Since the temperature generated by the laser diode in a single data bit can reach 400-700°C, designing an effective magnetic medium to that avoids affecting the substrate and adjacent data bits remains a significant challenge. Generally, heat sinks and interlayers are added to increase the heat dissipation rate of the medium. However, excessively rapid heat dissipation can lead to higher energy consumption. Achieving both efficient heat dissipation and low energy consumption simultaneously is a key challenge that requires further breakthroughs. Additionally, due to the diffraction limit, the spot size of light focused in the far field is difficult to reduce below 238 nm, significantly restricting storage density. Recently, Seagate has overcome the diffraction limit by utilizing surface plasmons in photonic crystals, achieving a storage density of 5 Tbpsi, with each data bit being heated and cooled within 1 ns.

MAMR technology uses microwave fields generated by a spin torque oscillator (STO) to assist in the switching of magnetic bits, enabling high-density data storage without the need for elevated temperatures. This technology primarily relies on two effects: spin transfer switching (STS) and microwave-assisted switching (MAS). The STS effect is a process in which the STO in MAMR controls the precession and switching of magnetization via STT. The STO typically consists of two magnetic layers: a field generation layer (FGL) and a spin injection layer (SIL). The working principle of the STO is as follows: when electrons reflect within the SIL, a spin current is generated. As the spin current flows into the FGL and STT is applied to the magnetization of this layer, the spin of the FGL undergoes oscillation, producing an alternating (AC) magnetic field. To ensure the STO generates a stable oscillating magnetic field with the necessary frequency and amplitude to flip magnetization, the design of the STO is particularly important. Furthermore, the MAS effect utilizes the AC magnetic field generated in the FGL to control the precession of magnetization, which is a key process in the magnetization switching of the MAMR system [22, 23]. Both MAMR and HAMR represent significant advancements in magnetic recording technology, offering promising pathways to meet the increasing demand for higher data storage capacities.

Nevertheless, the methods described above generally require a relatively prolonged timescale to achieve magnetization reversal. They face numerous technical challenges, including unavoidable magnetic interference, the necessity for specialized component design, and the requirement for assistance from other external excitation. In recent decades, femtosecond laser-induced ultrafast demagnetization in ferromagnetic films has garnered considerable attention and has given rise to an entirely new field of ultrafast spin dynamics. Ultrafast demagnetization results from complex interactions among the electron, spin, and lattice subsystems after the magnet absorbs ultrashort laser pulses. As the energy of the electrons is transferred to the spin and lattice systems, the angular momentum associated with the electron spins is also redistributed among these subsystems. Specifically, when electrons absorb energy from an ultrafast laser pulse, they experience rapid changes in their energy states. The excess energy can be transferred to the lattice through electron-phonon interactions, causing lattice vibrations. Concurrently, the angular momentum associated with the electron spins can be transferred to the lattice and the spin system through spin-lattice relaxation processes and spin-spin interactions, respectively. This results in the disorientation of the electron spins, reducing the magnetization on picosecond or shorter timescales [14, 24-27]. This enhances the understanding of the interactions between lasers and magnetic materials. With the assistance of ultrashort lasers, the magnetization reversal time can be reduced to the picosecond scale [28, 29]. A recent report [30] indicates that the magnetization switching time in GdCo alloys can be reduced to less than 500 fs by altering the sample concentration. This rapid reduction in magnetization represents a significant advancement in magnetic order manipulation, offering a path toward high-speed, low-energy magnetic switching applications.

In addition, researchers have utilized ultrashort laser pulses in rare earth–transition metal (RE–TM) ferrimagnetic alloys (predominantly Gd-based ferrimagnetic alloys) and multilayer film systems to achieve ultrafast magnetization reversal without the application of external magnetic fields or currents. This phenomenon is known as all-optical switching (AOS) [29, 31-36]. In terms of the process, when the laser fluence is lower than the material-specific threshold and the helicity of the light is important to determine the final orientation of the magnetization. On the contrary, if the applied laser fluence is chosen above a material-specific threshold the light needs not to be circularly polarized and the orientation of the magnetization can be toggled between two stable states with each pulse [37, 38]. The origin of magnetization reversal has been relatively well understood in ferrimagnetic system microscopically. Owing to the different response time of magnetization associated with $3d$ and $4f$ electrons in different sub-lattice [39], the ferrimagnetic system undergoes a transient ferromagnetic state. The magnetization reversal is the consequence of the angular momentum conservation during the demagnetization process [40].

In contrast to the ferrimagnetic systems, however, the fs laser-induced magnetization reversal is fundamentally different for ferromagnetic system. Phenomenologically, AOS in ferromagnetic system (such as FePt) exhibits an accumulative switching behavior [41] and the reversed pattern in the laser excited area is markedly different from that in the ferrimagnetic systems. A static picture has been given for the excited area [42], proposing that the window for helicity-dependent AOS (HD-AOS) is defined by two intensity thresholds. In the center of the excitation area, only demagnetization occurs due to the high fluence, while at the boundary of the excitation area mainly HD-AOS occurs. A significant modification of the AOS-induced magnetization of a ferromagnetic $[Co/Pt]_n$ multilayer system has been reported on a much longer timescale, and the AOS of the sample strongly depends on the ambient temperature [43]. To realize and fully leverage AOS across a broader range of material systems, it is crucial to develop a more precise and comprehensive understanding of the microscopic mechanisms driving ultrafast demagnetization.

Nowadays, numerous studies have been conducted on femtosecond laser-induced ultrafast demagnetization across various magnetic systems. This extensive body of research has further propelled the advancement of models and mechanisms governing ultrafast demagnetization processes. The subsequent sections will explore the development of femtosecond laser-induced ultrafast demagnetization in ferromagnetic materials.

**1.2 The state of the art in ultrafast demagnetization dynamics**

In 1996, Beaurepaire *et al.* observed ultrafast demagnetization in a ferromagnetic Ni film, 3*d* transition metal, using the time-resolved magneto-optical Kerr effect(TR-MOKE) [14]. Figure 1 (a) illustrates the variation in the MOKE signal with increasing delay time, which is defined as the time difference between the arrival of the pump pulse and the probe pulse at the sample. The MOKE signal provides information on the magnetization variation of the sample. When a laser with a 60 fs pulse irradiates the Ni film, the magnetization of the Ni film undergoes a rapid decrease, a phenomenon known as demagnetization. The duration required for the magnetization to reach its minimum is referred to as the demagnetization time, while the magnitude of this decline is termed the demagnetization degree. Afterward, the magnetization of the Ni film begins to gradually increase over time, corresponding to the relaxation process of the magnetization. In this study, they demonstrated that the demagnetization of the Ni film occurs within 2 ps and proposed a phenomenological three-temperature model (3TM) to describe the process. This initial discovery opened the door to the study of femtosecond laser-induced ultrafast spin dynamics. In the following year, two other research groups independently corroborated the phenomenon of ultrafast demagnetization in Ni films using second harmonic generation [44] and two-photon photoemission [45]. Later work on THz emission from laser-excited Ni films was also crucial in confirming the existence of ultrafast demagnetization [46].

For more than two decades, ultrafast demagnetization phenomena in 3*d* metal alloys and their multilayer systems have been extensively studied [15, 39, 47-59]. Subsequently, demagnetization phenomena distinct from those in 3*d* metal materials were observed in transition-rare earth (RE) metal alloys and their multilayer films, as well as in half-metallic magnetic oxides [39, 60-65]. In these materials, the pulse laser-induced demagnetization process can be divided into two stages, as shown in Fig. 1(b). The first stage mirrors phenomena observed in 3*d* metal films, where the material's magnetization decreases rapidly on the picosecond timescale, followed by a slow relaxation. However, this does not mark the end of the demagnetization process. The second stage of demagnetization (re-demagnetization) occurs almost three orders of magnitude slower than the first stage, and is also followed by a slow relaxation phase. To distinguish between the demagnetization processes occurring in 3*d* metal systems and RE metal systems, the former is termed "type I" or one-step demagnetization, while the latter is termed "type II" or two-step demagnetization.

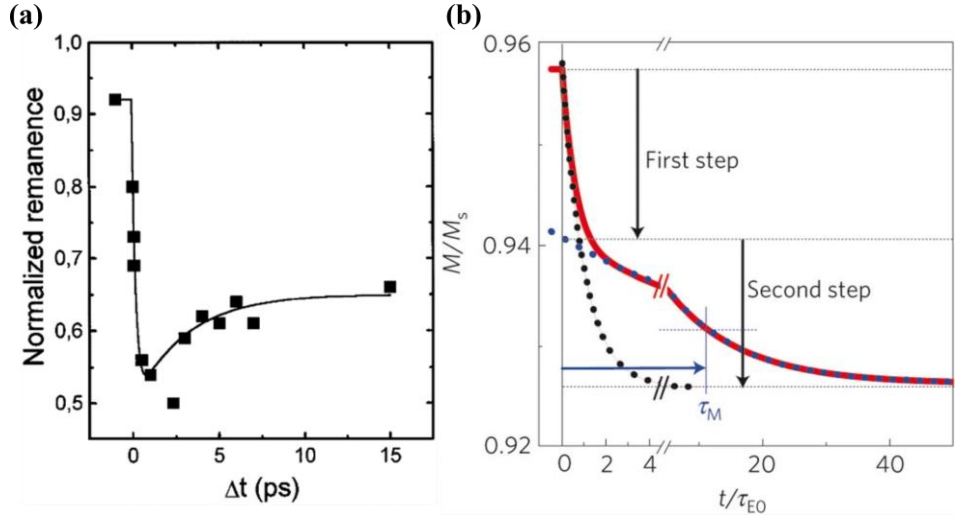

Fig. 1. (a) The ultrafast demagnetization in a ferromagnetic Ni film triggered by using a 60-fs laser pulse. It is observable that the normalized residual MOKE signal diminishes swiftly immediately after zero delay ($\Delta t$=0). (b) Time evolution of type II demagnetization. Fig. 1(a) is reproduced with permission from E. Beaurepaire *et al.*, Phys. Rev. Lett., **76**, 4250 (1996) [14]. Copyright (1996) by the American Physical Society. Fig. 1(b) is reproduced with permission from B. Koopmans *et al.*, Nat. Mater., **9**, 259-265 (2010) [39]. Copyright (2009) Springer Nature.

Moreover, another focus in this context is the influence of polarized pump light on ultrafast demagnetization. In 2007, Stanciu *et al.* [31] investigated the magnetization reversal in a GdFeCo system induced by ultrafast laser pulses in the absence of an external magnetic field. They confirmed that the direction of all-optical magnetization switching is determined by the helicity of the optical pulse. However, this study pertains specifically to the helicity dependence of AOS, rather than the helicity dependence of the demagnetization process. In the same year, Longa *et al.* [66] conducted a quantitative analysis of the impact of left-handed and right-handed circularly polarized pump light on the demagnetization process. Although the MOKE signal showed a strong dependence on the helicity of the pump light, this effect was found to be of non-magnetic origin (i.e., the transfer of angular momentum from circularly polarized photons to electron orbitals) and did not influence the demagnetization process. This study also ruled out the possibility of direct angular momentum transfer related to the demagnetization process, demonstrating that the contribution of photons to demagnetization can be negligible. Subsequently, numerous studies have focused on helicity-dependent magnetization switching behavior [30, 32, 42, 67-69]. Khorsand *et al.* [42] proposed that this helicity-dependent phenomenon is due to the differential absorption of light with varying polarizations by the material, a phenomenon known as magnetic circular dichroism (MCD). Currently, multiple reports suggest that the magnetization switching is dependent on the absorbed energy, which is determined by the polarization of the light rather than its helicity [32, 68-70].

The rapid dissipation mechanism of spin angular momentum is another challenging issue during ultrafast demagnetization [39, 48, 52-58, 64, 71-79]. Recently, Dornes *et al.* [52] proposed that the existence of the ultrafast Einstein-de Haas effect results in the majority of the angular momentum lost from the spin system being transferred to the lattice system within a sub-picosecond timescale. Tauchert *et al.* [79] observed a non-equilibrium population of anisotropic phonons produced within 150-750 fs in thin Ni films through ultrafast electron diffraction. These observations provide compelling evidence for the transfer of angular momentum between spin and phonons.

So far, researchers have proposed several models to comprehend the physical processes involved in femtosecond laser-induced ultrafast demagnetization. These models include the two-temperature model (2TM) [80], the three-temperature model (3TM) [14, 54], the extended three-temperature model (E3TM) [64], the microscopic multi-temperature model (MMTM) [57], the four-temperature model (4TM) [65], the microscopic three-temperature model (M3TM)[39], and the extended microscopic three-temperature model (EM3TM) [81]. This topic will be further expounded upon in Section 3.2.

The aforementioned models have been developed based on the physical origins underpinning ultrafast demagnetization, which have remained contentious for an extended period. Researchers have proposed several potential mechanisms responsible for the rapid dissipation of spin angular momentum, including spin-flipping (involving Elliott-Yafet scattering [39, 75, 82-85], electron-magnon scattering [86-90], photon-spin interaction [24, 91-94]), spin transport (involving superdiffusion transport [53, 95], electron magnon scattering-induced spin transfer [96, 296-299], and the optical inter-site spin-transfer (OISTR) effect[96-98]), non-thermal electronic distribution [48, 55, 64, 76, 99, 100], and the laser-induced lattice strain [101-103].

While Carva *et al.* [104] and Scheid *et al.* [105] have provided relatively comprehensive reviews of the progress in ultrafast spin dynamics, their works lack a comparative analysis of different ultrafast demagnetization models. Additionally, the discussion of various physical origins remains insufficient. Contradictions exist in the explanation of ultrafast demagnetization origins within the same or similar systems. This highlights the need for further empirical research and theoretical analysis to enhance this understanding. The current understanding of ultrafast demagnetization has advanced significantly, with several models and a plethora of advanced experiments providing detailed insights into the process. However, there remains a lack of comprehensive reviews on specific origins and progress, which continue to be subjects of active research and debate.

In this review, we provide a comprehensive summary of various experimental methods employed in the investigation of ultrafast demagnetization. Our primary focus are on summarizing and comparing different models and microscopic origins, as well as the corresponding research advancements in describing the ultrafast demagnetization process. Additionally, we present significant conclusions regarding the predominant origins of ultrafast demagnetization in various systems.

## 2. Experimental method

In this chapter, we introduce the experimental principles and schematics for the main techniques employed in the measurement of laser-triggered magnetization dynamics. These techniques include time-resolved magneto-optical Kerr effect (TR-MOKE), second harmonic generation (SHG), two-photon photoemission (TPPE), the X-ray techniques, terahertz range techniques, and ultrafast electron diffuse scattering (UEDS). TR-MOKE and SHG are primarily used to investigate spin dynamics, while TPPE is employed to study carrier dynamics. Techniques such as THz-TDs, OPOT, XMCD, TR-ARPES, TR-XRMS, TR-RIXS, and TR-X-PEEM—within the terahertz or X-ray ranges—are sensitive to both carrier and spin signals. UXDS and UEDS, on the other hand, are widely used to examine laser-excited lattice dynamics. These techniques provide comprehensive insights into ultrafast spin dynamics from various perspectives.

### 2.1 Time-resolved magneto-optical Kerr effect (TR-MOKE)

#### 2.1.1 Phenomenological description of MOKE

To well understand TR-MOKE, we first examine the phenomenological macroscopic description of MOKE. In 1845, Michael Faraday initially discovered the magneto-optical (MO) effect. However, his experimental results were not entirely reliable due to the limitations of experimental conditions at that time. In 1887, John Kerr made the significant discovery of MOKE on polished electromagnet poles. MOKE, which can detect variations in magnetization within a confined region, has since evolved into a crucial technique for characterizing the magnetic properties of magnetic films.

The MOKE phenomenon can be described as follows: when linearly polarized light impinges upon the surface of a magnetic material, the polarization of the reflected light undergoes a change due to magnetic birefringence. The polarization plane of the reflected light rotates by an angle relative to the polarization plane of the incident light, and there is also a change in the light's ellipticity during this process. These variations are commonly referred to as Kerr rotation ($\theta_k$) and Kerr ellipticity ($\eta_k$), respectively. Figure 2 illustrates a schematic diagram of the longitudinal MOKE (L-MOKE) measurement.

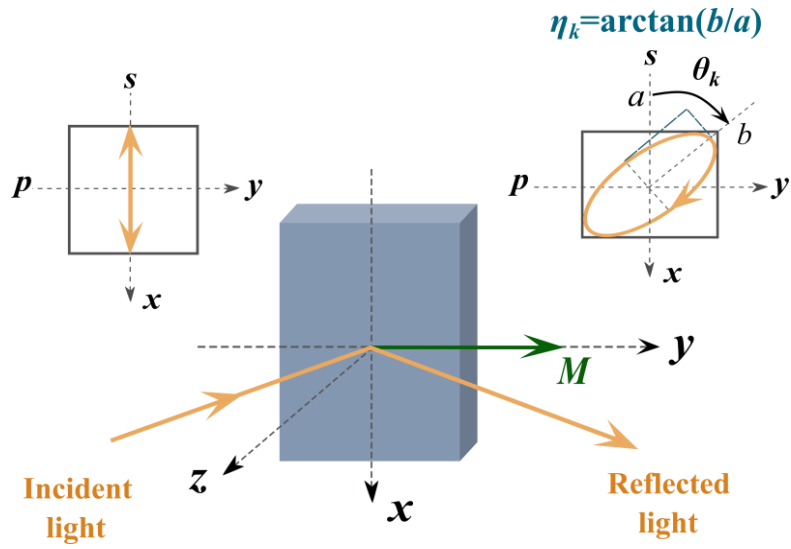

Fig. 2. Schematic diagram of the L-MOKE measurement. The case of $s$-polarized light (incident light) transforms into elliptically reflected light. The definitions of Kerr rotation ($\theta_k$) and Kerr ellipticity ($\eta_k$) are depicted in the upper right corner, where $a$ and $b$ represent the major and minor semi-axes of the ellipse, respectively.

Table 1 Geometric schematics of P-MOKE, L-MOKE, and T-MOKE measurements, along with their respective dielectric permittivity tensor $\varepsilon$ and polarization of the reflected light [106].

| | Polar | Longitudinal | Transversal |
|---|---|---|---|
| **Geometry** | (M along z) | (M along y) | (M along x) |
| $\hat{\varepsilon}$ | $N^2 \begin{pmatrix} 1 & iQm_z & 0 \\ -iQm_z & 1 & 0 \\ 0 & 0 & 1 \end{pmatrix}$ | $N^2 \begin{pmatrix} 1 & 0 & -iQm_y \\ 0 & 1 & 0 \\ iQm_y & 0 & 1 \end{pmatrix}$ | $N^2 \begin{pmatrix} 1 & 0 & 0 \\ 0 & 1 & iQm_x \\ 0 & -iQm_x & 1 \end{pmatrix}$ |
| **Polarization variation** | Linearly → Elliptically | Linearly → Elliptically | None |

In accordance with the orientation of the incident plane relative to the magnetization, as presented in Table 1, MOKE can be categorized into three distinct types: polar MOKE (P-MOKE), L-MOKE, and transverse MOKE (T-MOKE). In both P-MOKE and L-MOKE configurations, the polarization of the reflected light changes in comparison to the incident light. In the case of T-MOKE, if the incident light is $p$-polarized (where the polarization direction is parallel to the incident plane), there is a minor alteration in the light intensity of the reflected light while the polarization remains unchanged. If the incident light is *s*-polarized (where the polarization direction is perpendicular to the incident plane), both the intensity and the polarization of the reflected light remain nearly constant [107-109].

The variation of the MOKE process in a magnetic material can be mathematically described by employing the dielectric permittivity tensor $\varepsilon$ of the sample [108, 110], as shown in Equation (1). It is noted that only the effects linear in magnetization are included for the dielectric tensor used in this expression. Higher-order effects, such as the Voigt effect, are not included in this formulation. This simplification allows for a focused analysis of the primary magneto-optical interactions.

$$\hat{\varepsilon}=N^2 \begin{pmatrix} 1 & iQm_z & -iQm_y \\ -iQm_z & 1 & iQm_x \\ iQm_y & -iQm_x & 1 \end{pmatrix} \quad (1)$$

where the refractive index ($N$) describes the material's optical density and affects how light propagates through it. The magneto-optical coupling factor ($Q$) represents the strength of the interaction between the magnetization of the material and the electromagnetic wave. The imaginary unit $i$ denotes the phase shift due to the magneto-optical coupling. The symbols $m_x$, $m_y$, and $m_z$ represent the direction cosines of the magnetization along the *x*-, *y*-, and *z*-axis, respectively. According to the schematic diagram in Fig. 2, the magnetization direction of L-MOKE is arranged along the *y*-axis. Thus, the $\varepsilon$ for L-MOKE only needs to consider the antisymmetric terms of $m_y$. Similarly, the $\varepsilon$ for P-MOKE and T-MOKE include only the antisymmetric terms of $m_z$ and $m_x$, respectively. This asymmetry is a direct consequence of the magneto-optical interactions, where the magnetization direction introduces a preferential orientation, leading to a non-reciprocal optical response in the material. The directional dependence of these effects is crucial for the manipulation and detection of the magneto-optical properties in various applications, including optical isolators, sensors, and information storage technologies. The $\varepsilon$ expressions for the three geometrical configurations of MOKE, corresponding to particular directions of the magnetization, are presented in Table 1. These are specific cases derived from the general expression given in Equation (2).

According to the Jones matrix, one can also derive the relationship between the components of *s*- and *p*- polarized light [111, 112]

$$\begin{bmatrix} E_{r,p} \\ E_{r,s} \end{bmatrix} = \begin{pmatrix} r_{pp} & r_{ps} \\ r_{sp} & r_{ss} \end{pmatrix} \begin{bmatrix} E_{i,p} \\ E_{i,s} \end{bmatrix} \quad (2)$$

where $E_{r,p}$ and $E_{r,s}$ represent the reflected light vectors for *p*- and *s*- polarized light, respectively. $E_{i,p}$ and $E_{i,s}$ denote the incident light vectors for *p*- and *s*- polarized light, respectively. $r_{pp}$ and $r_{ss}$ are the reflection coefficients for *p*- and *s*- polarized light that remain their polarization upon reflection, respectively. Conversely, the reflection coefficients $r_{ps}$ and $r_{sp}$ describe the conversion of *p*- (*s*-) polarized incident light into *s*- (*p*-) polarized light upon reflection, respectively.

In the context of L-MOKE and P-MOKE, the complex Kerr angle of $\varphi_{k,s}$ is defined as $\varphi_{k,s}=\theta_k+i\eta_k=r_{ps}/r_{ss}$ for the *s*-polarized incident light, while $\varphi_{k,p}$ is defined as $\varphi_{k,p}=\theta_k+i\eta_k=r_{sp}/r_{pp}$ for the *p*-polarized incident light [108].

In the case of T-MOKE, only the terms $r_{pp}$ and $r_{ss}$ need to be considered due to the constant polarization of light. It is worth noting that a smaller angle of incident light in P-MOKE measurements enhances the Kerr signal due to the orthometric alignment between magnetization and the film plane. In L-MOKE measurements, where the direction of magnetization is parallel to the film plane, a larger angle of incident light enhances the Kerr signal.

MOKE is consistently employed in the assessment of static magnetic properties. It is utilized for the mapping of the hysteresis loop [113-115], detecting of the magnetic anisotropy [113], and imaging of the magnetic domains [116-119] within the magnetic film. These topics are not the focus of this paper, and we will not provide further elaboration.

### 2.1.2 Basic setup of TR-MOKE

The time-resolved MOKE (TR-MOKE) is typically conducted using a stroboscopic dual-beam pump-probe technique, which enables the comprehensive analysis of material magnetization dynamics at the sub-picosecond scale. The schematic diagram of TR-MOKE is illustrated in Fig. 3. The basic principle of TR-MOKE involves a pump beam with higher energy irradiating the sample, exciting the irradiated area to a highly non-equilibrium state. Subsequently, a probe beam with lower energy is directed at the same location on the sample to detect changes in the polarization of reflected light (including both *s*- and *p*-polarized light). By adjusting the optical delay stage (DS) between the corresponding pump and probe pulses, the magnetization dynamics are recorded over a time range from a few femtoseconds to several nanoseconds. The reflection geometry of MOKE makes it particularly sensitive to surface and interface magnetization, making it well-suited for studying thin magnetic films, multilayers, and nanostructures. The underlying principles of MOKE are detailed in Section 2.1.1. When the detector is placed on the back side of the sample to detect the transmitted probe beam, this experimental configuration is referred to as the magneto-optical Faraday effect [120-122]. Faraday effect occurs during transmission through a material and is bulk-sensitive. It is commonly used to measure bulk magnetization and magnetic field effects in transparent media, especially in materials that are transparent in the optical wavelength range. The transmitted light interacts with the magnetization along the same direction as the light propagation, meaning the magnetic field and the wave vector of the light are collinear (parallel). As light passes through the medium, the polarization is rotated (Faraday rotation) due to the interaction with the material's magnetization. The degree of Faraday rotation depends on the thickness of the material, the strength of the magnetization, and the Verdet constant, which is a material-dependent constant describing the strength of the Faraday rotation.

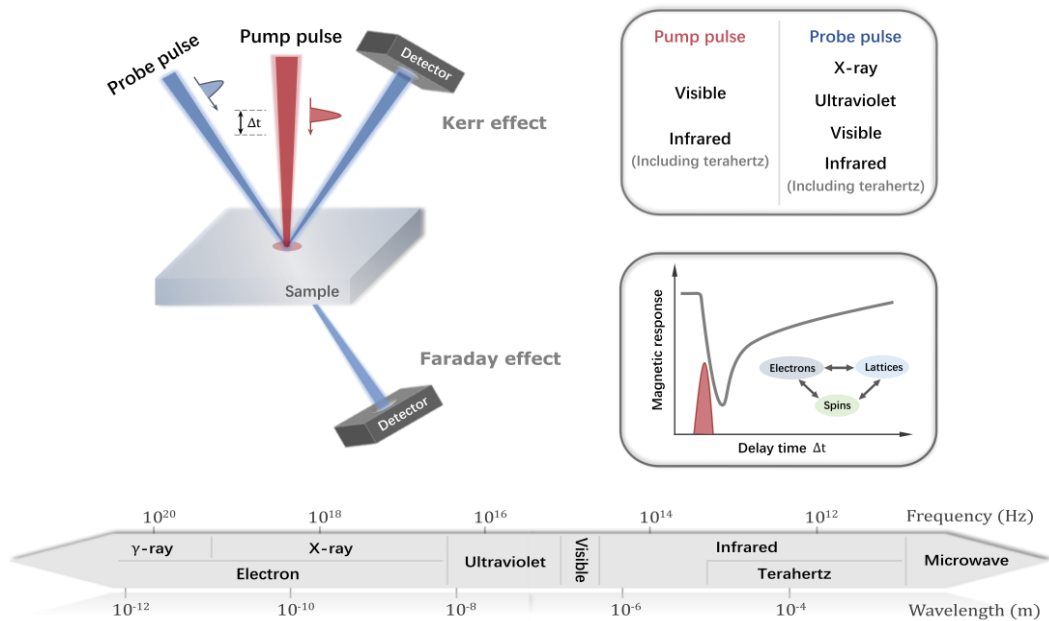

Fig. 3. The left side shows the schematic diagram of TR-MOKE (reflection geometry) and time-resolved magneto-optical Faraday effect (transmission geometry). The right panels display the typical pump and probe pulse light sources used in the measurement, along with the variation in magnetic response following pump light irradiation. The bottom section illustrates the frequencies and wavelengths corresponding to various segments of the electromagnetic spectrum. Different wavelength regimes can serve as probe sources for the two effects, depending on the material properties and the specific experimental setup. The Kerr effect is typically studied with visible and ultraviolet light, but infrared (including terahertz) and X-ray (EUV MOKE) can also be used in certain cases. The Faraday effect is most commonly observed in visible and infrared (including terahertz) regions but can also be probed using X-rays (XMCD) in specific cases.

Since different light sources exhibit varying sensitivities to spin and orbital degrees of freedom, selecting different sources for pump and probe light allows for the acquisition of more comprehensive information about the sample. The choice of pump pulses can lead to different electron distributions in the excited region of the sample, resulting in varying demagnetization behavior. Typical pump pulses include optical pumps with wavelengths ranging from 380 nm to 10 $\mu$m (i.e., visible, near-infrared, and mid-infrared light) [15, 123], and far-infrared pulses (including terahertz (THz) pulses [124, 125]) [126]. The specific information about the sample is primarily provided by the probe light. The selection of probe pulses influences the sensitivity in detecting spin and orbital degrees of freedom during the measurement. Current probe pulses may include optical pulses (380 nm-10 $\mu$m) [15, 48, 127], ultraviolet pulses [26, 47, 98, 128, 129], far-infrared pulses [79, 130], and X-ray pulses [54, 131]. The different probe light sources are discussed below. The far-infrared probe will be discussed in Section 2.2 Terahertz range techniques.

**I. Optical probe**

Optical probes are frequently employed in traditional TR-MOKE experiments, with the probe light typically sourced from the same laser system as the pump light. When the metal material is excited by the laser, the empty electronic states above the Fermi level are filled, and electrons below the Fermi level are removed. This leads to a decrease in the oscillator strength of the probe light photons propagating through the sample. Such pump-induced state filling effect is characterized by a non-thermal electron distribution. As a result, the sample becomes more transparent to the probe light, or dichroic bleaching occurs. Dichroic bleaching, in particular, affects the optical MOKE response [132].

The optical response induced by the probe in the medium primarily depends on electric dipole transitions. According to the electron spin conservation selection rule, no spin flip occurs during this process. Consequently, optical probes exhibit high sensitivity to magnetic order in samples with strong spin-orbit coupling. However, at higher energies, the optical spectral linewidth increases and overlaps, complicating the analysis of ultrafast dynamic behavior. In such cases, the MO signal and the magnetization are no longer linearly related, necessitating the consideration of higher-order terms. These details will be discussed at the end of this section.

**II. Extreme ultraviolet (EUV) and X-ray probe**

Among these probe pulses, both extreme ultraviolet (EUV) and X-ray radiation possess very high photon energy, leading to interband transitions of electrons within the metal. These two probe beams can be used for both element-selective TR-MOKE measurements and time-resolved magnetic domain imaging due to X-ray magnetic circular dichroism (XMCD). EUV and soft X-ray MOKE provide enhanced sensitivity and resolution, allowing for detailed investigation of spin dynamics on ultrafast timescales and at nanometer spatial resolution. These techniques have been instrumental in probing the interactions between electronic, spin, and lattice subsystems during demagnetization.

Among the three configurations of MOKE, T-MOKE is the most suitable for EUV and soft X-ray light sources. This is because: (i) T-MOKE can detect dielectric tensors and is particularly sensitive to the off-diagonal elements of the dielectric tensor; (ii) Circularly polarized light is difficult to achieve with high-order harmonics, and as a result, the polarization state of the reflected light remains unchanged. Researchers can therefore focus on analyzing the intensity of the Kerr signal, which is highly advantageous for high-energy photons [123, 133]; (iii) T-MOKE signals are primarily derived from pure-charge and pure-magnetic contributions. By adjusting the incident angle to the Brewster angle, charge scattering can be suppressed, enabling the detection of a purely magnetic signal [134]. Consequently, the combination of EUV and soft X-ray sources with the T-MOKE technique is highly regarded by researchers.

*EUV MOKE.* By utilizing EUV to probe transitions from atomic core shells to near the Fermi level, it is possible to obtain direct element specific magnetic dynamics. Additionally, by measuring the reflectivity at resonant and off-resonant conditions for different polarizations of the EUV probe, it can directly resolve and separate spin, charge, and acoustic dynamics in the sample after excitation by a femtosecond laser. Currently, EUV is already capable of analyzing information on individual absorption edges in a completely energy-resolved manner. However, compared to transmission-type X-ray XMCD measurements, the primary drawback of reflective EUV T-MOKE measurements is the substantial photon loss (>99%). This is due to the fact that the T-MOKE signal reaches its maximum near the Brewster angle. For instance, at a 40° grazing incidence, the reflectivity of pure Ni at 66 eV (the Ni $M$-edge) is only $10^{-3}$ [135]. Recently, EUV T-MOKE has been employed to confirm optically induced intersite spin transfer (OISTR) [98, 136-139], which will be discussed in detail in Section 4.2.3.

*Soft X-ray MOKE.* Unlike optical MOKE, soft X-ray MOKE requires careful selection of optical components. This is because, in the soft X-ray range, the transmission of optical materials is extremely low, or the extinction coefficient is very high, and the refractive index of soft X-rays is approximately 1. As a result, the intensity of reflected light is significantly reduced. Therefore, reflective optical components are typically employed. Furthermore, X-rays interact with the electronic states in the core levels of matter, such as the $2p$ or $3p$ absorption edges in $3d$ transition metals. The core electrons have a certain probability of transitioning to the valence hole bands, with the transition probability following Fermi's Golden Rule [132]. Compared to the electronic states near the Fermi level in magnetic metals, these core-level states are more localized to individual atoms, leading to a resonance effect in the MOKE signal.

However, it is essential to consider the influence of nonlinear effects and signal crosstalk, where signal crosstalk primarily occurs when EUV is used as a probe pulse. The following sections will introduce the nonlinear effects and signal crosstalk in detail.

*Influence of nonlinear effects.* As the light intensity increases, photons with higher energy interact nonlinearly with the sample. The nonlinear effects observed at the Co $L_3$ edge are attributed to stimulated emission [140-142]. In 2019, Higley *et al.* [143]

experimentally demonstrated that X-ray absorption of Co/Pd multilayers exhibits a nonlinear change with increasing energy. They further showed that the nonlinear response of X-ray spectra occurs at lower energies. Additionally, due to the sensitivity of T-MOKE to both the absorptive and dispersive properties of EUV photons, the relationship between the EUV T-MOKE signal and magnetization is more complex. Recently, Jana *et al*. [123] found that the reflectivity of Fe films displays a significant asymmetry when the Fe film is magnetized in opposite directions, and this asymmetric signal also exhibits a nonlinear relationship with the EUV photon energy. However, this phenomenon was not observed in Ni films. Therefore, they attributed the phenomenon to the magnetic excitations of the material (such as the type of excitation of the spin system) or to the photon energy of the probe pulse. In subsequent experiments [98], researchers preferred to conduct systematic studies using different photon probe energies. More information on nonlinear effects can be found in the work by Schwartz *et al*. [144].

***Signal crosstalk.*** For XMCD, signal crosstalk often occurs when the energy intervals of the absorption edges of different elements in an alloy are small. In such cases, the energy interval is narrower than the energy width of a single MO resonance spectrum, leading to the overlap of magnetic signals from different elements. For example, in a NiFe alloy, the absorption edge energies of Fe and Ni are very close, making it challenging to distinguish between the magnetic signals of these two elements [145, 146]. In the EUV range, the energies associated with spin-orbit and exchange interactions are comparable, leading to distortions in the 3*p* electronic states of Ni, Fe, and Co. This results in an overlap of the $M_{2,3}$ absorption edges, causing signal crosstalk. As a consequence, it becomes difficult to quantitatively analyze the magnetic moments of Fe and Ni using the XMCD method. This crosstalk issue can be resolved by using the well-separated soft X-ray $L_{2,3}$ resonances. Additionally, compared to optical probes, EUV and soft X-ray sources exhibit larger intensity fluctuations. The resulting low signal-to-noise ratio can be improved by increasing the pulse repetition frequency.

In addition to the choice of probe light source, the relationship between the transient MO signal and the magnetization in TR-MOKE measurements is a topic of intense debate. Initially, the MO signal obtained from static MOKE was assumed to be linearly proportional to the magnetization, and this relationship was applied to ultrafast demagnetization experiments in TR-MOKE. However, in 2000, Koopmans *et al*.[147] questioned the validity of this linear relationship under highly non-equilibrium conditions, such as those present during ultrafast demagnetization. They used TR-MOKE to compare the changes in Kerr rotation angle and ellipticity signals in wedge-shaped Ni films. The results indicated significant differences in the changes between these two signals within the first few hundred femtoseconds. This discrepancy was attributed to the distribution of non-equilibrium hot electrons following intense excitation of the material, resulting in dichroic bleaching effects. Kampfrath *et al.* [148] also confirmed the presence of a significant non-magnetic contribution to the MO signal in Fe films. They further demonstrated that this non-magnetic contribution, caused by the MO strain effect, could persist for tens of picoseconds [149].

In 2009, theoretical work by Zhang *et al.* [150] demonstrated that the ratio of the optical response to the magnetic response in the MO effect depends on the energy of the incident photon. However, Carva *et al.* [151] argued that this study only accounted for the magnetization changes directly induced by laser-excited electrons, without considering the perturbations caused by the laser itself. They posited that the first-order magnetic response of the simulated system (crystals with inversion symmetry) reported by Zhang *et al.* [150] should be zero.

In addition, Wieczorek *et al.* [99] demonstrated that, in the study of multilayer films, the MO signal cannot be simply considered as the average value within the laser penetration depth. This complexity arises from multiple light reflections at the interfaces and subsequent interference, which modifies the spectral line shape. As a result, the ellipticity of the light and the Kerr angle are weighted differently at various depths within the sample. Razdolski *et al.*[152] later confirmed that the MO signal is highly dependent on the thickness of the ferromagnetic layer and the properties of the substrate.

To address the above issues, several methods have been proposed, including (i) introducing Au nanosheets or nanoparticles to enhance the excitation of the plasmon mode in magnetic materials [153], or embedding the magnetic layer into an extreme anti-reflection platform (composed of two dielectric layers and a reflective layer) [117] to enhance the MO signal; (ii) employing multiple complementary pump-probe techniques simultaneously to differentiate and eliminate non-magnetic interference; (iii) utilizing generalized magneto-optical ellipsometry to extract the magnetic contribution from the MO response [154]. The combination of generalized magneto-optical ellipsometry and TR-MOKE measurements is anticipated to resolve this controversy.

## 2.2 Terahertz range techniques

Since the natural frequencies of many condensed matter systems fall within the terahertz range, THz waves are extensively employed to investigate spin dynamics and carrier behaviors in antiferromagnetic materials [155, 156], and multiferroic materials [155-158], such as the nonequilibrium carrier distribution, the conduction electron-photon interaction, and the low-energy magnetic excitations. The interaction of terahertz waves with samples can be described using conventional crystal optics, which outline the electromagnetic polarization properties of specific spins and their orientation relative to an external magnetic field. As a result, the THz probe technique has emerged as a novel method for studying carrier dynamics.

Initially, researchers employed pulsed lasers to irradiate lithium niobate crystals, generating terahertz pulses [159, 160]. Subsequently, Auston *et al*. (1984) [161] utilized the electro-optic sampling (EOS) method to achieve coherent detection of terahertz waves, successfully obtaining the time-domain waveform of these waves. This advancement led to the development of terahertz time-domain spectroscopy (THz-TDS) based on EOS, which was later applied to the study of spintronics.

In 2004, Beaurepaire *et al*. [46] first employed THz-TDS to investigate the ultrafast demagnetization process in magnetic metal films. Their findings demonstrated that when the surface of the film is irradiated by a femtosecond pulse laser, the magnetization decreases within a sub-picosecond time frame, leading to terahertz magnetic dipole emission. At this point, the electric field radiated in the far field is proportional to the second-order derivative of the magnetization, and this electric field can be described using Maxwell's equations:

$$E_y(t) = \frac{\mu_0}{4\pi^2 r} \frac{\partial^2 M_x}{\partial t^2}\left(t - \frac{r}{c}\right) \qquad (3)$$

Subsequently, a series of studies on THz radiation generated by metal films were conducted, such as the time evolution of THz conductivity and quasiparticle (e.g., phonon or magnon) population dynamics after light excitation. In recent years, Zhang *et al*. have focused on utilizing THz-TDS to reconstruct magnetization dynamics [57, 162-164]. They were the first to simultaneously observe ultrafast demagnetization driven by hot electrons and coherent phonons in Fe films. Rouzegar *et al*. (2022) [165] employed THz-TDS to measure laser-induced terahertz spin transport and ultrafast demagnetization in CoFe and NiFe films, evaluating the temporal evolution of these two processes.

Additionally, THz waves can be used as probe pulses in the pump-probe technique, specifically in the optical pump THz probe (OPTP) technique. OPTP combines the optical pump-probe method with THz time-domain spectroscopy (THz-TDS). The primary difference between OPTP and THz-TDS is that the OPTP system includes an additional sample excitation pump pulse. In OPTP, the ultrafast laser pulse is split into three beams: the sample excitation pump, the THz generation pump, and the THz probe beam. This setup allows for scanning THz time-domain spectroscopy at a fixed delay between the optical pump and the THz probe, as well as observing changes in THz transmission intensity at different delay times between the optical pump and the THz probe [166]. By analyzing THz emission spectroscopy from OPTP and THz-TDS, THz conductivity and dielectric function can be determined [167]. OPTP provides transient THz conductivity on timescales ranging from sub-picoseconds to nanoseconds, offering insights into carrier properties, mobility, and time-dependence. Additionally, THz emission spectroscopy can provide information on spin-to-charge conversion, allowing for the indirect study of the inverse spin Hall effect in heterojunctions [168].

Although the interpretation of THz spectra can be challenging and these techniques may require complex setups for the generation and detection of THz radiation, THz-TDS and OPTP undeniably serve as a bridge between terahertz spintronics and femtomagnetism. This technique is anticipated to be utilized for detecting spin transport and ultrafast demagnetization in antiferromagnetic materials in the future.

**2.3 Second-harmonic generation (SHG)**

Unlike MOKE, the second-harmonic generation (SHG) is a nonlinear optical effect. SHG was first discovered by Franken *et al.* in 1961, marking the beginning of nonlinear optical research [169]. In nonlinear optics, when an electromagnetic field (i.e., a light field) with sufficiently high electric field intensity interacts with matter, the electric polarization *P* of the material is no longer proportional to the electric field *E*. In this scenario, the polarization vector can be described as [170]:

$$\vec{P} = \varepsilon_0 \left( \chi^{(1)} \vec{E} + \chi^{(2)} \vec{E}\vec{E} + \chi^{(3)} \vec{E}\vec{E}\vec{E} + \chi^{(4)} \vec{E}\vec{E}\vec{E}\vec{E} + \cdots \right) \quad (4)$$

where $\varepsilon_0$ represents the permittivity in vacuum, and $\chi^{(n)}$ denotes the *n*-th order magnetic susceptibility tensor of the material. The first term on the right-hand side of the equation captures the difference between the speed of light in the medium and the vacuum, while the second term accounts for radiation sources with frequencies distinct from the driving electric field.

In SHG experiments, when strong coherent light interacts with a non-centrosymmetric medium (such as a surface or an interface), the frequency of the reflected light at the interface doubles due to the non-zero second-order nonlinear magnetic susceptibility. This process generates a second harmonic signal. SHG is highly sensitive to surfaces and interfaces, allowing for the direct detection of ultrafast magnetization dynamics at these locations [44, 72, 171, 172]. The spin-dependent effects in SHG can be classified into even and odd SH fields. The sign of the even SH field is independent of the magnetization and sensitive to the structural properties of the sample, while the sign of the odd SH field is directly dependent on the orientation of the sample's magnetization [173]. In metallic ferromagnets, the even SH field is typically larger than the odd SH field [174]. Consequently, to accurately extract the signal of the odd SH field, it is necessary to measure the second harmonic intensity of the sample under opposite magnetization conditions. This allows for the analysis of magnetization dynamics using the SHG data. It is worth noting that the SHG signal can be significantly influenced by surface and interface effects, necessitating meticulous attention during data analysis and interpretation. Chen *et al*. [175] employed the interference effects from various interfaces in the system to examine the spatially inhomogeneous magnetization dynamics induced by the propagation of spin currents within the system.

The combination of SHG and X-ray techniques demonstrates significant potential. Researchers can obtain element-specific information about samples and investigate element-specific ultrafast magnetization dynamics [176].

### 2.4 Two-photon photoemission (TPPE)

Two-photon photoemission (TPPE) is a powerful technique for investigating the electronic structure and electron dynamics following laser excitation. Based on the photoelectric effect, when the first pulse laser (pump) is incident on the material, electrons below the Fermi level are excited to an unoccupied state in the TPPE experiment. If the lifetime of this excited state is sufficiently long, the second pulse laser (probe) will excite the electrons from the excited state to above the vacuum energy level. By detecting the emission angle and kinetic energy of the emitted electrons, information about the excited (intermediate) state can be obtained. Ultraviolet pulses are generally used for the probe pulse of TPPE, and this technique is sensitive to the surface of the sample and is suitable for studying the dynamic behavior of carriers on surfaces and interfaces. However, a disadvantage of this technique is its requirement for high surface quality.

Initially, Scholl *et al*. [45] used TPPE to investigate demagnetization in Ni thin films induced by Stoner excitations on a sub-picosecond timescale and by phonon-magnon scattering on a timescale of hundreds of picoseconds. In the experiment, utilizing a narrow pulse laser enabled researchers to obtain information about the depopulation time of the intermediate state of the sample, thereby determining the average lifetime of excited electrons in a specific energy band state [177, 178]. Currently, TPPE is widely used to study ultrafast electron dynamics, including electronic bands, surface states, and bonding and anti-bonding states generated by the chemisorption of atoms and molecules on solid surfaces [179]. With the development of the new generation of synchrotrons, the combination of XMCD and TPPE techniques holds significant potential for detecting both spin and orbital momentum. This technique has the potential to distinguish between electron and spin dynamics.

### 2.5 The X-ray techniques

Since X-ray pulses have low mass energy and cause minimal impact on the energy levels and structure of the sample, this technique can non-destructively reveal the internal structure of the object. Utilizing X-rays as a probe can provide element-specific resolution measurement and allow for the detection of intrinsic quantities such as spin transfer between different atoms and contributions from both spin and orbital moments. Common X-ray techniques are introduced below.

***X-ray Magnetic Circular Dichroism (XMCD)***. When a magnetic field is applied to a sample, the incidence of the circularly polarized X-rays excites core electrons into the valence hole bands. At this point, the sample exhibits different absorption characteristics for left-handed and right-handed circularly polarized X-rays. This phenomenon is referred to as XMCD. XMCD is element-sensitive, allowing researchers to track the magnetic moments of specific atomic species (e.g., Fe, Co, Ni) in materials. This capability is critical in studying the distinct behavior of different elements during ultrafast demagnetization, especially in complex materials like alloys and multilayers.

In addition, XMCD is sensitive to both spin and orbital magnetic moments. It can reveal the changes in the spin and orbital contributions, and track how spin-orbit coupling contributes to ultrafast demagnetization. For ferromagnetic materials, a relatively weak saturation field is often sufficient to achieve the necessary magnetic alignment for XMCD measurements. The magneto-optical sum rule [180, 181] links the integration of the magneto-circular dichroism (MCD) signal over the relevant absorption edge with the spin magnetic moment ($m_s$) and the orbital magnetic moment ($m_o$). When applied to XMCD or X-ray absorption spectroscopy (XAS) spectra, this rule enables the quantitative analysis of $m_s$ and $m_o$ of the sample. In 1995, Chen *et al.* [182] confirmed the applicability of this rule to itinerant magnetic systems, specifically defining the sum rules for 3d transition metals as follows:

$$m_o = -\frac{4q(10-n_{3d})}{3r},$$

$$m_s = -\frac{6p-4q}{r} \times (10-n_{3d})\left(1+\frac{7\langle T_z \rangle}{2\langle S_z \rangle}\right) \qquad (5)$$

$$p = \int_{L_3} (\mu_+ - \mu_-)d\omega,$$

$$q = \int_{L_3+L_2} (\mu_+ - \mu_-)d\omega, \qquad (6)$$

$$r = \int_{L_3+L_2} (\mu_+ + \mu_-)d\omega,$$

where $m_o$ and $m_s$ are the orbital and spin magnetic moments in units of $\mu_B$/atom, and $n_{3d}$ is the *3d* electron occupation number of the specific transition metal atom. The $L_3$ and $L_2$ denote the integration ranges. <$T_z$> is the expectation value of the magnetic dipole operator, and <$S_z$> is half of $m_s$ in Hartree atomic units. The sum rules for RE magnetic systems could be found in the Refs. [180, 183, 184]. However, whether the magnetic moments derived from these response functions are consistent with the fundamental magnetic moments of the material remains a topic for further discussion. [185, 186].

XMCD is also highly effective in magnetic imaging technology, achieving spatial resolutions down to tens of nanometers. Based on synchrotron radiation facilities, the development of femtosecond X-ray sources [187, 188] and X-ray free-electron lasers [189, 190] has advanced X-ray techniques. It is now used to investigate the ultrafast spin dynamics in multicomponent magnetic materials and heterostructures, capturing the transient behaviors of various elements or layers [59, 191, 192]. This technique is also exceptionally sensitive to variations in electron density distribution, allowing for the determination of material parameters such as the shape, size, and spacing of microscopic structural units [191].

*Time- and angle-resolved photoemission spectroscopy* (*TR-ARPES*). TR-ARPES is sensitive to ferromagnetic order on the surfaces of materials. It can directly extract transient electron temperatures by detecting the ordered electronic structure (including band shifts and broadening caused by electron-electron, electron-phonon, and electron-spin interactions) and the electron dispersion relations of crystalline materials. Additionally, it provides information related to the coupling between different degrees of freedom. Tengdin *et al.* [26] employed this technique to observe the collapse of exchange splitting in Ni thin films, which occurred almost simultaneously with the demagnetization of Ni on a timescale of approximately 176 fs. They demonstrated that the energy absorbed by the electrons was transferred to the spin system within about 20 fs, elucidating the timescale of the non-equilibrium phase transition of the material. This technique has also advanced the research on ultrafast spin dynamics in antiferromagnetic materials. Lee *et al.* [193] combined TR-ARPES and Time-resolved X-ray resonant magnetic scattering (TR-XRMS) techniques to investigate the femtosecond dynamics of surface ferromagnetic order and bulk antiferromagnetic order in the two-dimensional antiferromagnetic material $GdRh_2Si_2$. Their results showed that the similar dynamical behaviors of these orders indicate a strong Ruderman-Kittel-Kasuya-Yosida (RKKY) exchange interaction between the localized 4*f* electrons of Gd and the 5*d*6*s* itinerant conduction electrons of Rh, Si, and Gd. However, as the development of ultrafast spin dynamics research on antiferromagnetic materials is not the primary focus of this review, it will not be discussed in detail here.

*Time-resolved X-ray resonant magnetic scattering* (*TR-XRMS*). TR-XRMS combines the capabilities of electron absorption spectroscopy and electron diffraction. This technique offers extremely high sensitivity to electronic order and can be utilized to study the temporal evolution of the phase space of spin, orbital, and electronic order. It is possible to separate the spin dynamics determined by the electronic structure from the contributions related to the lattice dynamics [194]. TR-XRMS is frequently used to detect magnetic structures associated with specific elements in Fourier space. However, this technique typically offers only average information on the overall magnetic structure. Vodungbo *et al.* [54, 195] were the first to use soft X-ray femtosecond pulses to image magnetic structures during ultrafast demagnetization, achieving spatial resolutions at the nanometer scale. Leveille *et al.* [50] observed a transient process in which homochiral Néel-type domain walls in [Pt(3 nm)/Co(1.5 nm)/Al(1.4 nm)]$_{\times 5}$ multilayer film structures transformed into a mixed Bloch-Néel-Bloch state after ultrafast demagnetization. They attributed this transformation to a transient spin current within the domain, which flowed towards the domain wall and exerted torque on the spins within it. It is worth noting that small-angle X-ray scattering (SAXS) is also a type of X-ray magnetic scattering (XRMS), which is sensitive to variations in electron density over a range of 1 to 100 nm in the material. In SAXS, the angle of elastic X-ray scattering is typically very small, ranging between 0.1° and 10°. When SAXS is used for magnetic imaging, it can achieve a spatial resolution of 9-11 Å. Numerous studies have employed SAXS to investigate ultrafast spin dynamics [59, 131, 196, 197].

***Time-resolved resonant inelastic X-ray scattering (TR-RIXS)***. TR-RIXS is an emerging spectroscopic technique designed to probe nonequilibrium collective modes with momentum resolution, such as magnons (spin waves). Due to its long probing depth (> 100 nm), TR-RIXS is a bulk sensitive photon-in, photon-out technique that is insensitive to the surface conditions of the sample. This method investigates non-equilibrium dynamics by scattering ultrashort X-ray pulses, which are tuned to a specific atomic absorption edge. When an X-ray photon is absorbed, the incident energy is tuned to resonate with a specific element. Simultaneously, the material excited by the pump transitions to an intermediate state, wherein a core-level electron is promoted to or above the valence orbitals. This excited state is unstable and decays within a few femtoseconds, during which a valence electron fills the core hole, emitting an X-ray photon in the process. By analyzing the scattered X-rays in terms of momentum and energy, the dynamic behavior of the material can be obtained. When the incident photon energy, momentum, and polarization are set to the appropriate values, the resonance condition of photons can enhance the scattering process. At this point, the energy transferred to the atom is sufficient to flip the electron spin. Since the energy required for spin flipping is proportional to the strength of the exchange field, the magnitude of the exchange interaction can be determined using the linearized spin wave theory of the Heisenberg model [198]. In general, TR-RIXS is sensitive to various excitations such as charge, orbital, and spin changes in the valence electrons [199].

***Time-Resolved X-ray Photoemission Electron Microscopy (TR-X-PEEM)***. TR-PEEM is a photon-in, electron-out technique derived from time-resolved two-photon photoemission (TR-TPPE). After the sample is excited by photons (pump pulse), electrons are excited from the ground state to the excited state. Subsequently, the probe pulse causes electrons to be emitted from the surface, which are then focused by an objective lens. Although TR-TPPE does not directly measure optical transitions, it offers sub-femtosecond temporal resolution, nanometer spatial resolution, and tens to hundreds of millielectron volt energy resolution. This enables the simultaneous capture of the dynamic evolution of electrons in terms of time, space, and energy, providing a direct method for imaging electron dynamics at the nanoscale [200]. The information obtained in a TR-TPPE experiment strongly depends on the choice of probe energy. When X-ray pulses are used as the probe light, the energy range in TR-X-PEEM typically falls between 100 and 2000 eV (soft X-ray range), which is sufficient to cover the absorption edges of transition metals (Fe, Co, Ni) and rare earth elements (Gd and Tb). The XMCD effect at the $L_{2,3}$ edge of transition metals enables direct imaging of magnetic domain dynamics, making this one of the important applications of the technique [201-203].

*Ultrafast X-ray diffuse scattering* (*UXDS*). UXDS can detect changes in lattice disorder and structural dynamics during the demagnetization process, providing insight into how lattice vibrations (phonons) evolve as the material absorbs energy. Experiments have shown that the lattice remains in a nonequilibrium state for hundreds of picoseconds [204]. To distinguish the nonequilibrium dynamics of electrons and phonons in the Brillouin zone, it is necessary to accurately measure nonthermal, momentum-dependent phonon populations on a sub-picosecond timescale. X-ray scattering is particularly sensitive to short-wavelength phonons, which reduce the intensity of Bragg peaks and enhance weak diffuse scattering between these peaks. Additionally, short-wavelength phonons can create structures in the diffuse scattering background [205]. By analyzing the energy and momentum of the inelastically scattered photons in UXDS measurements, researchers can obtain more comprehensive information on nonequilibrium phonon dynamics, which also aids in understanding the ultrafast Einstein–de Haas effect [52]. However, when selecting the X-ray energy and crystal orientation, it is important to avoid Bragg reflections on the detector [205].

## 2.6 Ultrafast electron diffuse scattering (UEDS)

In addition to the diffuse ultrafast X-ray diffraction described in Section 2.4, ultrafast electron diffuse scattering (UEDS) can also be used to obtain information about nonthermal, momentum-dependent phonon populations. Compared to X-rays, electrons have a larger scattering cross-section, making UEDS particularly advantageous for studying thin film samples, layered 2D materials, surfaces, and radiation-sensitive samples.

UEDS is a recently emerging technique that provides momentum-resolved information on electron-phonon and phonon-phonon coupling strength across the entire Brillouin zone, aiding in the study of phonon dynamics [206-208]. UEDS, like other pump-probe techniques, uses femtosecond lasers as the pump light, allowing for the selective excitation of specific degrees of freedom, such as carriers or phonons. After excitation, an ultrashort electron pulse is scattered through the sample, forming a scattering pattern. By analyzing the time-dependent diffuse scattering signals between the diffraction peaks in the UEDS pattern, one can obtain instantaneous crystal structure information and track the evolution of phonon mode amplitude at all momenta throughout the Brillouin zone.

According to the Debye-Waller effect [209], the incoherent motion of atoms near their equilibrium positions after laser excitation alters the diffraction intensities. An increase in lattice temperature reduces the intensity of the Bragg scattering peaks. Thus, the time evolution of scattering intensity provides momentum-resolved information about non-equilibrium phonon groups. The diffuse intensity in reciprocal space can be described by the following equation [207, 208]:

$$I(\boldsymbol{q}) = \sum_j \frac{n_j(\boldsymbol{q}) + 1/2}{\omega_j(\boldsymbol{q})} |F_j(\boldsymbol{q})|^2 \tag{7}$$

$$|F_j(\boldsymbol{q})|^2 = \left| \sum_s \exp(-W_s(\boldsymbol{q})) \frac{f_s(\boldsymbol{q})}{\sqrt{\mu_s}} (\boldsymbol{q} \cdot \boldsymbol{e}_{j,s,k}) \right|^2 \tag{8}$$

where $n_j(\boldsymbol{q})$, $\omega_j(\boldsymbol{q})$, $F_j(\boldsymbol{q})$, and $f_s$ represent the occupancy of phonon modes, the mode frequencies, the one-phonon structure factor, and the atomic scattering factor, respectively. $\boldsymbol{e}_{j,s,k}$ denotes the polarization vectors, and $s$, $j$, and $k$ correspond to the basis atom, the phonon mode in branch, and the momentum/wavevector, respectively. $k$ depends on the position of the nearest Bragg peak, $\mathbf{H}$, where $\boldsymbol{q} = \mathbf{H} - \boldsymbol{k}$. $\mu_s$ represents the atomic mass, and $W_s$ is the Debye-Waller factor. Therefore, the above formula allows for the direct determination of changes in phonon amplitude $n_j(\boldsymbol{q})/\omega_j(\boldsymbol{q})$ for all phonon wavevectors in the Brillouin zone.

Although UEDS does not have energy-resolving capabilities, the information it provides on ultrafast phonon dynamics is similar to that provided by TR-ARPES. However, TR-ARPES is primarily used to observe electron dynamics.

**2.7 Brief summary of the pump-probe techniques**

Chapter 2 primarily introduces several pump-probe techniques used to study the ultrafast dynamics of ferromagnetic materials, including TR-MOKE, terahertz range techniques (THz-TDs and OPTP), SHG, TPPE, X-ray techniques (XMCD, TR-ARPES, TR-XRMS, TR-RIXS, TR-X-PEEM, and UXDS), and UEDS. Each of these pump-probe techniques plays a distinct role in investigating ultrafast dynamics, as illustrated in Fig. 4 (a). The corresponding dynamic processes [210] occurring at different timescales are also shown in Fig. 4 (b).

TR-MOKE is a powerful technique for analyzing magnetization dynamics in materials on ultrafast timescales. By using a probe source with different wavelengths, researchers can extract various types of information from the sample. For instance, TR-MOKE with an optical probe can detect spin dynamics following laser excitation, whereas using ultraviolet or X-ray probes offers higher resolution and enables element-specific spin dynamics analysis. Due to its ability to probe various intrinsic properties, EUV and X-ray TR-MOKE have been employed to study the optical intersite spin-transfer (OISTR) effect [136].

Since the intrinsic frequencies of many condensed matter systems lie in the THz range, both OPTP and THz-TDS are used to probe quasiparticle (e.g., carrier or magnon) population dynamics. THz waves can be used as probe pulses in the optical-pump terahertz-probe (OPTP) technique, which combines pump-probe methods with THz-TDS. By analyzing the THz conductivity obtained from OPTP and THz-TDS, one can gain deeper insights into carrier dynamics. This technique also provides information related to spin-to-charge conversion, enabling the indirect study of the inverse spin Hall effect in heterostructures.

Unlike the MOKE, SHG is a nonlinear optical effect sensitive to surfaces and interfaces, SHG can reveal ultrafast magnetization dynamics through analysis of SH fields, though surface and interface effects can influence the signal. Furthermore, TPPE allows researchers to obtain information about the excited (intermediate) state by detecting the emission angle and kinetic energy of the emitted electrons. As a result, TPPE is widely used to study carrier dynamics at surfaces and interfaces of samples. In the future, the combination of XMCD and TPPE techniques is expected to enable the detection of both spin and orbital momentum.

X-ray techniques such as XMCD, TR-ARPES, TR-XRMS, TR-RIXS, and TR-X-PEEM offer significant advantages, including element specificity, high penetration depth, and nanometer spatial resolution in the femtosecond regime. These methods provide detailed information on charge, orbital, and spin excitations with high sensitivity. Notably, XMCD is sensitive to both spin and orbital magnetic moments. It can measure how these moments evolve over time, revealing changes in the spin and orbital contributions to the overall magnetization as the system undergoes demagnetization. In spin-polarized materials, TR-ARPES can track changes in the spin-dependent electronic band structure and capture the transient population of excited electrons (non-thermal distributions) immediately after optical excitation. TR-XRMS is capable of tracking the spatial and temporal evolution of the magnetic domain structure during ultrafast demagnetization. TR-RIXS is sensitive to collective spin excitations (magnons) and can detect changes in magnon spectra and their relaxation timescales. This technique also provides insights into how energy is transferred between the electron and lattice systems during demagnetization. TR-PEEM is a surface-sensitive technique that probes the spatial distribution of magnetization with nanometer-scale spatial resolution and femtosecond temporal resolution. It captures the dynamic evolution of electrons in terms of time, space, and energy, and when combined with X-rays (TR-X-PEEM), can be used to image magnetic domain structures.

In recent years, laser-induced lattice strain effect has gained increasing interest from researchers, with UXDS and UEDS emerging as powerful tools for studying lattice dynamics. UXDS measures nonthermal phonon populations on a sub-picosecond timescale by analyzing diffuse scattering between Bragg peaks, while UEDS, a more recent technique, analyzes momentum-dependent phonon populations and electron-phonon interactions across the Brillouin zone. Both UXDS and UEDS enhance our understanding of phonon dynamics and ultrafast lattice effects, providing valuable insights into the nonequilibrium states of materials.

Overall, these advanced techniques provide comprehensive tools for probing ultrafast dynamics in materials, providing critical insights into electron, spin, and phonon dynamics that are essential for understanding fundamental processes in condensed matter physics.

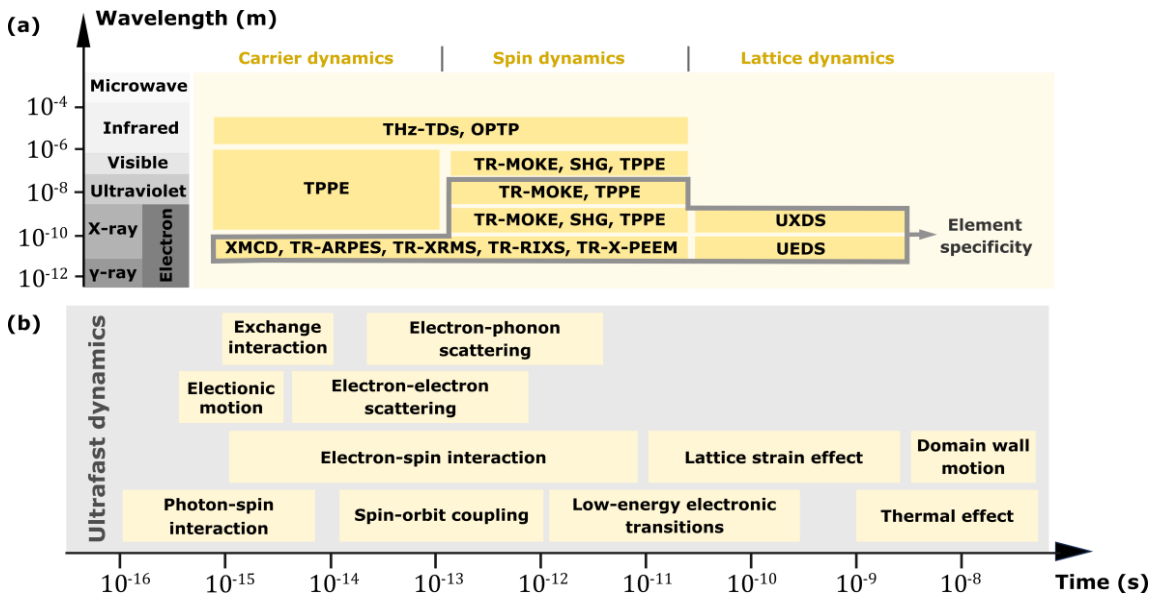

Fig. 4 (a) The wavelength range of probes used in various pump-probe techniques and the corresponding carrier, spin, or lattice dynamics they can detect. (b) The timescale of ultrafast dynamics in ferromagnetic materials.

## 3. Magnetization dynamics and ultrafast demagnetization model

In this chapter, we primarily explore the physical processes related to ultrafast magnetization dynamics. Section 3.1 elaborates on the magnetization precession model of magnetic materials, while Section 3.2 introduces seven models of microscopic ultrafast demagnetization.

### 3.1 The process of magnetization precession

When a laser irradiates a magnetic material, the magnetic moment in the material undergoes precession due to the disturbance of the pulsed laser. The Landau-Liftshitz-Gilbert (LLG) equation, which accounts for both precessional motion and damping effects in magnetization dynamics, is essential for comprehending the fundamental origins underlying magnetic behavior at the microscale. The equation accurately models the time evolution of magnetization in materials, which is important for understanding how magnetization responds to ultrafast laser pulses. Ultrafast demagnetization processes are significantly influenced by damping mechanisms, which describe how quickly a magnetic system returns to its equilibrium state after being disturbed. The damping term in the LLG equation is crucial for modeling these relaxation processes accurately, enabling a deeper understanding of how energy and angular momentum are dissipated in magnetic materials following excitation.

However, the LLG equation is suited for low-temperature regimes where thermal effects are negligible, and the magnitude of magnetization is conserved. It is important to note that the effective field in the LLG equation can include contributions from thermal fluctuations, exchange interactions, and other quantum effects relevant at ultrafast timescales. This flexibility makes the LLG equation a versatile tool for studying demagnetization processes that are influenced by a complex interplay of physical effects.

The LLG equation is fundamental to the study of magnetization dynamics, providing a foundation for our exploration of thermal effects and microstructure which are more comprehensively described by the Landau-Lifshitz-Bloch (LLB) equation. The LLB equation extends the LLG model to include thermal effects and is particularly useful for describing magnetization dynamics near and above the Curie temperature, where thermal fluctuations become significant, and the magnitude of magnetization can change.

The introduction of the LLG and LLB equations is provided below.

### 3.1.1 Landau-Liftshitz-Gilbert (LLG) equation

The relation between the magnetization $M$ and angular momentum $J$ in a ferromagnet can be expressed as follows:

$$\mu_0 M = -\gamma J \tag{9}$$

where $\gamma = g\dfrac{\mu_0 e}{2m}$ is the gyromagnetic ratio and $g$ is the Landé factor.

When an external magnetic field $H_{app}$ or internal field exerts on the magnetization $M$ of a system, the magnetization precesses around the direction of the effective magnetic field $H_{eff}$. The effective field $H_{eff}$ should typically composed of several contributions, such as the exchange field $H_{ex}$, anisotropic field $H_k$, and demagnetization field $H_d$, and possibly other internal fields,

$$H_{eff} = H_{app} + H_{ex} + H_k + H_d + \cdots \tag{10}$$

These contributions of $H_{eff}$ are physical fields that directly influence the magnetization. Thus, the torque $L$ is a result of the cross product between the magnetization $M$ and the effective field $H_{eff}$, described by the equation:

$$L = \mu_0 M \times H_{eff} \tag{11}$$

The $L$ in equation (11) can be expressed more precisely as the rate of change of angular momentum over time:

$$\frac{dJ}{dt} = \mu_0 M \times H_{eff} \tag{12}$$

According to equation (9), equation (12) can be further written as the variation ratio of magnetization over time:

$$\frac{dM}{dt} = -\gamma M \times H_{eff} \tag{13}$$

Equation (13) states that once the equilibrium configuration is disturbed, the magnetization $M$ precesses around the effective magnetic field $H_{eff}$ at a constant cone angle without reaching the equilibrium position (see Fig. 5a). However, this contradicts experimental observations, which indicate that the magnetic moment reaches equilibrium after a finite period of time. To account for this observation, an additional component needs to be incorporated into the precessional term, i.e., the damping term, which aligns with the effective field (see Fig. 5b). The damping term was independently formulated by Landau-Liftshitz [211] and Gilbert [212]. The final expression is commonly referred to as the Landau-Lifshitz-Gilbert (LLG) equation.

$$\frac{dM}{dt} = -\gamma M \times H_{eff} + \frac{\alpha}{M} M \times \frac{dM}{dt} \tag{14}$$

where $\alpha$ represents the damping factor. In femtosecond laser-induced magnetization dynamics, the laser pulse initiates the perturbation of the effective field of the ferromagnetic material, subsequently causing the precession of the magnetic moment around its effective field. By incorporating effective magnetic fields that account for various physical effects, the LLG equation allows researchers to connect fundamental magnetic interactions with the ultrafast demagnetization behavior observed in experiments.

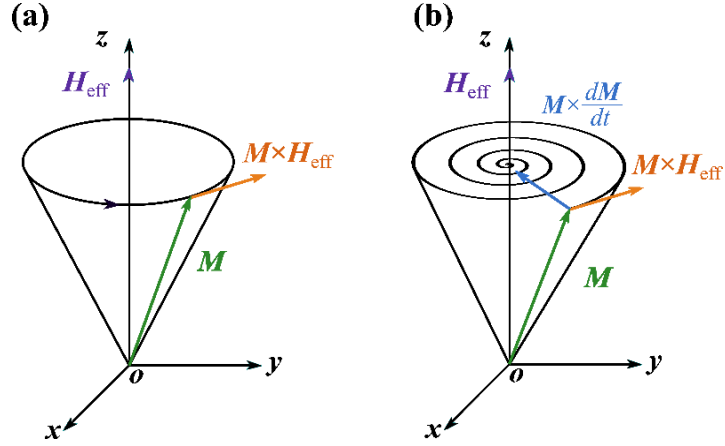

Fig. 5. (a) The magnetic moment *M* precesses around the effective magnetic field Heff without any damping. The torque acting on the moment is given by Equation (13). (b) When introducing damping, the magnetization follows a helical trajectory back to its equilibrium position, as described by Equation (14).

### 3.1.2. Landau-Lifshitz-Bloch (LLB) equation

The classical LLG equation provides a robust framework for understanding the magnetization dynamics of magnetic materials, especially at low temperatures. However, when considering laser-induced magnetization dynamics, thermal fluctuations become significant and cannot be neglected. In such cases, the LLG equation struggles to accurately describe longitudinal relaxation processes. Therefore, it is essential to extend the LLG equation to address the complexities of high-temperature scenarios. The Landau-Lifshitz-Bloch (LLB) equation emerges as a crucial alternative for characterizing micromagnetic phenomena.

Combining the low-temperature LLG model with the high-temperature Bloch equation [213], Garanin [214] proposed the LLB equation to account for both the transverse and longitudinal components of magnetization relaxation. This approach encompasses magnetization dynamics in ferromagnetic materials across the entire temperature range. In contrast to the LLG equation, the LLB equation considers temperature-dependent changes in magnetization and includes the previously neglected longitudinal magnetization relaxation [215].

To simulate the changes in magnetization direction and strength during magnetization dynamics, Garanin proposed a semi-phenomenological model applicable across the entire temperature range [214, 216, 217]. In the following sections, we will derive this model from the perspective of micromagnetism.

First, the spin dynamics of a single magnetic moment *s* under an applied magnetic field is described by:

$$\frac{dS}{dt} = -\gamma S \times (H + \zeta) - \gamma\lambda [S \times (S \times H)] \tag{15}$$

where $S = s/s_0$ is the normalized spin vector, $s_0$ is the atomic magnetic moment. $\gamma$ is the gyromagnetic ratio, $\lambda$ is a damping parameter characterizing the coupling between spins and the heat bath, and $\zeta = (\zeta_1, \zeta_2, \zeta_3)$ is the Langevin field represents thermal fluctuations in the local field $H$ where the spin *s* moves. The term $\zeta(t)$ is given by the following equation:

$$\langle \zeta_i(t)\zeta_j(t') \rangle = \frac{2\lambda k_B T}{\gamma s_0} \delta(t-t')\delta_{ij}, \tag{16}$$

where *i* and *j* are Cartesian components. The basis of this equation is the separation of timescales, assuming that the spin system acts slower than the phonon or electron system. In this case, the degrees of freedom of the phonon or electron system can be averaged out and replaced by a stochastic field with white noise ($\langle \zeta_i(t) \rangle = 0$) correlation functions. By using stochastic methods and combining the Langevin field fluctuations described in Eq. (16), the Fokker-Planck equation corresponding to Eq. (15) is derived. The magnetization of the material is defined as $M = s_0 \langle S \rangle$. According to Eq. (15), the LLB equation in a magnetic material can be written as follows [218-220]:

$$\frac{dM}{dt} = -\gamma [M \times H_{eff}] + \frac{\gamma\alpha_{\|}}{M^2}(M \cdot H_{eff})M - \frac{\gamma\alpha_{\perp}}{M^2}[M \times (M \times H_{eff})] \tag{17}$$

where the first term corresponds to magnetization precession, while the second and third terms respectively denote longitudinal and transverse magnetization relaxation. The parameters $\alpha_{\|}$ and $\alpha_{\perp}$ represent the temperature-dependent longitudinal and transverse damping, respectively. Atxitia *et al.* [218] demonstrated that when the spin value approaches infinity ($S \to \infty$), $\alpha_{\|}$ and $\alpha_{\perp}$ can be denoted as $\alpha_{\|} = \frac{2\lambda T}{3m_e T_c}$ and $\alpha_{\perp} = \frac{\lambda}{m_e}\left(1 - \frac{T}{3T_c}\right)$, and the LLB equation is then equivalent to an ensemble of exchange-coupled atomistic spins modeled by stochastic LLG equations [220, 221]. In this context, $\lambda$ is directly proportional to the spin-flipping rate. The choice of this parameter $\lambda$ significantly influences the model of ultrafast demagnetization and re-magnetization [218]. If the spin value is $S=1/2$, the LLB equation is equivalent to the M3TM model (which will be discussed in Section 3.2.6) proposed by Koopmans *et al.* [39].

Since the LLB equation can elucidate magnetization dynamics at any temperature and bridges atomic simulation with micromagnetic simulation, it has found applications in the field of multiscale simulations [220]. Consequently, the LLB equation offers a crucial micromagnetic simulation approach for research on ultrafast demagnetization. Nevertheless, a notable shortcoming of both the LLG equation and the LLB equation is the absence of due consideration of spin-electron coupling.

## 3.2 Microscopic model of ultrafast demagnetization

The process of laser-induced ultrafast demagnetization involves multiple mechanisms, including photon-electron interactions, electron-electron scattering, electron-phonon coupling, electron-magnon scattering, non-thermal electronic distribution, spin transport, and laser-induced lattice strain. Over the years, various models and theories have been proposed to elucidate the physical phenomena underlying ultrafast demagnetization.. Among these, temperature-based models have been widely employed. When combined with micromagnetic simulations or density functional theory (DFT), these models have successfully replicated the ultrafast demagnetization dynamics in numerous systems and provided reasonably accurate explanations. However, these models still face challenges in offering a fundamental understanding of ultrafast demagnetization. Therefore, it remains necessary to utilize diverse experimental methods (refer to Chapter 2) to observe the dissipation of angular momentum. Nonetheless, it is essential to discuss several temperature models.

Current temperature models arise from the inseparable interaction among three subsystems: electrons, lattices, and spins. Nevertheless, each model considers different specific interactions. For instance, the 2TM, 3TM, E3TM, MMTM, and 4TM models predominantly focus on energy conversion among the three subsystems. Among these models, the 2TM, 3TM, and 4TM models assume that the temperature of each subsystem is always in an internal thermalization state, whereas the E3TM and MMTM models account for the energy transfer of non-thermal electrons and non-thermal phonons, respectively. Additionally, the M3TM and EM3TM models consider both energy and angular momentum transfer between these subsystems. Consequently, employing different models for the reconstruction of the ultrafast demagnetization process may yield varying outcomes. In this section, we will introduce seven common models and compare the differences among them. Finally, we will briefly discuss the angular momentum transfer process in the experiment in Section 3.2.8.

### 3.2.1 Two-temperature model (2TM)

As early as 1960, Lifshitz *et al*. [222] proposed the theory of energy transfer between the electron system and the lattice system. In 1977, Zarzycki *et al*. [223] introduced the Two-temperature model (2TM) based on this theory. However, this model is not applicable when the laser pulse width is much larger than the electron mean free path and relaxation time. When an ultrafast laser pulse (pulse width < 10 ps) excites the metal, the electron movement speed approaches Fermi velocities. In this regime, heat transfer between electrons occurs without collisions and can be approximately described as ballistic transport. Anisimov *et al*. [224] modified the 2TM to approximately capture the equilibrium state between electrons and phonons. In 1993, Qiu *et al*. [225] investigated the effects of ultrafast lasers on materials and improved the 2TM using the Boltzmann transport equation. Subsequently, Chen *et al*. [226, 227] extended the model to include the impact of lattice heat conduction and electron drift velocity on heat transfer. Currently, the integration of the 2TM with numerical simulations has allowed for more comprehensive research into ultrafast lasers and non-magnetic metal materials [228-231].

The 2TM model was later extended to magnetic materials to describe the ultrafast demagnetization after laser excitation. When ultrashort laser pulses irradiate magnetic materials, the duration of these pulses is longer than the electron relaxation time but shorter than the lattice relaxation time. Consequently, the electrons are initially excited by the laser pulse, leading to a rapid increase in their temperature. Due to the interaction between the excited electrons and the lattice, energy is transferred to the lattice system, resulting in an increase in lattice temperature. The 2TM assumes that the spin temperature $T_s$ is equal to the electron temperature $T_e$ and that the electron and lattice systems are always in internal thermal equilibrium. Therefore, during ultrafast demagnetization, only the energy conversion between the electron and lattice systems needs to be taken into consideration. The electron temperature $T_e$ and the lattice temperature $T_l$ can be considered separately [232, 233], as depicted in Fig. 6. The temporal evolution of the subsystem in the 2TM can be described by a set of two coupled differential equations [224, 234]:

$$\begin{aligned} C_e(T_e)\frac{dT_e}{dt} &= -G_{el}(T_e - T_l) + P(t) \\ C_l(T_l)\frac{dT_l}{dt} &= G_{el}(T_l - T_e) \end{aligned} \quad (18)$$

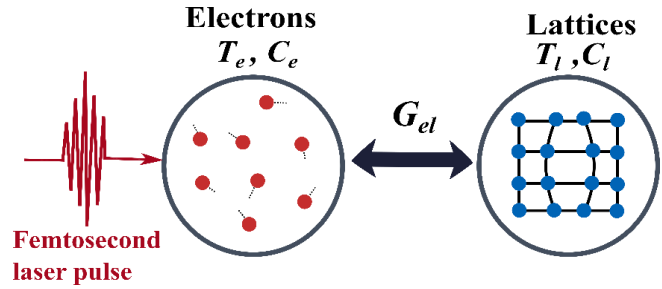

Fig. 6. Energy conversion between electron and lattice systems within the 2TM.

2TM offers a framework for understanding energy exchange between subsystems and describes the non-equilibrium dynamics of a sample following laser excitation. Due to its flexibility for extensions, researchers have continuously modified and proposed various versions of the modified 2TM [235-237] to capture the primary characteristics of non-equilibrium electron or phonon [238, 239] and non-thermal electronic distribution [240].

In the traditional 2TM, the heating of thermal electrons is dependent on the laser source. However, in the modified 2TM, which considers the non-thermal electronic distribution, the laser initially excites the non-thermal electrons. These non-thermal electrons are subsequently converted into thermal electrons through electron-electron scattering. Although the 2TM proposed by Carpene *et al.* (2006) [240] incorporates non-thermal electronic distribution, it assumes that the density of states (DOS) at the Fermi edge is constant. This assumption was refined by Tsibidis *et al.* (2018) [241], and their work accurately considered the DOS in real-world scenarios. The limitation of Tsibidis *et al.*'s work [241] is that, despite accounting for hot electron transport, it overlooks non-thermal electron transport, similar to previous studies.

In recent years, Uehlein *et al.* [242] have expanded upon the research conducted by Tsibidis *et al.* [241] regarding non-thermal electron transport. Although the temperature of non-thermal electrons cannot be defined, their energy density can be quantified. This model enables a comparative analysis of the dynamics between non-equilibrium and equilibrium systems. Moreover, it accounts for the entire energy range, rather than being limited to the vicinity of the Fermi edge.

Beyond examining non-thermal electronic distribution during the initial stages of laser excitation, it is essential to carefully consider the interactions between phonons and other phonons or systems in the later stages. The previously proposed 2TM by Sadasivam *et al.* (2017) [243] assumed that in each "time step", the relaxation process only depended on the lifetime of a few individual scattering processes. Additionally, they also did not adequately account for phonon-phonon interactions, resulting in the absence of any phonon mode exhibiting a temperature higher than the electron temperature. Consequently, 2TM does not account for the spatial distribution of the laser within the system nor the thermal diffusion during the magnetization recovery process. The 2TM assumes that all phonons share a common temperature $T_l$. However, the lattice subsystem of any material consists of multiple independent phonon branches that remain in thermal equilibrium. Therefore, using this model to reconstruct the ultrafast demagnetization process may result in inaccurate interpretations [244]. This limitation is improved in the MMTM proposed by Maldonado *et al.*[238] (see Section 3.2.4).

The 2TM can also be used to analyze the demagnetization process in insulator systems. However, unlike the previous case, the 2TM in insulators primarily considers the spin and lattice temperatures[245]. Laser irradiation predominantly leads to an increase in lattice temperature, which is subsequently followed by demagnetization on a picosecond timescale through phonon-magnon interactions.

### 3.2.2 Three-temperature model (3TM)

In 1996, Beaurepaire *et al.* [14] initially proposed the 3TM to elucidate the ultrafast demagnetization phenomenon observed in a nickel film. As depicted in Fig. 7, they explained that, following the incidence of a femtosecond laser on the surface of the ferromagnetic film, the temperature of the electrons in the irradiated region initially increases rapidly. Subsequently, energy is transferred among the three subsystems: electrons, lattices, and spins.

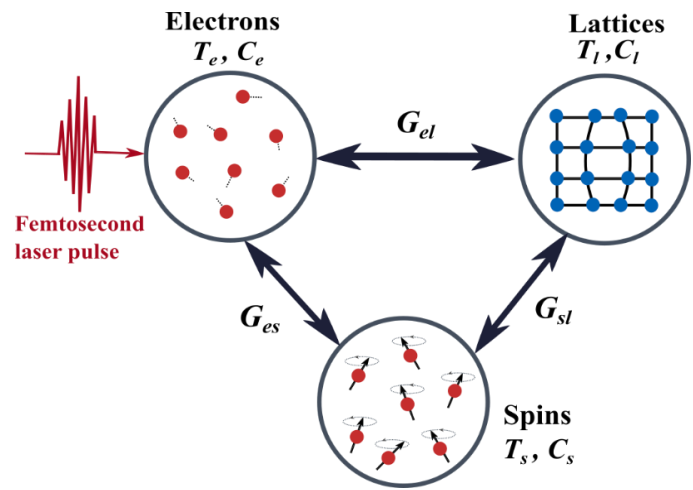

Fig. 7. The interaction processes among the three subsystems of electrons, spins, and lattices within the 3TM.

The 3TM is a phenomenological model, and the temporal evolution of the subsystem can be delineated through the use of three interconnected differential equations [14]:

$$C_e(T_e)\frac{dT_e}{dt} = -G_{el}(T_e - T_l) - G_{es}(T_e - T_s) + P(t)$$

$$C_s(T_s)\frac{dT_s}{dt} = -G_{es}(T_s - T_e) - G_{sl}(T_s - T_l) \tag{19}$$

$$C_l(T_l)\frac{dT_l}{dt} = -G_{el}(T_l - T_e) - G_{sl}(T_l - T_s)$$

In equation (19), $P(t)$ represents the laser energy absorbed by the electronic system. $T_e$, $T_s$, and $T_l$ denote the temperatures of electrons, spins, and lattices, respectively. $G_{el}$, $G_{es}$, and $G_{sl}$ are the coupling constants for electron-lattice, electron-spin, and spin-lattice interactions, respectively. $C_e$, $C_s$, and $C_l$ represent the heat capacities of electrons, spins, and lattices, respectively. The time-dependent behaviors of $T_e$, $T_s$, and $T_l$ are illustrated in Fig. 8. It is evident that the temperature of electrons increases rapidly following laser excitation, primarily due to their lower heat capacity.

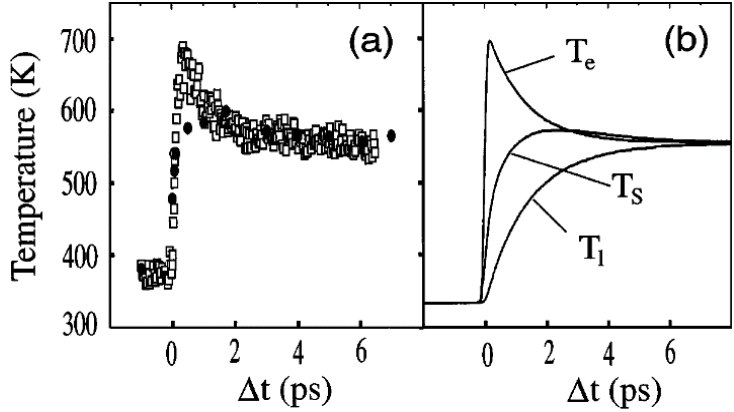

Fig. 8. (a) Experimental results dependence on delay time. The open squares and solid circles represent the electron temperature $T_e$ and spin temperature $T_s$, respectively. (b) Simulation results depicting the dependence of electron temperature $T_e$, spin temperature $T_s$, and lattice temperature $T_l$ on time [14]. Reprinted with permission from E. Beaurepaire *et al.*, Phys. Rev. Lett., **76**, 4250 (1996) Copyright (1996) by the American Physical Society.

During the ultrafast demagnetization process, the electron distribution in the irradiated region of the sample undergoes significant changes [147, 234, 246]. Figure 9(a) shows the electronic density of states of sample prior to laser irradiation [83]. When the laser irradiates the magnetic film, as illustrated in Fig. 9(b), some electrons in the irradiated area become excited above the Fermi level due to the absorption of laser energy. This results in a highly non-equilibrium electron distribution within a short timeframe. When a pump pulse with frequency $v$ and laser fluence is used in the measurement, the relative occupancy of the DOS in different spin bands near the Fermi level ($E_F \pm hv$) can be determined according to the model proposed by Oppeneer *et al.* [246], as shown in Fig.

9(b). Since spin-up and spin-down electrons exhibit asymmetrical densities of occupied and unoccupied states, laser excitation induces an imbalance between the two spin populations. These non-equilibrium electrons rapidly achieve thermal equilibrium through electron-electron interactions and assume a Fermi-Dirac distribution, as illustrated in Fig. 9(c). Simultaneously, spin-up electrons are relaxed into the minority empty states via spin-flip transitions mediated by SOC, leading to a reduction in the magnetization of ferromagnet on timescales of less than 20 fs [247]. In the following few hundred femtoseconds, high-energy state majority electrons are relaxed into minority empty states via the localized spin-flip scattering [39], as well as the generation of spin currents [53]. These processes further reduce spin polarization near the Fermi level and a further reduction of magnetization (Fig. 9(d)). After 1 ps, the magnetization begins to recover towards its initial state, as shown in Fig. 9(e). We note that, during this relaxation, energy is transferred from the electron and spin subsystems to the lattice. Although the electron, spin and lattice subsystems reach thermal equilibrium after a few picoseconds, the Fermi energy level is a relatively blurred region and the overall temperature remains elevated [248].

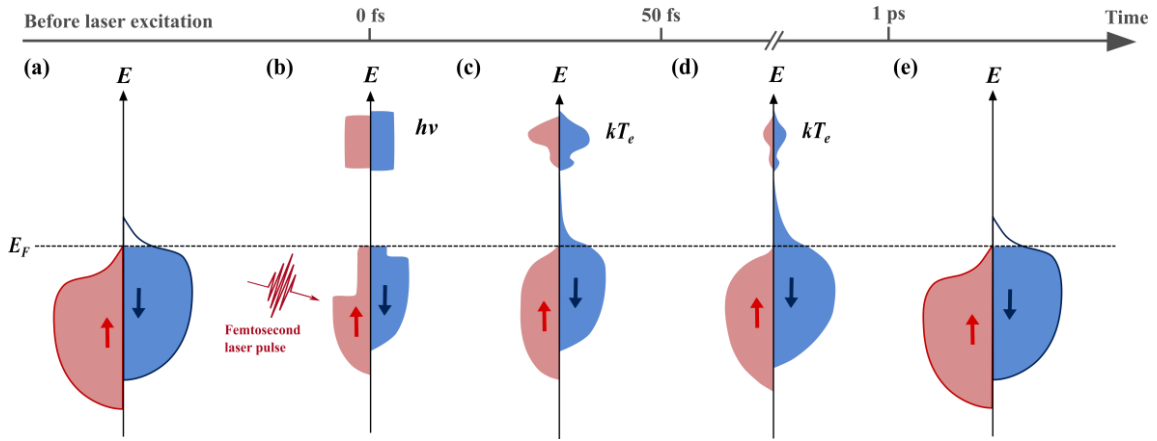

Fig. 9. Schematic representation of the ultrafast demagnetization process in ferromagnets. (a) The density of states in a ferromagnet before laser excitation. (b) Laser excitation leads to unequal excitation of spin-up and spin-down electrons due to the asymmetric distribution of occupied and unoccupied state densities. (c) The electronic system rapidly thermalizes, restoring the Fermi-Dirac distribution, where demagnetization begins as indicated by changes in spin populations. This spin-flip events lead to a decrease in magnetization within 20 fs [247]. (d) The spin-flip scattering of electrons with other quasiparticles, as well as the generation of spin currents, further reduces the magnetization. (e) After 1 ps, the magnetization relaxes, and the DOS of system nearly returns to its initial state.

Initially, it was difficult for the researcher to obtain the corresponding analytical solution from Eq. (19), necessitating further simplification of the 3TM. At low pump laser fluence, due to the similar variation in the behavior of electron temperature $T_e$ and spin temperature $T_s$, Kampen proposed a simplified 3TM (S3TM) [234]. In this model, it is assumed that the electron-spin coupling of the system is very strong, the variation of $T_s$ follows $T_e$ but lags behind it with a delay time $\tau_1$. When $\tau_1$ approaches zero, due to the very fast electron-spin scattering, $T_s$ equals $T_e$. The S3TM reduces computational complexity by simplifying the equations governing the energy exchange between electrons, phonons, and spins. This enables faster simulations and broader parameter space exploration. This simplification facilitates the derivation of analytical solutions or approximations, providing clearer insights into the fundamental mechanisms and timescales of ultrafast demagnetization processes. Furthermore, by comparing the results of the S3TM with those from the original 3TM or other more sophisticated models, researchers can better evaluate the influence of various assumptions and approximations. Although the S3TM is similar to the basic 2TM [224], S3TM is extended 2TM with a spin bath. The temporal evolution of the subsystem in the S3TM can be described by the following three differential equations:

$$C_e(T_e)\frac{dT_e}{dt} = -G_{el}(T_e - T_l) + P(t)$$
$$C_l(T_l)\frac{dT_l}{dt} = G_{el}(T_l - T_e) \quad (20)$$
$$\frac{\partial T_s(t)}{\partial t} = \tau_1^{-1}\left[T_e(t) - T_s(t)\right]$$

The 3TM has been effectively employed to elucidate the phenomenon of ultrafast demagnetization in a significant body of research [39, 54, 191, 249]. Assuming that the experiment is conducted in the low-laser-fluence limit, the electron and lattice heat capacities ($C_e$ and $C_l$) remain relatively stable within the laser-induced thermal field, and the heat capacities of spins $C_s$ are neglected. The electron temperature rises instantaneously under laser excitation [250]. Based on phenomenological 3TM, the analytical solutions to Equation (19) typically manifest in two forms, represented by Equations (21) and (22). These solutions are predominantly utilized to fit ultrafast demagnetization curves, such as the $\Delta\theta_k$-$t$ curves derived from TR-MOKE [25, 114] or the temporal magnetization evolutions indirectly extracted from TR-XRMS [54, 191].

One is the three-exponential function [54]:

$$-\frac{\Delta M(t)}{M} = \left[A_1(1-e^{\frac{-t}{\tau_M}})e^{\frac{-t}{\tau_E}} + A_2\left(1-e^{\frac{-t}{\tau_E}}\right)\right]\Theta(t)\otimes G(t) \quad (21)$$

Another is the two-exponential function [250]:

$$-\frac{\Delta M(t)}{M} = \left[ A_2 - \frac{(\tau_E A_1' - \tau_M A_2)e^{\frac{-t}{\tau_M}} + \tau_E(A_2 - A_1')e^{\frac{-t}{\tau_E}}}{\tau_E - \tau_M} \right] \Theta(t) \otimes G(t) \quad (22)$$

where $\Theta(t)$ is the Heaviside step function when the sample is excited by a laser at delay time $t=0$. $G(t)$ is the Gaussian convolution. $\tau_M$ and $\tau_E$ are the demagnetization time and magnetization recovery time, respectively. $A_1$ is the amplitude of the demagnetization, which is proportional to the electron temperature. $A_2$ represents the value $-\frac{\Delta M(t)}{M}$ once the electrons, spins, and lattice have returned to equilibrium. In the above equations, the temperature of the highly excited (non-thermal) electrons is not clarified. However, in Eq. (22), it is assumed that the spin dynamics depend mainly on the temperature of the non-thermal electrons and the lattice system. The temperature of non-thermal electrons is defined approximately by Eq. (20) through the excess energy term, $E_{ex} = \frac{1}{6}\pi^2 D_F (k_B T_{e,nonthermal})$. In addition, $\tau_M^{-1} = \tau_{M,e}^{-1} + \tau_{M,l}^{-1}$, $A_1' = A_1 \tau_M / \tau_{M,e}$. $\tau_{M,e}$ and $\tau_{M,l}$ represent the contributions of the spin-electron and spin-lattice channels, respectively. By fitting parameter $A_1'/A_1$, the values of parameters $\tau_{M,e}$ and $\tau_{M,l}$ can be determined [250]. Although the 3TM does not account for angular momentum transfer, the use of Eqs. (21) and (22) enables a quantitative analysis of the demagnetization and relaxation times in ultrafast demagnetization processes.

While the 3TM has found widespread application in the context of ultrafast demagnetization, there are four significant shortcomings that should not be overlooked:

i) The 3TM presupposes that electrons, spins, and lattice systems are perpetually in an internal thermalization state, making it challenging to accurately determine the temperature of non-thermal electrons. Research has demonstrated the considerable significance of non-thermal electronic distribution in ultrafast demagnetization [48, 64]. In addition, the model proposed by Beaurepaire et al. [14] suggests an approximate relationship: $C_e = \gamma T_e$. This approximation is applicable only in cases of low electron temperature. Consequently, this will lead to a significant underestimation of the subsequent demagnetization process.

ii) The 3TM exclusively focuses on energy conversions between the three thermal reservoirs but overlooks angular momentum dissipation between subsystems.

iii) The fundamental premise of the 3TM is the absence of microscopic mechanisms in the ultrafast demagnetization process. Consequently, this model struggles to effectively assess contributions stemming from various microscopic mechanisms during the demagnetization process.

iv) The model inadequately considers the thermalization process of the electron-lattice system, resulting in a distortion of the magnetization recovery process reconstruction. Recent findings have shown that when a material is excited by a laser, it produces a non-thermal phonon distribution [239, 251], which can prolong the thermalization of the lattice system for more than 20 picoseconds [238]). Hence, the non-equilibrium processes initiated by the lattice system significantly impact both the electron-lattice coupling strength and the overall magnetization recovery process [238, 239, 252]. Outcomes solely derived from the thermal phonon approximation possess limitations. Although the 3TM has been successfully applied to explain the ultrafast demagnetization of $3d$ transition metals [15, 54], its coverage of the delay time range is relatively narrow (< 100 ps). This model also does not account for the magnetic field dependence of magnetization relaxation over long-timescales. The spin heat capacity in the 3TM is evaluated at the spin temperature $T_s$ of the magnetization reservoir, which is significantly lower than electron temperature $T_e$ during ultrafast demagnetization [253]. It is still a challenge for 3TM to accurately capture the demagnetization dynamics of both the itinerant ($5d6s$) conduction electrons and the localized $4f$ electrons in heavy rare earth metals, which also remains a topic of current research debate [254, 255].

### 3.2.3 Extended three-temperature model (E3TM)

It has been demonstrated that the contribution of non-thermal electronic distribution to ultrafast demagnetization cannot be disregarded [48, 64, 240, 256]. As early as 1994, Sun *et al.* [257] incorporated the consideration of nonthermal electrons into the 2TM model. In 2003, Koopmans *et al.* [149] proposed that the presence of non-thermal electrons could lead to dichroic bleaching in the material, thereby affecting its magneto-optical response. However, they did not describe them as a driving force for demagnetization. To explain the excitation of non-thermal electrons within the first picosecond and the subsequent energy transfer from non-thermal electrons to thermal electrons, lattice, and spin subsystems, Zhang *et al.* (2006) [258] introduced the non-thermal electronic distribution based on the 3TM and proposed the extended three-temperature model (E3TM). Building on this work, Kim *et al.* (2009) [64] employed the E3TM to account for the step-like demagnetization process observed in $Tb_{35}Fe_{65}$ thin films. The energy conversion of general E3TM is illustrated in Fig. 10. The temporal evolution of the subsystem in E3TM can be described by four coupled differential equations [48, 64, 248, 258]:

$$\frac{\partial N}{\partial t} = P(t) - P_e(t) - P_l(t) - P_s(t)$$

$$C_e(T_e)\frac{\partial T_e}{\partial t} = P_e(t) - G_{el}(T_e - T_l) - G_{es}(T_e - T_s)$$

$$C_l(T_l)\frac{\partial T_l}{\partial t} = P_l(t) - G_{el}(T_l - T_e) - G_{sl}(T_l - T_s) - K_l(T_l - 300)^3 \quad (23)$$

$$C_s(T_s)\frac{\partial T_s}{\partial t} = P_s(t) - G_{es}(T_s - T_e) - G_{sl}(T_s - T_l)$$

$$P_i(t) = \frac{C_{ei}}{C_e} N, \quad i = e, l, s$$

where $N$ represents the energy density in a non-thermal electron system and $P(t)$ denotes the laser energy. $P_e(t)$, $P_l(t)$, and $P_s(t)$ stand for the energy flow from the non-thermal electrons to thermal electrons, lattices, and spins, respectively. $K_l(T_l-300)^3$ describes the thermal diffusion of the lattice system.

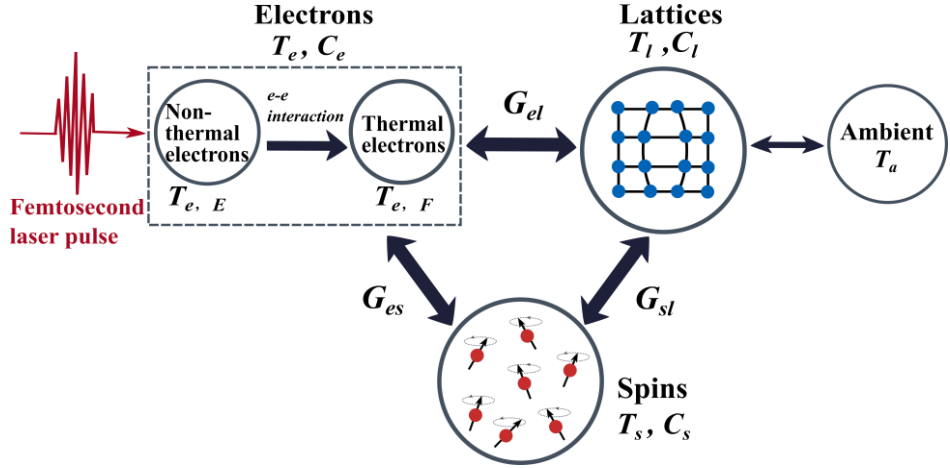

Fig. 10. The interaction processes among the non-thermal electrons, thermal electrons, spins, and lattices in the E3TM.

In 2020, Shim *et al.* [48] employed E3TM model to demonstrate that non-thermal electronic distribution could be another crucial mechanism for understanding ultrafast laser-induced spin dynamics. They introduced a new energy exchange coefficient, $G_{ee}$, to describe the energy exchange between non-thermal and thermal electrons. At low energy fluence, the interaction governed by $G_{ee}$ dominates, indicating that the energy relaxation from non-thermal to thermal electrons is rapid. In this regime, the contribution of non-thermal electronic distributions is negligible, and the temperature evolution of both the spin and electron systems remains consistent, making 2TM model sufficient to effectively describe the demagnetization process. However, as the energy fluence increases, the interaction of $G_{ee}$ decreases, and the contribution from non-thermal electronic distributions becomes significant, further influencing the temperatures of other subsystems. Since thermal electrons and the lattice must first acquire energy from non-thermal electrons before interacting with the spin system, the spin relaxation time is extended. Additionally, the E3TM proposed by Shim *et al.* [48] incorporated thermal diffusion from the lattice to the external environment, describing how the pump laser pulse energy flows to the substrate, ultimately allowing the system to relax back to its pre-excitation state. In 2021, based on the E3TM model, Wang *et al.* [248] introduced additional phenomenological time constants ($\tau_f^*$ and $\tau_s^*$) to describe both the fast (ps) and slow (ns) magnetization dynamics, respectively, and examined the magnetization relaxation dynamics after demagnetization in Co/Pt multilayers on timescales exceeding 100 ps. Wang *et al.* [248] demonstrated that if the thermal diffusion coefficient $K_l$ in the E3TM is held constant, the model fails to reconstruct spin dynamics beyond 200 ps. As a result, compared to the 3TM model, E3TM provides a more comprehensive energy conversion process involving the three thermal reservoirs. The timescale of E3TM could involve wide range (up to 200 ps), depending on which is adapted or extended to incorporate additional physical mechanisms based on specific experimental observations and theoretical needs. However, it still overlooks the incorporation of angular momentum conversion.

### 3.2.4 Microscopic multi-temperature model (MMTM)

In order to provide a comprehensive description of the non-equilibrium process in ultrafast demagnetization, Waldecker *et al.* (2016) [239] proposed a model that incorporates the non-thermal phonon distribution within the framework of the 2TM. They described the non-thermal phonon distribution as the sum of the thermal distributions of three phonon branches. This model, proposed by them, is regarded as the prototype for the microscopic multi-temperature model (MMTM). Subsequently, the MMTM was further discussed and revised in ultrafast dynamics [57, 238, 239, 259-261]. Maldonado *et al*. extended this model to investigate the nonequilibrium dynamics in FePt film (2017) [238] and Ni film (2020) [207] induced by ultrashort laser pulses. Their results demonstrate that the lattice remains in a non-equilibrium state throughout the process. Initially, the laser energy is transferred to the electron system, which then shares this energy unevenly with the lattice through electron-phonon coupling. This interaction generates entangled energy flows between different phonon modes and the electron system. Over a longer timescale, thermal phonon modes with higher energy densities begin to lose energy, while phonon modes with lower energy densities continue to gain energy. The energy exchange between electrons and the lattice is influenced by the system's magnetism, leading to lattice dynamics driven by changes in magnetization.

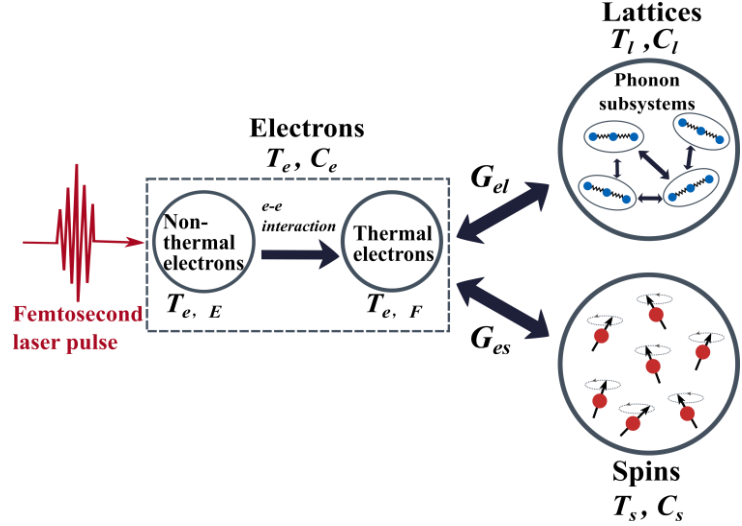

Fig. 11. The interaction processes among the non-thermal electrons, thermal electrons, spins, and lattices in the MMTM. The lattice system is subdivided into several phonon branches and the couplings are considered separately.

In 2020, Zhang *et al.* [57] utilized MMTM for the first time to analyze ultrafast demagnetization phenomena. Unlike previous models (2TM, 3TM, and E3TM), MMTM takes into account the temperatures of independent phonon subsystems within the lattice, thereby addressing nearly all fundamental non-equilibrium processes. As a result, this model provides a comprehensive description of the electron and lattice systems dynamics, including energy exchange between electrons and spins (Fig. 11). The temporal evolution of the MMTM subsystem can be elucidated through a set of three interrelated differential equations [57]:

$$C_e \frac{\partial T_e}{\partial t} = G_{es}(T_s - T_e) + \sum_Q C_Q \gamma_Q I(T_e)(T_p^Q - T_e)\left[1 + J(\omega_Q, T_p^Q)(T_p^Q - T_e)\right] + \frac{\partial U_{e-e}}{\partial t}$$

$$C_Q \frac{\partial T_p^Q}{\partial t} = -C_Q \gamma_Q I(T_e)(T_p^Q - T_e)\left[1 + J(\omega_Q, T_p^Q)(T_p^Q - T_e)\right]$$
$$- \sum_k C_Q \Gamma_{Qk}(T_p^Q - T_p^k)\left[1 + J(\omega_Q, T_p^Q)(T_p^Q - T_p^k)\right]$$
$$+ \frac{\partial U_{e-ph}}{\partial t}, \quad Q = Q_1, Q_2, \ldots, Q_N$$

$$C_s \frac{\partial T_s}{\partial t} = -G_{es}(T_s - T_e) \tag{24}$$

where $T_e$, $T_s$, and $T_p^Q$ are the electron, spin, and phonon mode-dependent temperatures, respectively. $Q$ and $k$ are the number $N$ of distinct and independent phonon subsystems. Each phonon subsystem corresponds to a specific branch $v$ and momentum $\boldsymbol{q}$. The interaction between phonon subsystems and other phonons or electrons depends on the $v$ and $\boldsymbol{q}$, and these interactions result in different temperatures $T_p^Q$. Phonon-phonon scattering causes the phonon subsystems to reach a common lattice temperature, eventually achieving thermal equilibrium. This approach of MMTM overcomes the limitations of 3TM and 2TM. $C_e$, $C_Q$, and $C_s$ are the heat capacities of electron, phonon mode-dependence, and spin, respectively. $C_e = \int_{-\infty}^{+\infty}(\partial f_k / \partial T_e)g(\varepsilon)\varepsilon d\varepsilon$, where $g(\varepsilon)$ is the electron density of states at energy $\varepsilon$ and $\varepsilon_F$ is the Fermi energy. $C_Q$ is expressed as

$$C_Q = \hbar\omega_Q \frac{\partial n_Q}{\partial T} = \frac{\partial E_Q}{\partial T},$$

$E_Q$ is the phonon-mode dependent internal energy. $\gamma_Q$ and $\Gamma_{Qk}$ denote the phonon mode-dependent electron-phonon and phonon-phonon-induced line width, respectively. $\gamma_Q$ can be calculated using the density functional perturbation theory. The relation between $\Gamma_{Qk}$ and phonon lifetimes $\tau_Q$ are related as follows: $\tau_Q = \left[2\sum_k \Gamma_{Qk}\right]^{-1}$. $G$ is the coupling constant of electron-spin coupling. $I(T_e)$,

$$I(T_e) = -\int_{-\infty}^{+\infty} d\varepsilon \frac{\partial}{\partial \varepsilon} \frac{g^2(\varepsilon)}{g^2(\varepsilon_F)},$$

represents the electron temperature dependence of electron-phonon coupling, computed via numerical integration. Among this, $\frac{\partial U_{e-ph}}{\partial t}$ and $\frac{\partial U_{e-e}}{\partial t}$ represent the energy conversion rates of non-thermal electron-lattice and non-thermal electron-thermal electron, respectively. $J(\omega_Q, T_p^Q)$ is the second-order Taylor expansion of the out-of-equilibrium phonon populations around the mode-dependent phonon temperatures [238], the specific form is as follows:

$$J(\omega_Q, T_p^Q) \equiv \frac{\hbar\omega_Q}{k_B T_p^2}\left(\frac{\exp\left(\frac{\hbar\omega_Q}{k_B T_p^Q}\right)+1}{\exp\left(\frac{\hbar\omega_Q}{k_B T_p^Q}\right)-1} - \frac{2k_B T_p^Q}{\hbar\omega_Q}\right), \tag{25}$$

where $\hbar\omega_Q$ is the phonon energy. To obtain the full solution of MMTM, the material specific quantities, $C_e$, $C_Q$, $\gamma_Q$ and $\Gamma_{Qk}$ are the key quantities. For a detailed derivation process, please refer to the work of Maldonado *et al.* (2017) [238].

It is worth noting that the nonequilibrium *ab initio* theory proposed by Maldonado *et al.* [238] provides a detailed description of the interactions between electrons and phonons, building on the foundations laid by the 2TM model. This model encompasses the nonthermal states of both electrons and phonons. Although the temperatures of nonthermal electrons and phonons cannot be directly measured, they can be tracked in reciprocal space using TR-ARPES [262] and ultrafast electron diffraction (UED) [207, 263]. Maldonado *et al.*(2020) [207] employed diffuse UED to investigate momentum-resolved phonon occupation dynamics across the Brillouin zone. The experimental results indicated that within the first 4.9 ps after laser excitation, the phonon occupation was markedly different from the thermalized occupation, consistent with their prior theoretical findings [238]. In addition, they also found that energy can flow back to electrons from phonon modes at the Brillouin zone boundary.

Although the MMTM proposed by Zhang *et al.* [57] incorporates the contributions of the spin system, it still has limitations. For instance, the effective spin temperature $T_s$ is equivalent to the magnon temperature, which characterizes the incoherent excitation of multiple magnon modes in 3TM. Moreover, this model presently lacks the capability to provide the electron-magnon coupling constant that depends on momentum $q$; it only approximates the effective electron-magnon coupling by solely thermalizing the spin system alone and uses this coupling to describe the ultrafast demagnetization process. In recent years, Sharma *et al.* [264] demonstrated that specific phonon modes dominate the ultrafast demagnetization of FePt samples, promoting the necessity of using the MMTM for analyzing ultrafast demagnetization.

### 3.2.5 Four-temperature model (4TM)

In 2013, Mekonnen *et al.* [65] proposed a 4TM model to elucidate the three-step demagnetization process (demagnetization, relaxation, and re-demagnetization) following laser excitation in GdCo and GdCoFe alloys. This model segregates the spin system into two separate subsystems, namely the 4*f* spin and 3*d* spin, as illustrated in Fig. 12. The figure depicts the interactions among four thermal reservoirs, including the electron, lattice, 4*f* spin, and 3*d* spin.

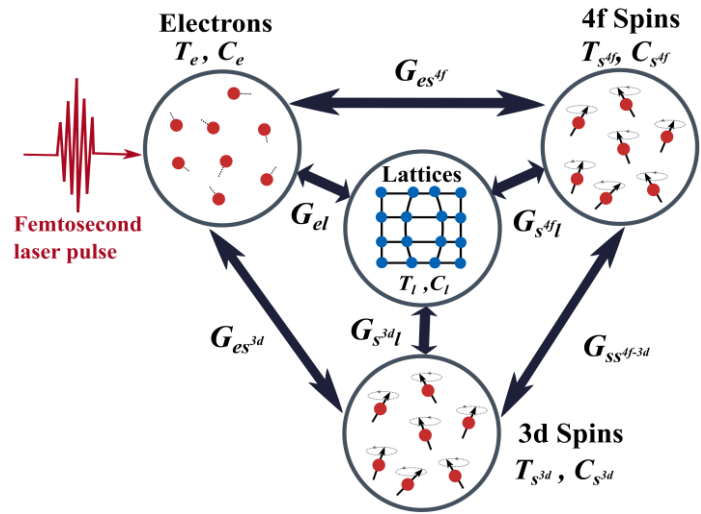

Fig. 12. The interaction processes among the electron, lattice, 4*f* spin, and 3*d* spin in the 4TM.

The temporal evolution of the subsystem for 4TM can be described by four coupled differential equations:

$$C_e(T_e)\frac{dT_e}{dt} = -G_{el}(T_e - T_l) - G_{es^{3d}}(T_e - T_{s^{3d}}) - G_{es^{4f}}(T_e - T_{s^{4f}}) + P(t)$$

$$C_l(T_l)\frac{dT_l}{dt} = -G_{el}(T_l - T_e) - G_{s^{3d}l}(T_l - T_{s^{3d}}) - G_{s^{4f}l}(T_l - T_{s^{4f}})$$

$$C_{s^{3d}}(T_{s^{3d}})\frac{dT_{s^{3d}}}{dt} = -G_{es^{3d}}(T_{s^{3d}} - T_e) - G_{s^{3d}l}(T_{s^{3d}} - T_l) - G_{ss}^{4f-3d}(T_{s^{3d}} - T_{s^{4f}})$$

$$C_{s^{4f}}(T_{s^{4f}})\frac{dT_{s^{4f}}}{dt} = -G_{es^{4f}}(T_{s^{4f}} - T_e) - G_{s^{4f}l}(T_{s^{4f}} - T_l) - G_{ss}^{4f-3d}(T_{s^{4f}} - T_{s^{3d}})T_{s^{3d}}$$

(26)

where $C_e$, $C_l$, $C_{s^{3d}}$, and $C_{s^{4f}}$ represent the heat capacity of the electron, lattice, 3$d$ spin, and 4$f$ spin, respectively. $G_{el}$, $G_{es^{3d}}$, and $G_{es^{4f}}$ denote the coupling constant of the electron with lattice, 3$d$ spin, and 4$f$ spin, respectively. $G_{s^{3d}l}$ and $G_{s^{4f}l}$ are the coupling constants of the lattice with 3$d$ spin and 4$f$ spin, respectively. $G_{ss}^{4f-3d}$ indicates the coupling constant between the 3$d$ and 4$f$ metal sublattice. The exchange interaction between the 3$d$ and 4$f$ metal sublattices provides an additional energy conversion path that differs from other models.

Without considering the Gd sublattice in CoGd, the demagnetization dynamics obtained from the 4TM model are consistent with those observed in pure 3$d$ metal films. Therefore, Mekonnen *et al.* [65] suggested that the Gd sublattice is the primary cause of Type II demagnetization. Currently, the 4TM model is considered the most suitable for studying pure RE metals, RE- TM alloys, and half-metallic ferrimagnets [255, 265-268]. In this framework, a distinction is made between the 4$f$ moment and the 5$d$ moment on the 4$f$ ion (typically Gd or Tb). This description is necessary as the 5$d$ electrons are excited by the laser but not the 4$f$ electrons. The contribution of 5$d$ electrons requires further discussion in future studies.

### 3.2.6 Microscopic three-temperature model (M3TM)

In addition to 3TM, M3TM is another model widely employed for the investigation of ultrafast demagnetization. This model was initially proposed by Koopmans *et al.* in 2010 [39] and elucidated the coupling of subsystems through the microscopic Hamiltonian. M3TM not only encompasses interactions among the three thermal reservoirs (electron, lattice, and spin) but also accounts for the dissipation of angular momentum. Figure 13 illustrates the interactions among the three subsystems of electrons, spins, and lattices. In accordance with the conservation of angular momentum in the system, the total angular momentum *J* of the system comprises four terms:

$$J = L_e + S_e + L_{phonon} + L_{photon} \tag{27}$$

where $L_e$, $S_e$, $L_{phonon}$, and $L_{photon}$ are the electron orbital angular momentum, electron spin angular momentum, phonon angular momentum in the lattice system, and laser field-induced angular momentum, respectively.

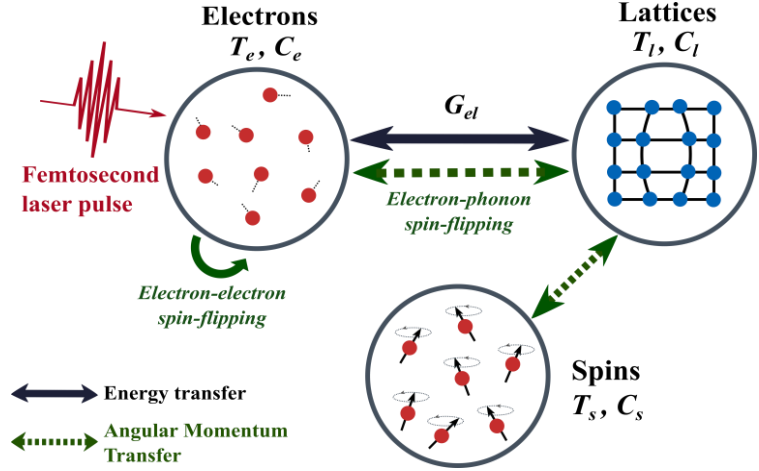

Fig. 13. The interaction processes among the three subsystems of electrons, spins, and lattices in the M3TM.

The M3TM acknowledges that the transfer of angular momentum and energy arises from microscopic scattering, with the Elliott-Yafet scattering (electron-phonon and electron-electron scattering) mechanism playing a predominant role in the ultrafast demagnetization process [269]. Elliott-Yafet scattering will be introduced in Section 4.1.1. In addition, the M3TM includes the gradient of electronic temperature.

There are three assumptions for the M3TM. i) The electron-phonon scattering rate is not influenced by the spin system. ii) The ambient temperature exceeds the Debye temperature ($T_D$). iii) The electron spin $S$ is $S=1/2$, i.e., the spin-up and spin-down states possess equal probabilities. Based on these three assumptions, the temporal evolution of the subsystem for M3TM can be described by three coupled differential equations [39]:

$$\begin{aligned}
&\gamma T_e \frac{dT_e}{dt} = \nabla_z (\kappa \nabla_z T_e) + G_{el}(T_l - T_e) \\
&C_l \frac{dT_l}{dt} = G_{el}(T_e - T_l), \quad G_{el} = \frac{3\pi D_F^2 D_P k_B^2 T_D \lambda_{el}^2}{2\hbar} \\
&\frac{dm}{dt} = Rm\frac{T_l}{T_c}\left(1 - m\coth\left(\frac{mT_c}{T_e}\right)\right), \quad R = \frac{8a_{sf}G_{el}T_c^2}{k_B T_D^2 D_S}
\end{aligned} \tag{28}$$

where $\nabla_z$ represents the derivative with respect to *z*. The initial (*t*=0 s) electron temperature distribution under transient heating is assumed to be $T_e(z,0) = \Delta T_e(0,0)\exp(-z/\lambda)$, $\lambda$ represents the laser penetration depth. This assumption regarding the electron temperature distribution is also applicable to other models. $\kappa$ represents the coefficient of thermal conductivity and $\gamma$ denotes the electronic specific heat constant. $\lambda_{el}$ signifies an electron-lattice matrix element. $D_F$ is the material property parameter. $D_P$ represents the number of potential polarization states, while $D_s$ denotes the spin density. $k_B$ refers to the Boltzmann constant. $a_{sf}$ indicates the probability of spin flipping. $T_c$ and $T_D$ are the Curie temperature and Debye temperature, respectively. The coupling constant of the electron-lattice interaction $G_{el}$ determines the electron relaxation time. The third equation pertains to the magnetization speed of the system, with $m$ representing the normalized magnetic moment. The factor $R$ depends on the material and is proportional to $a_{sf}$. Since the M3TM model considers spin flips caused by electron-phonon scattering in an averaged manner, the probability of spin flips depends on $G_{el}$ [39], and the average magnetization is influenced by the temperature of the electrons and phonons, as well as the angular momentum transfer between the electrons and the lattice. Most of the electron and phonon systems are in local thermal equilibrium (refer to Supplementary Information 1 of Ref. [270]). This also implies that a small fraction of electrons and phonons remain in a non-thermal state due to the absorption of the majority of the laser energy. The M3TM model, however, does not fully address the non-thermal electrons and phonons.

When employing M3TM for the analysis of experimental data, these parameters are typically derived from ab initio calculations. Since the concept of spin-flipping within the E-Y scattering mechanism was originally postulated based on non-magnetic conductor materials, M3TM cannot be directly utilized for the analysis of ferromagnetic substances. The excitation spectrum of magnetic materials is considerably more intricate compared to non-magnetic conductors. Currently, M3TM is extensively employed not only to analyze type I demagnetization in TR metals [55, 82, 271, 272] but also to investigate type II demagnetization in RE metals [39, 253, 273-276]. For the analysis of M3TM in rare earth metal systems, at least one of the following conditions should be met: i) The ferromagnet possesses a large magnetic moment; ii) The contribution of electron-spin scattering is weak; iii) The sample temperature during the experiment is close to the Curie temperature $T_c$. Koopmans et al. [39] predicted that as $T_c$ is approached, the disorder in magnetic order will lead to the weakening of exchange splitting. Since the demagnetization rate in M3TM is dependent on exchange splitting, the experimental temperature near $T_c$ will necessarily result in a slower demagnetization rate. As the ambient temperature increases, the spin temperature $T_s$ changes much slower. Therefore, in rare earth metals, the time required for electron-spin thermal equilibrium is longer than that for electron-lattice thermal equilibrium. After the electrons and the lattice reach thermal equilibrium, the angular momentum of the lattice system is transferred to the spin system, resulting in type II demagnetization.

However, there are several contentious aspects within the M3TM. i) It postulates that the electron, lattice, and spin systems maintain internal equilibrium and presupposes that electrons consistently adhere to the Fermi-Dirac distribution. ii) In M3TM, defining the phonon temperature during electron-phonon scattering poses challenges, and Fermi's golden rule, employed for calculating electron-phonon and electron-phonon-spin scattering within this framework, relies on phonon energy. Consequently, the reasonableness of the approximate results yielded by M3TM necessitates further examination and verification. Compared with 3TM, M3TM considers the influence of external magnetic fields [86], but both the effectiveness of 3TM and M3TM are limited to shorter timescales (< 100 ps).

### 3.2.7 Extended microscopic three-temperature model (EM3TM)

In 2019, Bonda *et al.* [81] modified the M3TM by incorporating a medium thermal reservoir, thus proposing an extended M3TM. The primary aim was to explain the phenomenon in which the timescale of the de- and re-magnetization processes in the Heusler alloy ($Ni_{54.3}Mn_{31.9}Sn_{13.8}$) film is significantly extended when the test temperature approaches the Curie temperature. As illustrated in Fig. 14, the EM3TM incorporates a lattice cooling process. The temporal evolution of the subsystem in the M3TM can be elucidated through the following set of differential equations:

$$\begin{aligned}
\frac{dT_e}{dt} &= \frac{(T_l - T_e)}{\tau_{le}} + \frac{P(t)}{C_e} \\
\frac{dT_l}{dt} &= \frac{(T_e - T_l)}{\tau_{el}} + \frac{(T_a - T_l)}{\tau_a} \\
\frac{dm}{dt} &= Rm \frac{T_l}{T_c} \left[ 1 - m \coth\left( m \frac{T_c}{T_e} \right) \right] \\
\frac{dT_a}{dt} &= 0
\end{aligned} \quad (29)$$

where $P(t)$ is the laser energy. $m=M/M_s$ is the normalized magnetic moment. $\tau_{le}=C_e/G_{el}$ and $\tau_{el}=C_l/G_{el}$ are the respective relaxation times. $\tau_a$ is the lattice relaxation time, and $T_a$ is the temperature of the ambient thermal reservoir. $R$ is the prefactor controlling the demagnetization rate [39, 127].

To elucidate the anomalous ultrafast demagnetization behavior involving demagnetization, relaxation, and remagnetization in NiMnSn Heusler alloys, in 2022, Bonda *et al.* [277] incorporated the weak exchange coupling effect into the Ni and Mn sublattices, using the EM3TM framework [81]. The anticipated outcomes are in good agreement with the experimental data. However, $\tau_a$ is treated as a constant as assumed in the EM3TM model, only the magnetization dynamics within the first 200 ps can be accurately reconstructed [248].

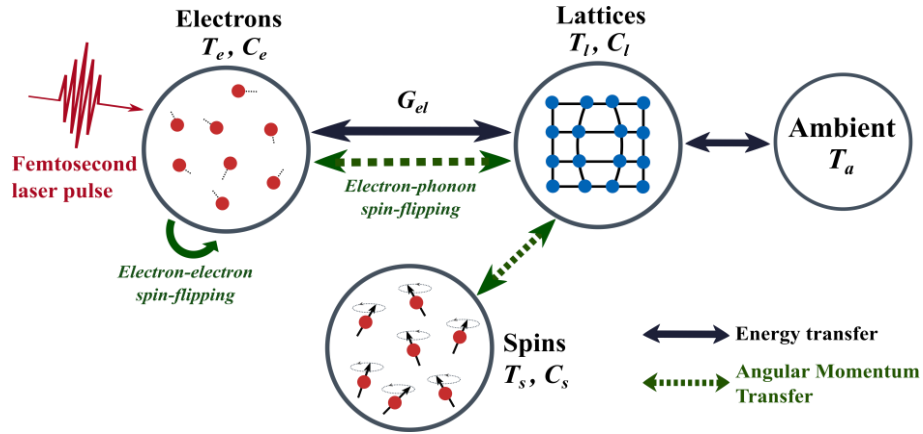

Fig. 14. The exchange of energy and angular momentum between different systems, as well as the conversion of energy between lattice systems and the external environment in the EM3TM.

Since the M3TM does not include the spin coupling with electrons and phonons, the magnetization flipping mechanism in RE-TM materials[278], along with the delayed demagnetization observed in Ni and Fe in permalloy alloys [47], has long puzzled researchers. This has further prompted several attempts to modify this model [270, 275, 276, 279, 280]. Unlike above EM3TM model proposed by Bonda *et al.* [81], which introduces an ambient reservoir, another EM3TM model developed by Schellekens *et al.* (2013) [280] extended the spin excitations in the M3TM from $S = 1/2$ to $S = N/2$. Similar to the 4TM model, this approach introduces multiple spin systems ($m_i$, $i = 1, 2, …$), which are not in internal equilibrium nor in equilibrium with each other. Each spin system, $m_i$, is characterized by $2S+1$ discrete energy levels, where $S$ represents the spin quantum number, and the spacing between these energy levels is given by $J_{ex}$. Additionally, the model incorporates the exchange scattering mechanism. Through exchange scattering, the angular momentum can be transferred between spin systems via electron-electron scattering. Simultaneously, the angular momentum can be also exchanged between the spin system and the lattice system through electron-phonon scattering. The energy and angular momentum transfer processes within this model are depicted in Fig.15.

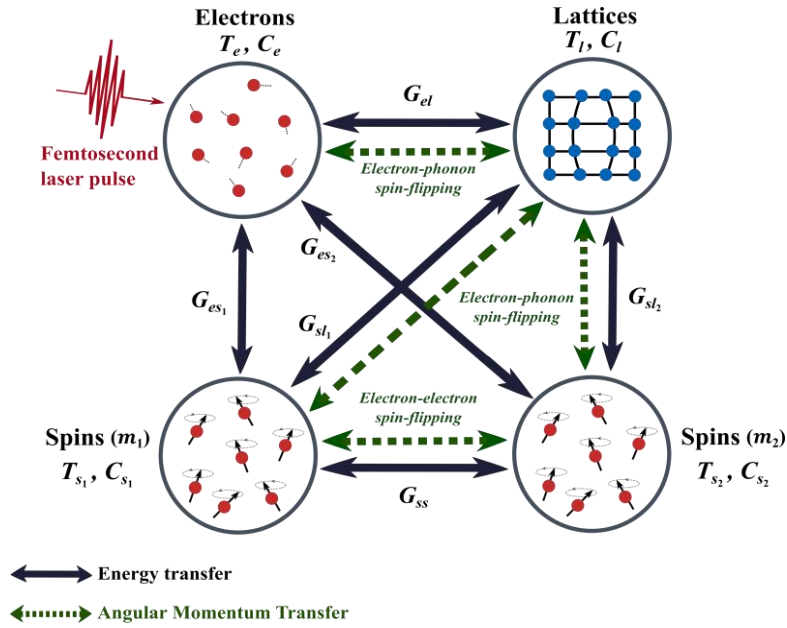

Fig. 15. The exchange of energy and angular momentum between electrons, spins ($m_1$, $m_2$), lattices systems in the EM3TM. The model divides the spin system into two or more components (with two spin systems illustrated in the schematic diagram) and considers the interactions between the spin systems and the subsystems.

Based on this EM3TM model, this approach has been effectively used to reproduce the AOS and demagnetization dynamics of ferromagnetic materials. Beens *et al.* (2019) [275] extended the investigation of the AOS process accurately replicating the all-optical switching phenomena observed in Co/Gd bilayer films and GdCo alloys. Their results demonstrate that, with the driven of exchange scattering, the switching mechanism at the Co/Gd interface propagates through the Co layer, ultimately inducing magnetization flip in Co. However, similar to conventional M3TM, in the model of Beens *et al.* [275], they assume separate spin subsystems coupled with single electron and phonon subsystems. The electrons are considered as spinless free electron gas, and the specific heat of the spin system is neglected in this EM3TM. In 2022, Chatterjee *et al.* [279] extended this EM3TM based on the work of Beens *et al.* [275, 276], and applied it to [Co/Pt]$_n$/CoGd multilayers. In this work, they considered two spin angular momentum transfer channels in CoGd, i.e., the electron-electron exchange scattering and Elliott-Yafet spin-flip scattering caused by electron-phonon interaction between Co and Gd sublattices. Additionally, due to the long-range Ruderman–Kittel–Kasuya–Yosida (RKKY) exchange coupling with the Co/Pt multilayers, they suggested that there is an additional angular momentum transfer channel for this system model. Whether the model is equally effective for other systems still needs to be verified through additional studies.

**3.2.8 Brief comparison of the microscopic model of ultrafast demagnetization**

We briefly introduced seven models (2TM, 3TM, E3TM, MMTM, 4TM, M3TM, and EM3TM) for the ultrafast demagnetization process and discussed the distinctions between these models. The 2TM, 3TM, and M3TM models do not strictly account for non-thermal electrons and phonons, which has prompted various extensions based on these models [81, 149, 238-240]. The E3TM includes non-thermal electrons explicitly and accounts for the energy transfer among non-thermal, thermal electrons, spins and phonons [48, 258]. This provides a better understanding of the initial stages of demagnetization. It is important to note that neglecting the distribution of non-thermal phonons leads to an underestimation of the coupling between electrons and lattices as well as the magnetization recovery time [238]. The MMTM model more comprehensively considers the distribution of non-thermal phonons. However, the MMTM proposed by Maldonado *et al*. [238] does not account for the interaction between the spin system and other subsystems. Although Zhang et al. [57] improved upon MMTM by considering the coupling between electrons and spins, they still ignored the interaction between spins and lattices.

The most significant difference between M3TM, EM3TM, and other models is their consideration of angular momentum transfer. M3TM discusses angular momentum transfer only between electrons and lattices. Some studies have shown that the angular momentum transfer between spins and lattices cannot be ignored [52, 79, 281]. Consequently, Schellekens *et al.* (2013) [280] expanded upon M3TM by introducing multiple spin systems. Through exchange scattering, angular momentum can be transferred between spin-spin and spin-lattice systems via electron-electron and electron-phonon scattering.

Table 2 Energy conversion or angular momentum transfer between different systems considered by various models.

| Model | $N_e$ | $N_p$ | $\Delta T_{es}$ | $\Delta T_{el}$ | $\Delta T_{sl}$ | $\Delta T_{ss}$ | $\Delta T_{la}$ | $\Delta J_{el}$ | $\Delta J_{sl}$ | $\Delta J_{ss}$ | References |
|---|---|---|---|---|---|---|---|---|---|---|---|
| 2TM | | | | √ | | | | | | | [224] |
| 3TM | | | √ | √ | √ | | | | | | [14] |
| E3TM | √ | | √ | √ | √ | | √ | | | | [64, 258] |
| MMTM | √ | √ | √ | √ | √ | | | | | | [57, 238] |
| 4TM | | | √ | √ | √ | √ | | | | | [65] |
| M3TM | | | | √ | | | | √ | √ | | [39] |
| EM3TM | | | | √ | | | √ | √ | √ | | [81, 277] |
| | | | √ | √ | √ | √ | | √ | √ | √ | [280] |

$N_e$: Non-thermal electronic distribution.
$N_p$: Non-thermal phonon distribution.

| | $\Delta T_{esl}$: Energy conversion among electrons, spins, and lattices. |
| --- | --- |
| | $\Delta T_{ss}$: Energy conversion between different spin subsystems. |
| | $\Delta T_{la}$: Lattice cooling to ambient temperature. |
| | $\Delta J_{esl}$: Angular momentum transfer among electrons, spins, and lattices. |
| | $\Delta J_{ss}$: Angular momentum transfer between different spin subsystems. |

In summary, when analyzing the laser-induced ultrafast demagnetization process, it is crucial to consider the non-thermal state of electrons and phonons and the angular momentum transfer between various subsystems. For rare earth metals, the 3$d$, 4$f$, and 5$d$ spin systems must also be discussed separately. The seven models mentioned above each have their own strengths and weaknesses. Table 2 summarizes the consideration of energy and angular momentum by the various models. Even with the same experimental data, employing different models can yield different interpretations. Consequently, the selection of an appropriate model and the acquisition of dependable information through that model remain the future objectives of researchers' endeavors.

## 4. Origins of ultrafast demagnetization dynamics

After comprehending the various models of ultrafast demagnetization, this text delves into the origins of ultrafast demagnetization dynamics. As indicated in the preceding section, divergent interpretations may arise when employing different models to discuss and analyze experimental data. However, the transfer of angular momentum at the sub-picosecond timescale remains a topic of ongoing debate. This issue has been subsequently addressed in various ways in the literature [24, 39, 75, 191, 282].

As described in Section 2.4, X-ray techniques can determine the total magnetic moment of each element using XMCD. These techniques can separate the spin and orbital contributions through the application of magneto-optical sum rules. Stamm *et al.* (2007) [282] and Boeglin *et al.* (2010) [191] employed this method to investigate the temporal evolution of orbital and spin angular momentum. They observed that the changes in the orbital and spin angular momentum of electrons occur nearly simultaneously. From these findings, they inferred that angular momentum might be transferred to the lattice via SOC during the spin-flip scattering process, leading to the dissipation of angular momentum.

Subsequently, Dornes *et al.* [52] tracked the structural dynamics of Fe films through time-resolved X-ray diffraction measurements. They reported observing an ultrafast demagnetization induced by strain waves in Fe. The authors concluded that the angular momentum transferred to the lattice due to this effect accounted for 80% of the spin angular momentum lost during the demagnetization process. In 2022, Tauchert *et al.* [79] critically examined the work of Dornes *et al.* [52]. Their study demonstrated that laser excitation of Ni leads to a rapid increase in lattice temperature through isotropic atomic displacements, while also inducing additional, long-lived, and anisotropic lattice motion relative to the initial magnetization direction. These experimental results suggest that the observed non-thermal phonon dynamics have a magnetic origin. Combined with molecular dynamics simulations, they posited that prior to demagnetization, the magnetic moment of each Ni atom was $\mu_{Ni} = 0.6\mu_B$. To achieve 50% demagnetization, each Ni atom needs to dissipate a spin angular momentum of $\Delta L = D\frac{\mu_{Ni}\hbar}{g'\mu_B} \approx 0.16\hbar$, where $g'$ is the gyromagnetic ratio of the electron. On femtosecond and picosecond timescales, angular momentum cannot escape from the excitation region. If the lattice is to absorb this spin angular momentum, each atom must oscillate around its equilibrium position within the unit cell with an average radius $R \approx \sqrt{\frac{\Delta L}{M_{Ni}\omega}}$ (where $M_{Ni}$ is the mass of the Ni atom and $\omega$ is the angular frequency). Their study provided an important validation by showing that the estimated angular momentum transfer is consistent with experimental observations. This finding is significant as such consistency underscores the model's relevance and reliability in capturing the essential physics of angular momentum transfer in ultrafast demagnetization processes. Furthermore, Tauchert et al. suggested that, due to the conservation of angular momentum, the circularly polarized phonons cannot reach isotropic thermal equilibrium. Consequently, the total angular momentum within the lattice may last longer than the initially excited specific subset of high-frequency phonons. The transfer of angular momentum outside the detection region can only occur through phonon-phonon scattering into low-frequency sound and strain waves. This result further emphasized the significance of the lattice as an angular momentum sink. Tauchert et al. [79] also demonstrated that the conclusion of Dornes et al. [52], which attributed the transfer of angular momentum from spin to lattice to the ultrafast Einstein–de Haas effect, requires further revision.

The situation in RE-TM alloys is slightly different. In 2014, Bergeard et al.[283] observed differences in the changes in angular momentum of Gd and Tb in $Co_{0.8}Gd_{0.2}$ and $Co_{0.74}Tb_{0.26}$, respectively. For Gd in CoGd, the spin angular momentum decreases slightly, while the orbital angular momentum remains unchanged. In contrast, for Tb in CoTb, both the spin and orbital angular momenta decrease simultaneously. They attributed this to the coupling between the 5d and 3d orbitals. They also observed that the total angular momentum in CoTb is conserved during the early stages of demagnetization (<140 fs) before it starts transferring to other subsystems. This phenomenon aligns with previous theoretical studies [267]. Hennecke et al.[281] observed slower demagnetization of Gd in GdFeCo, which is consistent with Bergeard et al. [283] and an earlier report [284]. Their study indicated that since the time constant for orbital angular momentum transfer to the lattice is approximately 1 fs -faster than the spin-orbit mediated angular momentum transfer (tens of femtoseconds)- the accumulation of orbital angular momentum cannot be observed. Furthermore, spin-orbit mediated angular momentum transfer is the rate-limiting step for angular momentum flow to the lattice, and the lattice is the ultimate sink for angular momentum.

However, discrepancies remain in these conclusions. Considerable research is still required to elucidate the contributions of various microscopic origins to the ultrafast demagnetization process and the specific mechanisms of angular momentum transfer during demagnetization. Chapter 4 will delve into the several microscopic origins of ultrafast demagnetization. The prevailing consensus suggests that ultrafast demagnetization stems from localized spin-flipping scattering [39, 52] or spin transport [53, 54]. Additionally, researchers have also posited that non-thermal electronic distribution [48, 55, 64] and laser-induced lattice strain [27, 102, 103, 206, 254] may significantly influence the demagnetization process. In this chapter, we aim to elucidate the disparities among the existing origins of demagnetization and provide an overview of the current state of research in this area. This endeavor will possess a substantial bearing on subsequent studies in the realm of ultrafast demagnetization.

## 4.1 Spin-flipping processes

Currently, researchers have found that Elliott-Yafet scattering (which includes electron-phonon scattering [82, 83, 129] and electron-electron Coulomb scattering [84, 85, 285]), electron-magnon interaction [87, 89, 90], and the photon-spin interaction [91, 286] can lead to spin-flipping. The rapid occurrence of these interactions on short timescales provides a reasonable explanation for ultrafast demagnetization observed within a few hundred femtoseconds. This section primarily introduces several instances of spin-flipping caused by different scattering processes and further explores their contributions.

### 4.1.1 Elliott-Yafet scattering

The demagnetization induced by spin angular momentum transfer occurs within hundreds of femtoseconds in metallic magnetic materials during the laser-induced ultrafast demagnetization process. In 2005, Koopmans *et al.* [39, 75] proposed the principles of E-Y scattering to elucidate this phenomenon. In fact, the E-Y related exchange scattering effect facilitating angular momentum transfer was included in the original M3TM, and E-Y scattering was considered an important mechanism of demagnetization [82, 287]. Subsequently, E-Y scattering-induced spin-flipping has been extensively studied as one of the most probable mechanisms for the rapid dissipation of spin angular momentum in ultrafast spin dynamics. Traditional E-Y scattering refers to

**57 / 122**

electron-phonon scattering. However, since electron-electron Coulomb scattering also affects the spin mixing character of materials with significant spin-orbit coupling (SOC), Krauß et al. [288] were the first to categorize electron-electron scattering within the framework of the E-Y scattering mechanism. Unlike electron-phonon scattering, the evaluation of the contribution of electron-electron scattering requires using the band structure as an input parameter. This approach enhances our understanding of demagnetization phenomena in materials with unique band structures.

In the presence of SOC within the material system, the electron spin is no longer in a pure state (i.e., exclusively spin-up or spin-down). Instead, it exists in a mixed state where spin-up $\psi_{k,\uparrow}$ and spin-down $\psi_{k,\downarrow}$ coexist. This mixed state leads to spin flip events during electron-electron scattering and electron-phonon scattering, as shown in Fig. 16. For electron-phonon scattering, the spin-flipping probability is denoted as $a_{sf}$ [Fig. 16(c)].

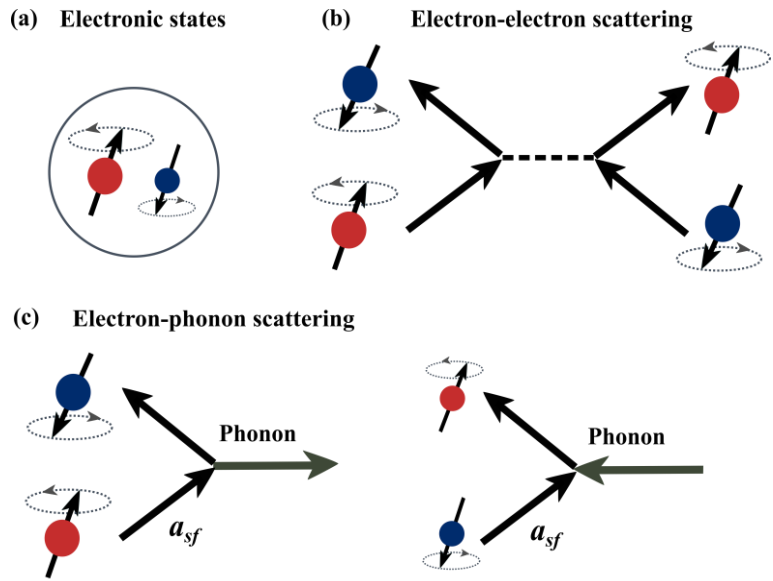

Fig. 16. The Elliott-Yafet spin-flipping mechanism. (a) Due to the spin-orbit coupling, the electron spin exhibits a blend of both spin-up and spin-down states. (b) Spin-flipping event due to electron-electron scattering. (c) Spin-flipping event due to electron-phonon scattering through emission or absorption of a phonon.

The electron state with a spin-up configuration can be expressed as follows:

$$\psi_{k,\uparrow} = \left[ a_k(\mathbf{r}) |\uparrow\rangle + b_k(\mathbf{r}) |\downarrow\rangle \right] e^{i\mathbf{k}\cdot\mathbf{r}} \tag{30}$$

where $\mathbf{k}$ represents the wave vector. Elliott and Yafet propose that the probability of spin-flipping $a_{sf}$ in a scattering event is linked to the spin mixing parameter $<b^2>$, and the value of $<b^2>$ can be formulated as follows:

$$\langle b^2 \rangle = \overline{\min\left( \langle \psi_k | \uparrow \rangle \langle \uparrow | \psi_k \rangle, \langle \psi_k | \downarrow \rangle \langle \downarrow | \psi_k \rangle \right)} \tag{31}$$

Equation (31) defines an appropriate average over the states involved around the Fermi level. When <$b^2$>=0, the electron spin is in a pure spin state. When <$b^2$>=0.5, it indicates that the spin states of 'spin up' and 'spin down' are equally distributed. The relation between $a_{sf}$ and <$b^2$> can be expressed as follows:

$$a_{sf} = p\langle b^2 \rangle \tag{32}$$

Here, $p$ represents a parameter specific to the material, with its typical value ranging from 1 and 10.

At the beginning, Elliott (1954) [289] suggested that the momentum-dependent spin mixing in the wave functions contributed additionally to spin relaxation. However, the Elliott approximation only considered the spin mixing caused by SOC, without taking into account the electron-phonon matrix elements and the real phonon dispersion spectrum. Subsequently, Yafet (1963) [290] introduced the concept of electron-phonon scattering, building upon the degeneracy of electron states in non-magnetic materials. Steiauf *et al.* (2010) [271] extended this theory to encompass magnetic material systems. This extension enhances the accuracy of spin relaxation calculations within the ultrafast spin dynamics of magnetic materials, while also considering the conservation of the total number of electrons. It is worth mentioning that Carva *et al.* (2013) [55] calculated E-Y electron-phonon scattering from first principles using a generalized spin-flip Eliashberg function and *ab initio* computed phonon dispersions. Their results show that the $a_{sf}$ obtained using the Elliott approximation is two to three times larger than that obtained from actual phonon calculations. This discrepancy may be because the Elliott approximation was originally derived for paramagnetic metals.

In addition, the E-Y scattering model, as applied to magnetic materials, still requires further modification. This includes additional considerations such as electron-electron scattering [288, 291, 292], the feedback effect between exchange splitting and spin-flip scattering [82], and the role of SOC in demagnetization dynamics dominated by E-Y scattering [83, 293], among other factors.

In 2011, Mueller *et al.* [285] combined the M3TM model to elucidate the respective roles of electron-electron scattering and electron-phonon scattering in laser-induced ultrafast demagnetization. They indicated that electron-electron spin-flipping scattering primarily contributes to the demagnetization phase, whereas electron-phonon spin-flipping scattering enhances the demagnetization effect and predominates in the relaxation phase. As shown in Fig. 17, the simulation illustrates the influence of four scenarios on ultrafast magnetization dynamics: electron-phonon scattering, electron-electron scattering, the coexistence of both scatterings, and the omission of the phonon system (phonon bath). Previous studies have shown that neglecting the contribution of one type of scattering in simulations leads to an underestimation of the demagnetization intensity [76, 294-296]. In addition, Mueller *et al.*(2013) [82] subsequently combined

Boltzmann scattering dynamics with the Stoner criterion model and introduced a relevant E-Y scattering spin-flipping model for magnetic materials. In this model, the disparity in chemical potential between majority and minority spins results in Stoner exchange splitting. This Stoner exchange splitting subsequently engenders an asymmetry in chemical potential, amplifying the demagnetization effect. This model circumvents the constraints of band structure at absolute zero temperature [95], thereby facilitating a quantitative examination of ultrafast demagnetization dynamics.

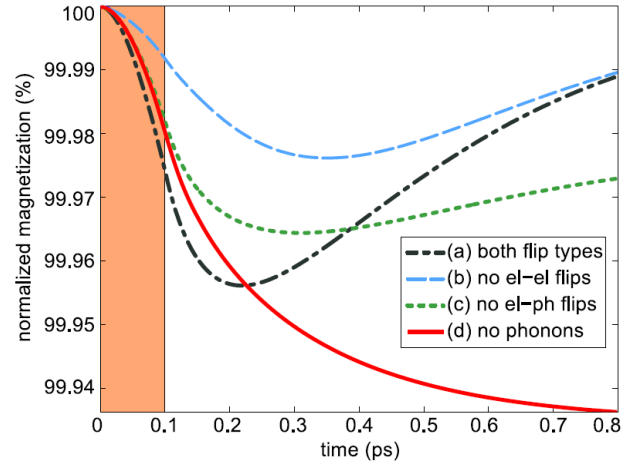

Fig. 17. Magnetization dynamics for four distinct scattering processes. (a) Simultaneous occurrence of electron-electron and electron-phonon scattering, (b) exclusive electron-phonon scattering, (c) exclusive electron-electron scattering, and (d) the omission of the phonon system. The computation is conducted using the spin-resolved Boltzmann transport equation within the context of a free electron gas. Reproduced with permission from Ref. [285].

In 2019, Chen *et al.* [84] further confirmed the important role of electron-electron scattering in the ultrafast demagnetization process of Ni film. They demonstrated that electron-electron interactions primarily govern the ultrafast demagnetization. Although electron-phonon scattering also plays a role, it is no longer the most critical factor. Chen *et al.* achieved an approximate simulation in line with experimental values, using a real-time time-dependent density functional theory (rt-TDDFT) combined with phenomenological LLG model. The rt-TDDFT could be used to capture the effect of light-spin, light-orbital, spin-orbit, electron-electron, and electron-phonon interactions, as well as the effects of finite-temperature spin disorder in larger systems during the ultrafast demagnetization, while the LLG model was applied to describe the subsequent spin dynamics. However, as discussed in Section 3.1, the limitations of the LLG equation hinder a comprehensive description of the magnetization precession process at high temperatures. Consequently, there is a need for further improvement in this method.

In 2020, Acharya et al. [85] investigated the role of electron correlations in femtosecond-scale ultrafast demagnetization dynamics. They proposed that electron correlations are the primary mechanism driving these dynamics, rather than photon-orbital momentum or electron-phonon interactions. Using a theoretical model based on noncollinear spin-density time-dependent density functional theory (TDDFT) and incorporating non-Markovian dynamics, they derived the exchange-correlation (XC) kernel from dynamical mean-field theory (DMFT) appropriate for transition metals with partially filled $d$ orbitals. This method enables the tracking of electron correlations on timescales where lattice effects can be neglected, such as during the initial ultrafast spin dynamics of Ni within the first 0–20 fs. Their simulations of Ni using different XC kernels revealed that demagnetization rates were significantly influenced by electron correlation-induced spin-flip events. It is worth noting that since the demagnetization time is closely correlated with the pulse duration, and the timescale of electron correlations and memory effects is expected to be in the range of 1–10 fs, Acharya et al. employed a short pulse duration of 7.2 fs. As shown in Fig. 18(a), the demagnetization rates for the scenarios with full memory (time dependence of electron-electron interaction) effects, without memory effects, and with memory effects only in the spin-flip part were 56%, 25.8%, and 50.6%, respectively. These findings indicate that the primary channel of demagnetization is the transition from the spin-up state to the spin-down state, with significant effects observed within the first 0.1 fs, dissipating within 1 fs [Fig. 18(b)]. This timescale is consistent with that of electron-electron scattering in related materials. This study demonstrates that electron correlations in a noncollinear spin system can lead to substantial ultrafast demagnetization in Ni.

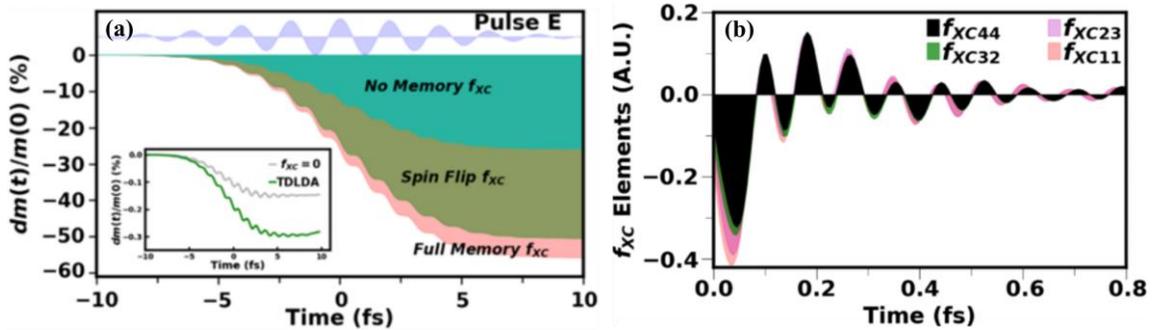

Fig. 18. Time dependency in ultrafast demagnetization using three different exchange-correlation (XC) kernels $f_{XC}$ through the TDDFT method in Ni. These are categorized as follows: (i) Pink, with full memory effects; (ii) Dark green, with no memory effects; (iii) Light green, considering only the spin-flipping component. (b) Time dependence of the nonzero components of the DMFT XC kernel of bulk Ni [85]. Reprinted with permission from S. R. Acharya et al., Phys. Rev. Lett., 125, 017202 (2020). Copyright (2020) by the American Physical Society.

In 2021, Chekhov et al. [125] proposed a model based on the spin-flipping mechanism. As shown in Fig. 19, this model accounts for almost all spin-conserving electron scattering events. When the electron density of states (DOS) is independent of electron energy, both spin-up and spin-down electrons exhibit nearly the same spin-flipping probability. The net magnetization remains unchanged [Fig. 19(a)]. As depicted in Fig. 19(b), when the DOS for spin-down electrons depends on electron energy, four spin-flipping events occur. Since the scattering probability of spin-up electrons in a high-energy state is higher than that of spin holes in a low-energy state, a significant number of spin-up electrons flip to a spin-down state, leading to a decrease in magnetization. As laser fluence increases, the number of spin-down electrons also rises, and causing a further reduction in magnetization. Although this model can be employed for the quantitative analysis of ultrafast demagnetization induced by optical and terahertz pump light, it relies on a weight function that is linearly related to the electron energy near the Fermi level in the unexcited sample. The validity of this approximation is still under investigation.

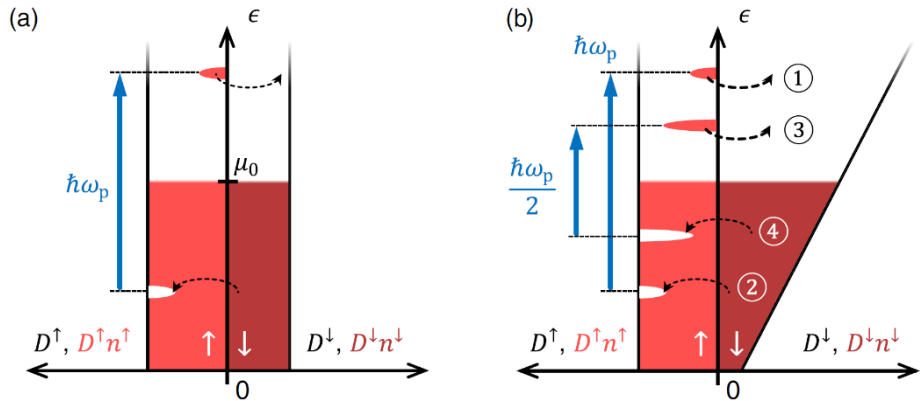

Fig. 19. Ultrafast demagnetization model based on the spin-flipping mechanism. (a) The model of the electronic DOS ($D^{\uparrow\downarrow}$) is independent of the electron energy ($\epsilon$). (b) A linear model of the electronic DOS ($D^{\uparrow\downarrow}$) and the electron energy ($\epsilon$) [125]. Reprinted with permission from A. L. Chekhov et al., Ultrafast Demagnetization of Iron Induced by Optical versus Terahertz Pulses, Phys. Rev. X, **11**, 16 (2021), DOI: 10.1103/PhysRevX.11.041055. CC BY 4.0.

Since E-Y scattering involves electron-phonon coupling, electron-electron interaction, and SOC, the influence of SOC on E-Y scattering has been thoroughly examined. In 2022, Zheng et al. [83, 293, 297] utilized time-domain *ab initio* nonadiabatic molecular dynamics (NAMD) to delineate the respective roles of electron-phonon coupling and SOC in the demagnetization dynamics of Ni films. Zheng et al. demonstrated that electron-phonon coupling is responsible for the direct relaxation channel of the same-spin states, while SOC provides the spin-splitting channel for opposite-spin states. As depicted in Fig. 20, based on the energy band structure of the ferromagnetic material, when the laser irradiates the Ni film, minority spin electrons will relax to the unoccupied same-spin states through electron-phonon coupling, and spin-flipping will not occur in these spins. Conversely, since majority spin electrons have no unoccupied same-spin states above the Fermi level, these spins relax to the minority spin state via SOC, resulting in spin-flipping. This study highlights the significant role of SOC in the ultrafast demagnetization of Ni films and offers a unique perspective on understanding the ultrafast demagnetization mechanism. However, the simulation overlooks the effects of photon excitation, electron-electron scattering, and magnon dynamics, resulting in a slightly weaker simulated demagnetization strength compared to the experimental value. Recently, Chen *et al*. [293] proposed a different perspective. They demonstrated that the enhancement of SOC reduces the effect of E-Y spin-flip scattering, thereby slowing down the demagnetization process. These seemingly contradictory conclusions suggest that SOC, to some extent, plays a significant role in the dynamics of demagnetization and cannot be overlooked.

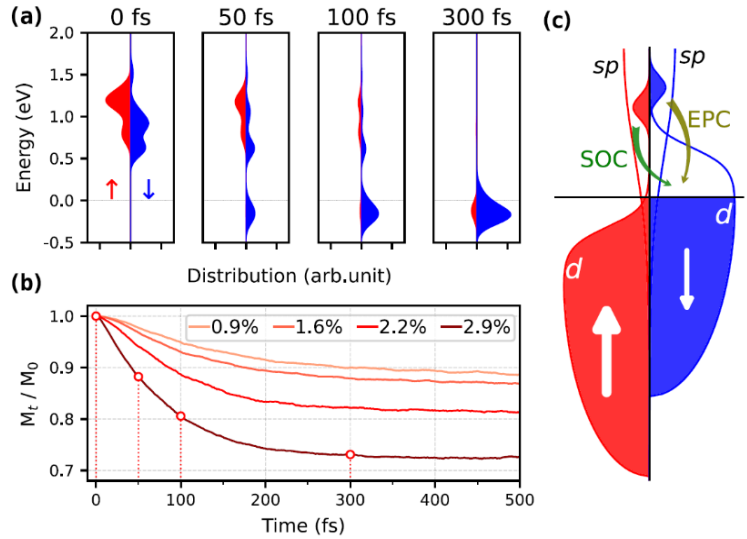

Fig. 20. (a) Electron distribution at various time points in a multi-electron NAMD simulation. (b) Changes in relative magnetic moments at different electron excitation rates over time. (c) A schematic representation of the mechanism of SOC-induced demagnetization in nickel. The electron-phonon coupling (EPC) leads to relaxation of the same-spin states without spin flip.[83]. Reprinted with permission from Z. F. Zheng *et al.*, Phys. Rev. B., **105**, 7 (2022). Copyright (2022) by the American Physical Society.

Elliott-Yafet spin-flip scattering also plays a significant role in the demagnetization dynamics of RE metals and TM-RE alloys [29, 61, 63, 268, 298-300]. However, this process is complex and requires careful consideration of various factors at different stages of demagnetization. These factors include spin-lattice coupling [301, 302], exchange splitting [299, 300], 4*f*-5*d* electronic coupling in rare earth metals [61, 300, 303], and 3*d*- 5$d$ hybridization [40] or 3$d$-4$f$ coupling [302] in TM-RE alloys.

In 2015, Teichmann *et al.* [299] demonstrated that in Gd and Tb, the exchange splitting-induced spin mixing increases the spin-flip probability of E-Y scattering. Recently, Frietsch *et al.*[268] investigated the demagnetization dynamics of the rare earth metals Tb and Gd. Their results indicated that due to the strong coupling between the 4$f$ orbital momentum and the lattice, as well as the coupling between spin and orbital moments in Tb, the excitation of spin waves can drive the demagnetization of 4$f$ magnetic moments. In contrast, the Gd specimen lacks this demagnetization channel, and its demagnetization is driven by the excitation of 5$d$ magnetic moments. This conclusion was corroborated by the work of Decker *et al.* [304]. They also revealed that ultrafast demagnetization in Gd is achieved through the spin flip of 5$d$ electrons, and E-Y spin-flip scattering occurs only when the 5$d$ electrons surpass a certain temperature threshold.

In 2022, Zhang *et al.* [302] investigated the demagnetization dynamics of GdFeCo using the E-Y spin-flip scattering theory. Their findings indicate that when the exchange coupling strength between the RE and TM element sublattices is strong, the 3$d$-4$f$ coupling leads to rapid heating of the 4$f$ spin. Consequently, the 4$f$ spin plays a dominant role in type II demagnetization. Conversely, when the exchange coupling strength is weak, the demagnetization is primarily governed by the 3$d$ spin, resulting in type I demagnetization.

### 4.1.2 Electron-magnon scattering

The electron-magnon scattering process involves the interaction between conduction electrons and magnons in a magnetic material. Magnons are quantized excitations of spin waves within a magnetic system, effectively representing collective oscillations of electron spins in the lattice. When an electron traverses a magnetic material, it can interact with these spin waves, leading to scattering. This interaction facilitates the exchange of energy and momentum between the electrons and the magnetic subsystem.

Previous studies have demonstrated that the demagnetization rate, as computed using the above E-Y scattering theory, is lower than the experimental value [76, 294, 295]. Consequently, some researchers infer that additional factors may be contributing to the ultrafast demagnetization process. Currently, researchers identify two primary sources of demagnetization related to electron-magnon interactions: i) E-Y spin-flip scattering or local Stoner excitations. These single-particle excitations quench the magnetic moment, thereby reducing exchange splitting and ultimately causing a shift in spin-polarized bands, which decreases the net magnetic moment of the material [82, 295, 299, 305]. ii) Collective spin excitation (transverse spin fluctuations or magnon generation). In this scenario, adjacent magnetic moments tilt towards each other, leading to a reduction in the overall magnetization [86, 87, 90, 306, 307]. As illustrated in Fig. 21, the first mechanism is derived from the Stoner model of itinerant ferromagnets, while the latter originates from the Heisenberg model. In the Heisenberg model, spin fluctuations can induce rapid changes in the spin-split electronic states, resulting in a band mirroring effect, which is a hallmark of magnon generation [306, 308, 309]. Electron-magnon interactions are associated with both of these demagnetization sources, and extensive research has been conducted to differentiate the two effects.

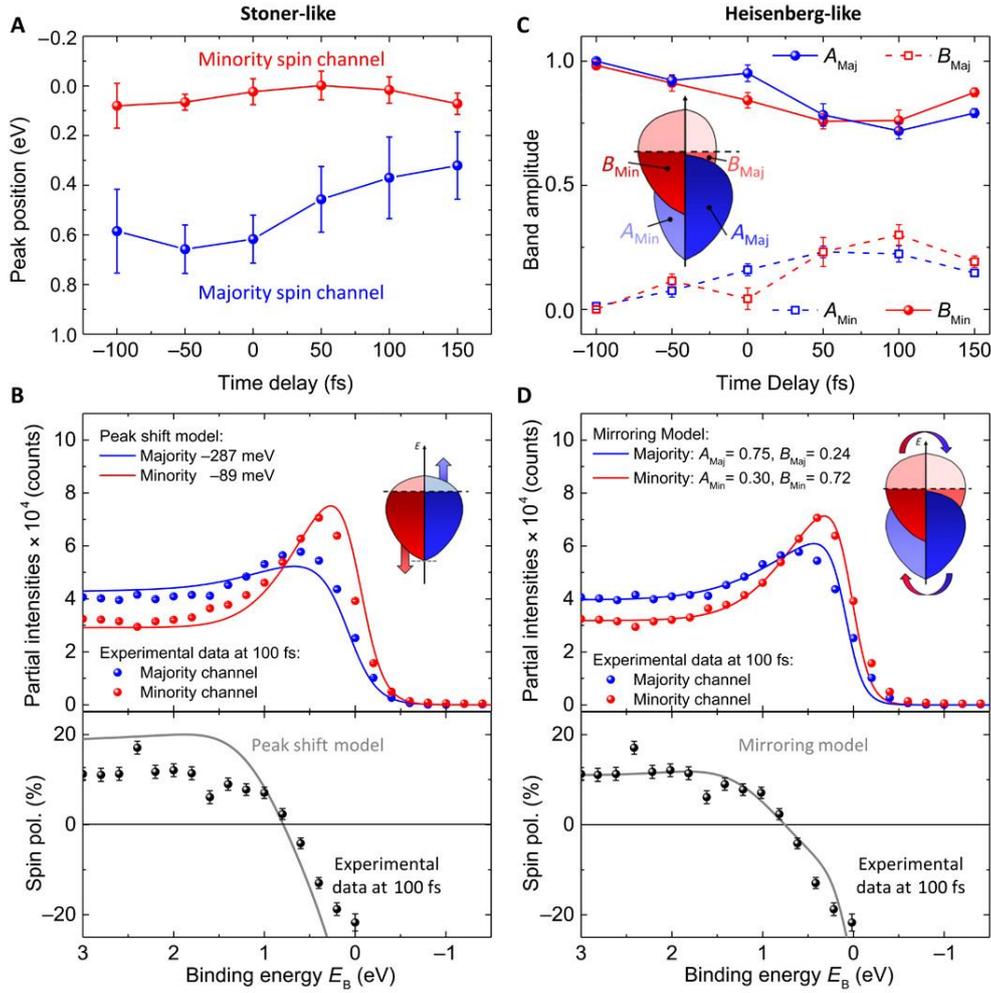

Fig. 21. Analysis of possible exchange collapse versus band mirroring by fitting the experimental data from time- and spin-resolved photoelectron spectroscopy. (a) Extracted energy shifts of the majority and minority bands as a function of time. The majority bands shift towards lower binding energies due to the reduction of the exchange splitting, while the minority bands remain unchanged. (b) Modeled majority and minority spectra (top) and spin polarization (bottom) in the Stoner-like picture via a potential collapse of the exchange splitting, and compared to the measured experimental data at $t = 100$ fs. The agreement between the modeled and experimental results is low. (c) Extracted amount of band mirroring. (d) Same as (b), but considering only band mirroring. The model is consistent with experimental results. Reproduced with permission from Ref. [306].

In 2003, Rhie *et al*. [305] demonstrated that the primary cause of the demagnetization of the Ni/W(100) sample was the reduction of exchange splitting. This conclusion was subsequently confirmed by numerous photoemission studies [26, 82, 295, 299]. However, in 2008, Carpene *et al*. [90] presented an opposite conclusion in their study of Fe/MgO by combining TR-MOKE and time-resolved reflectivity (TR-R) techniques, proposing that magnon generation is the primary mechanism driving demagnetization. They suggested that electron-magnon scattering facilitates the transfer of spin to orbital angular momentum, which is subsequently absorbed by the lattice. The dominant role of the magnon generation in the demagnetization process has also been demonstrated in various systems [86-88]. Schmidt *et al*. [86] utilized two *ab initio* calculation methods: i) considering only single-particle Stoner excitations within the GW approximation (used to calculate self-energy in multi-body systems); ii) incorporating multiple scattering of the Stoner electron-hole pairs and spin wave excitations in the GW scheme. They proposed that comparing these two approaches can elucidate the impact of magnon generation on the decay rates of spin-up and spin-down electrons. Haag *et al*. [87] reproduced these findings using *ab initio* calculations based on Fermi's golden rule. They quantitatively analyzed the emission and absorption of magnons, as well as the electron-magnon scattering rate in Fe and Ni. Despite the magnon emission rate being higher than the absorption rate, the calculated demagnetization rates for Ni and Fe were significantly lower than those observed experimentally. Therefore, neither electron-magnon scattering nor electron-phonon scattering alone can account for the observed demagnetization, contradicting the conclusions of Carpene *et al*. [90].

In 2017, Eich *et al*. [306] were the first to use time- and spin-resolved photoelectron spectroscopy with an XUV probe to capture the transient behavior of over the full energy range of the valence bands in 3*d* ferromagnetic materials (Fig. 21). Although they seemingly observed a shift in the spin-polarized bands intensities induced by Stoner excitations, the fitting results of the spin polarization collapse did not align well with this phenomenon. Conversely, the band mirroring model successfully replicated this shift. They demonstrated that the band mirroring effect is the primary cause of the rapid decay of spin polarization at high binding energies and plays a crucial role in the demagnetization of Co/Cu(001).

Since the demagnetization responses and magnon modes can be captured by using TR-RIXS, this technique allows researchers to directly study excitations in the time domain and investigate transient phases. Previous studies have provided theoretical insights into spin wave (magnon quasiparticle) excitations in RIXS [310-312]. RIXS has been extensively utilized to investigate spin-resolved valence band excitations and collective magnetic excitations in transition metal alloys [198, 313-315]. Notably, Elnaggar *et al.*[198] employed momentum-resolved-RIXS to directly observe spin ordering with element- and site-selective precision. Dean *et al.* (2016) [316] and Cao *et al.* (2019) [317] used TR-RIXS to directly measure the magnetization dynamics of the Mott insulator $Sr_2IrO_4$. Dean *et al.*'s results indicated that, although the magnetic Bragg peak was nearly obliterated in the transient state, the high-energy magnons remained observable at the momentum transfer position $Q = (\pi, 0)$ and these magnons exhibited negligible changes before and after pump injection. In contrast, the low-energy magnons at $Q = (\pi, \pi)$ showed significant changes. This phenomenon is likely due to the high-energy magnons re-establishing equilibrium within 2 ps and the higher-energy excitation decaying into lower-energy multiparticle excitations via multiple methods. The application of this technique will significantly advance the understanding of electron-magnon interactions.

In 2024, Weißenhofer and Oppeneer [307] developed an *ab initio*-based theory using a quantum kinetic approach to describe magnon generation caused by electron-magnon scattering. They found that after pulse laser irradiation of Fe film, the simple relation between magnetization and temperature calculated in the equilibrium state does not hold. This is because the process is primarily driven by high-energy (non-thermal) magnons. The laser-excited electrons produce fewer magnons when transferring energy, resulting in weaker demagnetization than expected. The calculations of Weißenhofer and Oppeneer showed that a substantial number of non-thermal magnons are excited under pulsed laser irradiation, inducing significant demagnetization within 200 femtoseconds. This finding highlights the crucial role of magnon generation in the demagnetization process.

Moreover, numerous studies have demonstrated that demagnetization arises from multiple sources [318, 319]. Turgut *et al.* [318] identified the band mirroring effect and a reduction in exchange splitting within the $Co/Ta/SiO_2/Si$ system. In 2019, Müller *et al.* [319] demonstrated that the calculation method employed by Schmidt *et al.* [86] was insufficient to account for the broadening observed in photoemission experiments and might have missed significant scattering events. They proposed that the contribution of Stoner excitations is comparable to that of the collective spin excitations of magnons during the demagnetization process.

Although these conclusions are contradictory, the contribution of electron-magnon interactions in demagnetization is of significant importance. We hope that these findings can be experimentally validated across a broader range of systems in the future.

### 4.1.3 Photon-spin interaction

The photon-spin interaction has garnered significant attention. In the early stages, the thermal excitation mechanism was employed to explain the ultrafast spin dynamics induced by laser pulses, while the rapid and nonthermal control of magnetism by lasers was disregarded. In 1966, Shen *et al.* [320] discussed photon-magnon coupling in ferromagnetic materials inducing the spin-Raman effect, resulting in spin-flipping. Subsequently, a question has perplexed researchers for over a decade: Does photon-magnon coupling contribute to the ultrafast demagnetization process? In 2000, Zhang and Hübner [91] proposed that neither the laser field nor the single SOC of the material alone could independently lead to significant ultrafast demagnetization of the sample on the femtosecond timescale.

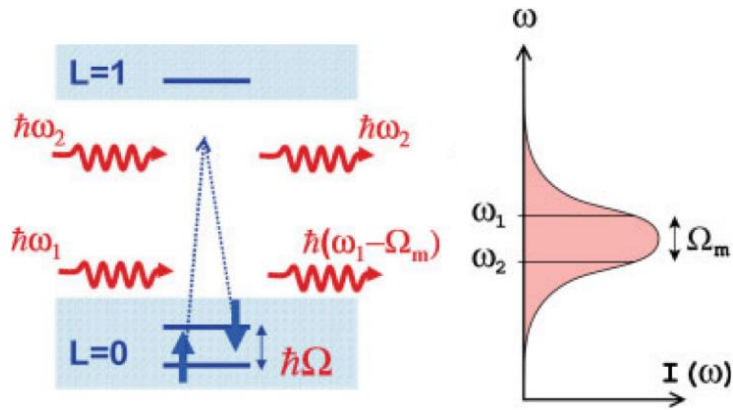

Fig. 22. Ultrafast spin-flipping via the process of the interaction between photons and spins [92]. Reprinted with permission from A. V. Kimel *et al.*, Laser Photonics Rev., 1, 275 (2007). Copyright (2007) by the WILEY‐VCH Verlag GmbH & Co. KGaA, Weinheim.

The photon-spin interaction process (stimulated Raman scattering) is depicted in Fig. 22. Initially, electrons reside in the ground state. When a laser beam impinges upon the film's surface, electrons within the irradiated region become excited to a higher energy state. The corresponding wave function becomes a superposition of several eigenstates at this point. Subsequent increments in electron orbital angular momentum contribute to an enhancement in SOC, thereby increasing the likelihood of spin-flip events. However, if the photon energy is insufficient to induce electron transitions to an excited state, electrons in the ground state can still undergo spin-flips and emit a photon with energy $\hbar\omega_2 = \hbar(\omega_1-\Omega_m)$ [92].

Since magnetization fundamentally involves changes or transfers in angular momentum, circularly polarized photons, which also carry angular momentum, can be utilized for control. By using the angular momentum of photons to directly alter the spin of electrons, it becomes possible to achieve ultrafast and non-thermal control of magnetization via laser. Kimel *et al.* [92] considered a case of non-dissipative light-matter interactions. These models consider the coherent interaction between photons and spins without accounting for energy dissipation, providing insights into the fundamental processes governing spin dynamics induced by ultrafast laser pulses. They posited that photon angular momentum remains conserved in laser-induced ultrafast spin dynamics, owing to the unaltered photon number and polarization in Raman scattering. However, in the this model, Woodford [93] presented a contrasting viewpoint regarding the conservation of photon angular momentum. Woodford demonstrated that SOC influences the direction of propagation for scattered photons, dependent on spin variations, which, in turn, modifies the photon's angular momentum. Moreover, Woodford demonstrated that photons can impart angular momentum in the absence of any other sources of angular momentum in the inverse Faraday effect.

In 2009, Bigot *et al.* [24] observed a nonlinear relationship in the thermalization time between electrons and spins, and raised doubts about the dominant spin-flipping mechanism (E-Y scattering) in laser-induced ultrafast spin dynamics. They also observed the existence of spin-independent electron dynamics, demonstrating coherent coupling between femtosecond lasers and magnetization in ferromagnetic Ni and $CoPt_3$ films. This coupling originates from relativistic quantum electrodynamics. The contribution of photon-spin interactions in ultrafast demagnetization is subsequently supported by theoretical work [150, 286, 321, 322]. Mondal *et al.* (2017) [322] examined ultra-relativistic photon-spin interactions based on the general relativistic Hamiltonian. They indicated that when the polarization directions of the pump laser and the probe laser are perpendicular, the overlap of the two beams generates a light-induced opto-magnetic field. The variation of the induced opto-magnetic field with pump fluence aligns with the variation in the amplitude of spin oscillations observed in magneto-optical measurements. The spin oscillations refer to the oscillatory behavior of magneto-optical signals occurring on a timescale of several hundred picoseconds following laser excitation. This phenomenon is attributed to the oscillation of magnetization. This research provides a theoretical explanation for the observations made by Bigot *et al.* [24] through the lens of relativistic light-spin interactions. In 2021, Mondal *et al*. [323] incorporated the relativistic optical spin-orbit torque (OSOT) into the LLG equation. This torque arises from the SOC between the spin and the external electromagnetic field (light field). The magnitude of OSOT depends on the helicity, intensity, and frequency of the ultrafast light pulse. Their study demonstrated that the effect of OSOT on the spin is similar to that of the Zeeman torque. This torque would not disappear even when the temperature of the ferromagnetic material surpasses the Curie temperature. In the same year, Popova-Gorelova *et al*. [323] demonstrated that fully controlling the magnetization dynamics in the material requires not only describing the magnetization dynamics following pump pulse excitation but also to considering the coupling between electron spins and pulsed lasers. This work considered the electronic structure which is far from that of a metal. However, the primary emphasis was on the antiferromagnetic system, which falls outside the scope of this discussion.

In comparison to other demagnetization mechanisms, photon-spin interactions are less prominent as they are unlikely to induce a significant demagnetization effect. The direct interaction of photons with materials presently leads to the rapid generation of a substantial number of excited electrons. This thermal effect seems sufficient to obscure the photon-spin coupling mentioned earlier. Studies have also indicated that the contribution of photon-spin interactions to demagnetization dynamics is negligible [66, 74, 84]. Notably, by combining TR-MOKE and time-resolved magnetization modulation spectroscopy (TIMMS), Longa *et al*. [66] observed a demagnetization phenomenon irrespective of whether the photon angular momentum was parallel or antiparallel to the magnetization. The TIMMS measurements revealed that the influence of photons on demagnetization is less than 0.01%. Their findings suggest that neither direct angular momentum transfer between the laser field and the spin nor helicity-dependent laser-enhanced spin-flip scattering are the primary mechanisms driving the demagnetization.

Therefore, the contribution of photon-spin interactions to demagnetization dynamics, as well as the influence of various pulsed laser parameters on these interactions, remain controversial topics. These discussions undoubtedly advance our understanding of magnetization dynamics under laser excitation.

### 4.2 Spin transport processes

In this section, three different spin transport processes are discussed, including superdiffusion transport, electron magnon scattering-induced spin transfer, and optically induced intersite spin transfer (OISTR).

### 4.2.1 Superdiffusion transport

In 2008, Malinowski *et al*. [71] discovered that laser-induced spin transport within a spin valve structure can significantly enhance the process of ultrafast demagnetization. Subsequently, Battiato *et al*. [53, 95] proposed a model to elucidate the superdiffusion transport of spin-polarized electrons. This model successfully explained the ultrafast demagnetization phenomenon in Ni/Al films. They also posited that ultrafast demagnetization induced by the superdiffusion spin transport mechanism occurs more rapidly than that resulting from the spin-flipping mechanism. Unlike the spin-flipping mechanism, the angular momentum transfer in the spin transport process is explained by the electron transport induced by laser excitation and the distinct lifetimes of majority and minority spin electrons.

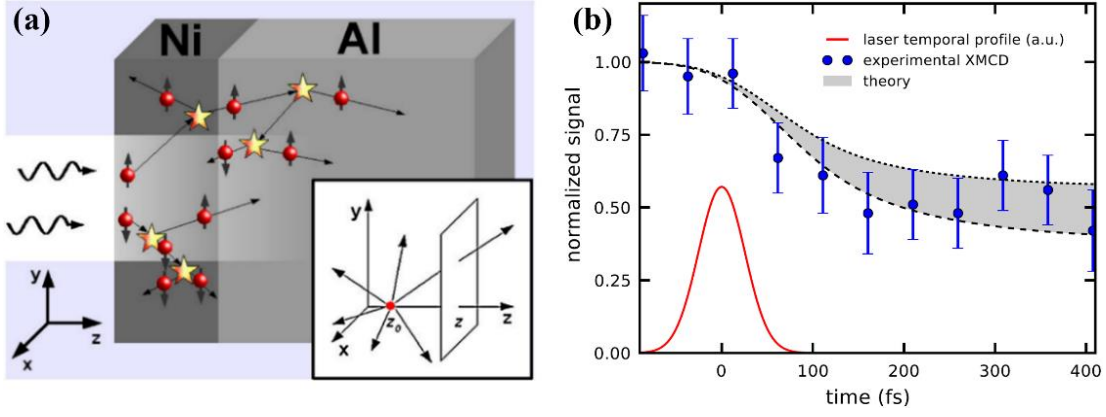

Fig. 23. (a) Schematic diagram of the laser-induced superdiffusion spin transport model. (b) Ultrafast spin dynamics of the Ni film. The shaded area represents the theoretical result, the blue points depict the experimental values obtained via XMCD, and the red solid line represents the laser pulse [53]. Reprinted with permission from M. Battiato *et al.*, Phys. Rev. Lett., **105**, 027203 (2010). Copyright (2010) by the American Physical Society.

Within the framework of the superdiffusion spin transport model [53, 95], the laser pulse excites electrons from the *d* orbital to the *sp* orbital. In comparison to *d*-band electrons, *sp*-band electrons exhibit higher mobility. These excited electrons move randomly in all directions. As illustrated in Fig. 23(a), when all electrons travel a certain distance *s*, the probability *P* of scattering remains constant:

$$P(s) = \exp\left[-\int_0^s ds' / \tau(z(s')) v(z(s'))\right] \quad (33)$$

where $\tau$ and $v$ represent the lifetime and speed of the electron, respectively. These parameters depend on the spin and energy of the electron as well as the electron's position within the material.

Assuming the initial position of the electron is $z_0$, the number of electrons passing through an infinite plane located at $z$ and perpendicular to the $z$-axis at time $t$ can be calculated by integrating over all possible emission angles of a single excited electron. This quantity is referred to as the average flux, denoted as $\phi(z,t;z_0,t_0)$. Considering a distributed source of excited electrons, the total flux can be expressed as:

$$\Phi(z,t) = \int_{-\infty}^{+\infty} dz_0 \int_{-\infty}^{t} dt_0 S^{ext}(z_0,t_0) \phi(z,t;z_0,t_0) \quad (34)$$

where $S^{ext}$ represents the electron source term. Based on the expression of the total flux, the continuity equation for the density of first-generation electrons can be formulated as follows:

$$\frac{\partial n^{[1]}}{\partial t} + \frac{n^{[1]}}{\tau} = -\frac{\partial \hat{\phi} S^{ext}}{\partial z} + S^{ext} \quad (35)$$

The second term on the left side of Eq. (35) represents the number of electrons that are scattered and subsequently transformed into the secondary generation of electrons. The operator $\hat{\phi}$ is defined by $\hat{\phi} S^{ext} = \Phi$. Similarly, the second, third, and fourth scattering events of electrons can be calculated. By combining all the resulting equations, the complete transport equations can be derived:

$$\frac{\partial n^{tot}}{\partial t} + \frac{n^{tot}}{\tau} = \left( -\frac{\partial \hat{\phi}}{\partial z} + \hat{I} \right)\left( \hat{S} n^{tot} + S^{ext} \right). \tag{36}$$

where $\hat{I}$ is the identity operator. Eq. (36) characterizes the dynamics of nonequilibrium electrons excited by lasers. In this model, laser-excited electrons experience a finite, yet nonzero, number of scattering events during transport. This is distinct from ballistic transport (where no scattering events occur) and diffusive transport (which involves countless scattering events). In the context of spin-dependent electron transport, there is a modification in magnetization. Typically, after thermalization, electrons transition to energy bands characterized by lower mobility, subsequently diminishing the impact of transport.

The process of angular momentum transfer in spin transport differs from that of spin-flipping. The former involves nonlocal superdiffusion of spin currents, leading to angular momentum dissipation. The latter, however, is characterized by the scattering of spins induced by phonons or lattice defects, resulting in angular momentum dissipation to the lattice system, and this dissipation is localized. Although the significant differences in the microscopic description of the spin-flipping and spin-transport mechanisms, the superdiffusion spin-transport model successfully accounts for most of the experimental results. Consequently, researchers initiated a series of investigations into the ultrafast spin dynamics of this mechanism [54, 72-74, 95, 99, 197, 324-329].

In a multi-domain structure, due to the different lifetimes and velocities of the majority and minority hot electrons, the majority hot electrons traverse the domain wall into the adjacent domain, where they become minority electrons. This process leads to the accumulation of minority electrons near the domain walls, which increases the softness of the domain wall and decreases the overall magnetization. Therefore, investigating the temporal and spatial evolution of the domain structure is a valuable method for studying spin transport [197, 326, 329].

In the work of La-O-Vorakiat *et al.* [330], it was predicted that XMCD in an HHG-based pump-probe spectroscopy experiment would enhance the temporal and spatial resolution of magnetic domain imaging. In 2011, Vodungbo *et al.* [195] first demonstrated the application of this method for imaging nanoscale magnetic domain structure and investigated ultrafast demagnetization in a periodic material system.

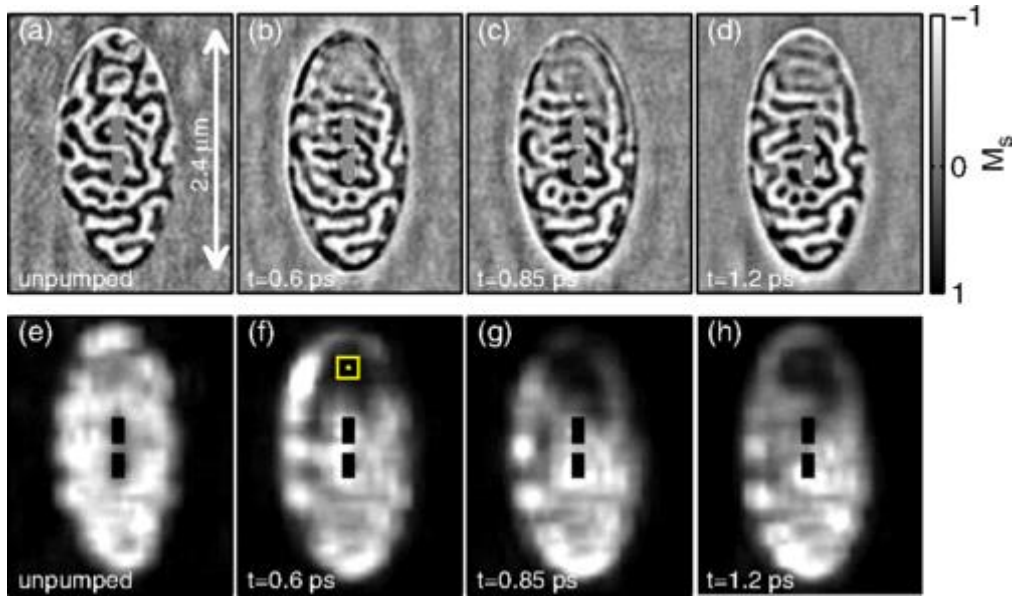

Fig. 24. Magnetic domain structures of Co/Pt samples without pump light are depicted in (a), and with pump light in (b) to (d) at selected delay times between optical excitation and holographic probing. The local magnetic contrast decreases due to laser excitation. The panels (e) to (h) show the calculated variance of XMCD magnetic contrast. [326]. Reprinted with permission from C. von Korff Schmising *et al.*, Phys. Rev. Lett., **112**, 217203 (2014). Copyright (2014) by the American Physical Society.

Subsequently, numerous researchers have employed this experimental method to investigate the ultrafast demagnetization and temporal evolution of magnetic domains in films of Co/Pt [197, 326], Co/Pd [54, 59, 126, 131], Co/Cu/Ni [331], Co/Fe$_{75}$Gd$_{25}$ [332], and GdFeCo [192] with inhomogeneous magnetic structures. Fig. 24 illustrates a representative instance of the temporal evolution of magnetic domains in a Co/Pt film. Since XMCD allows for the visualization of magnetic domain evolution during ultrafast spin dynamics, it has become a crucial method for probing ultrafast spin dynamics. This experimental technique enhances the credibility of our understanding of spin transport mechanisms.

In 2020, Baláž *et al.* [329] calculated the ultrafast demagnetization caused by superdiffusive spin transport between domains. Their findings indicate that spin transport leads to the blurring of the domain wall. Further calculations reveal that when a laser is precisely focused near the domain wall, a substantial spin current can be induced, resulting in STT. This STT prompts the domain wall motion with a velocity of $10^4$ m/s. Concurrently, partial ultrafast demagnetization occurs on the same timescale as the STT. The work of Baláž *et al.* demonstrated that laser-driven currents can induce both demagnetization and STT. However, this model presupposes that domain wall motion does not affect spin transport, and that the transport of spin currents is restricted by their mean free path [53, 197]. In recent years, Jangid *et al.*[51] utilized time-resolved ultrafast EUV magnetic scattering to observe that the rapid shift duration of the domain wall within the labyrinth domain is significantly slower than the demagnetization time, while the recovery speed is faster than the relaxation process. They attributed this phenomenon to an effective inertia-bound magnetic soliton of the domain wall, which cannot be driven by torque. Consequently, they concluded that this laser-induced domain wall movement is solely due to ultrafast demagnetization.

Investigating the impact of spin transport on demagnetization dynamics is not limited to magnetic domain imaging techniques. The methods discussed in Chapter 2 are also relevant. However, since spin transport and spin flipping effects generally coexist in demagnetization processes, it is essential to meticulously discern the contributions of other mechanisms. Researchers have investigated the ultrafast demagnetization induced by spin transport mechanisms through the elemental specificity of XMCD [74, 128, 192, 330, 333], manipulation of film thickness [71], analysis of the demagnetization signal from the substrate side [72, 325, 327], and examination of the ultrafast magnetization dynamics induced by laser-excited spin current in noncollinear magnetic multilayers [15, 325, 327, 334]. These approaches can reduce the contributions to demagnetization generated by direct photon excitation and spin-flipping mechanisms to some extent.

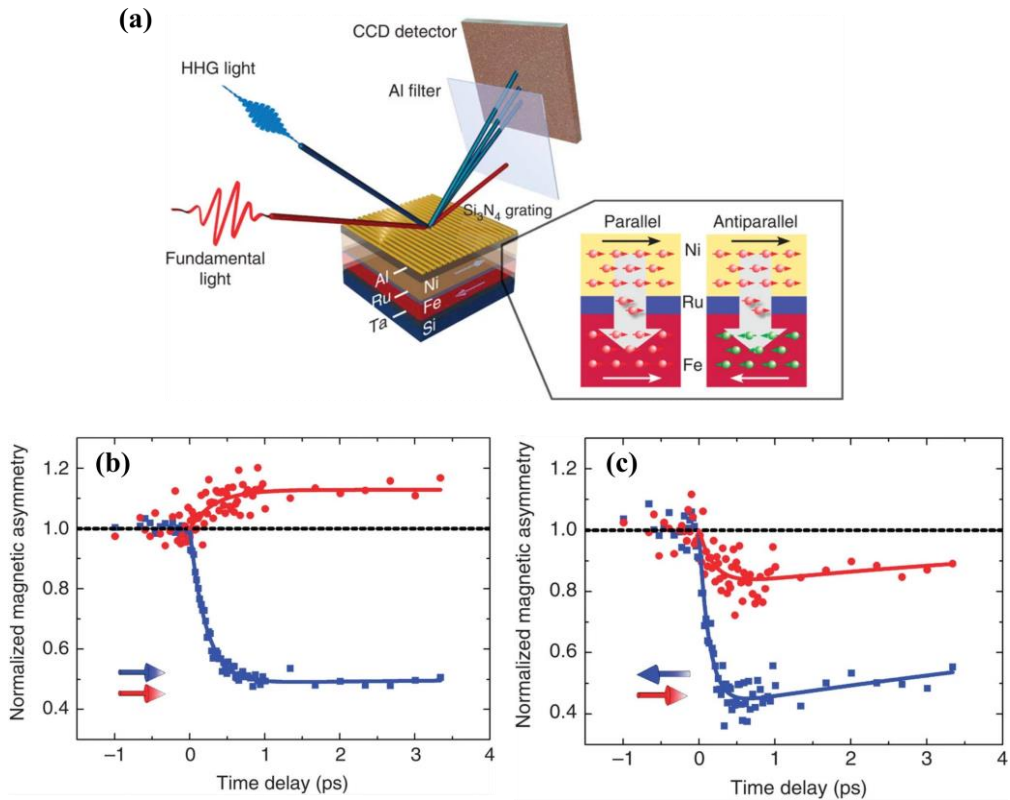

Fig. 25. (a) Experimental setup and schematic of the Al/Ni/Ru/Fe multilayer sample. Time-resolved magnetization measurements of the Fe and Ni layers in the Ni/Ru/Fe multilayer for parallel (b) and antiparallel (c) magnetization alignment [73]. Reprinted

with permission from D. Rudolf *et al.*, Nat. Commun., 3, 1037 (2012). Copyright (2012) Springer Nature Limited.

In 2009, La-O-Vorakiat *et al*. [330] were the first to employ soft X-rays generated through high-harmonic generation (HHG) as probing light to measure element-resolved ultrafast magnetization dynamics. Rudolf *et al*. [73] employed XUV T-MOKE to investigate the demagnetization dynamics in Ni/Ru/Fe multilayers, demonstrating that superdiffusive spin transport significantly influences the demagnetization process. They found that the demagnetization dynamics in the Fe layer depend on the magnetization orientations of both the Ni and Fe layers. Following laser excitation of the sample, the majority spin electrons in the Ni layer migrate to the Fe layer. When the magnetization directions of the Ni and Fe layers are aligned parallel [Fig. 25 (b)], the majority spin electrons from the Ni layer increase the population of majority spin electrons in the Fe layer, thereby enhancing its net magnetization. Conversely, when their magnetization directions are antiparallel [Fig. 25 (c)], the majority spin electrons from the Ni layer reduce the minority spin electron population in the Fe layer, leading to a decrease in its net magnetization. In 2013, Eschenlohr *et al*. [74] employed the element-selective XMCD technique to investigate the ultrafast spin dynamics of Ni layer in Au(30 nm)/Ni(15 nm)/Pt(3 nm) multilayers. By leveraging photon absorption within the Au layer, they excluded contributions from photon excitation and focused on examining the ultrafast demagnetization driven by superdiffusive electron currents in the Ni layer. Their findings indicated that direct photon excitation is not a prerequisite for demagnetization. Instead, the primary mechanism driving the demagnetization process is superdiffusive spin transport. These results provide compelling evidence for the significant role of spin transport mechanisms in the ultrafast demagnetization process. However, this conclusion was challenged by Khorsand *et al*. [335], who calculated the absorption profile of light in the multilayer structure used by Eschenlohr *et al*. [74]. Their results demonstrated that Eschenlohr *et al*. [335] overestimated the absorption in the Au layer and that direct photo excitation still occurred in the Ni layer.

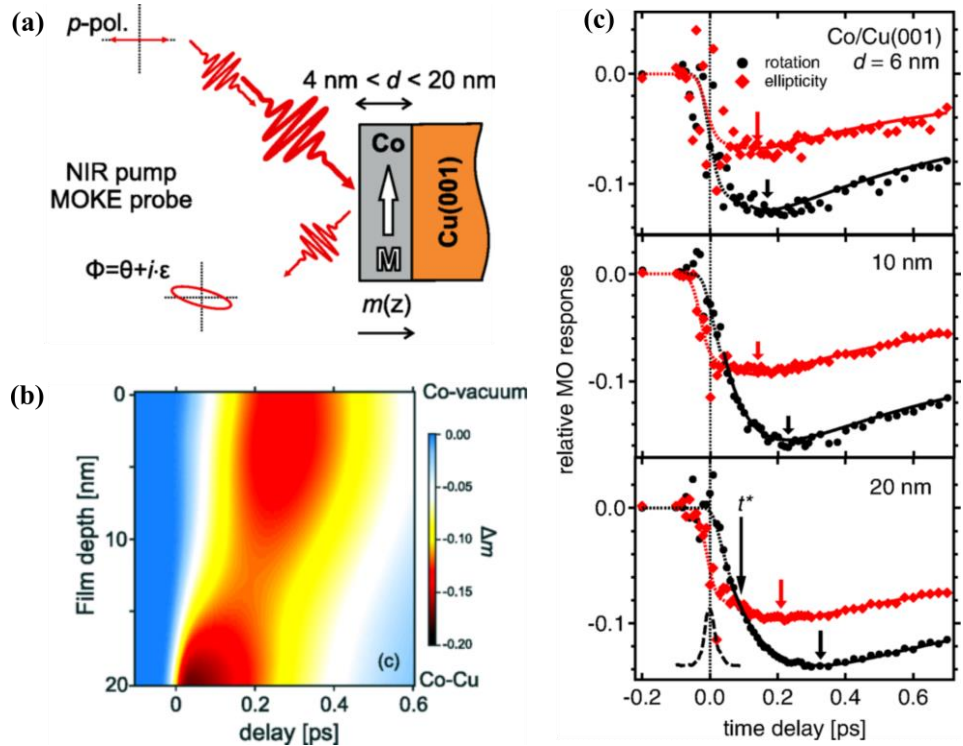

Fig. 26. (a) Kerr rotation and ellipticity of Co/Cu(001) measured using a pump-probe experiment. (b) Time-dependent relative magneto-optical (MO) responses of Kerr rotation and ellipticity for Co/Cu(001) with Co film thicknesses of 6, 10, and 20 nm. (c) Simulated spatiotemporal variation of the relative magnetization change for a 20 nm thick Co film on Cu(001), displayed in a false-color representation as a function of time delay and position within the film [99]. Reprinted with permission from J. Wieczorek et al., Phys. Rev. B, 92, 174410 (2015). Copyright (2015) American Physical Society.

In 2015, Wieczorek et al. [99] attempted to distinguish between spin-flipping and spin transport contributions in a Co/Cu(001) film by analyzing the Kerr rotation angle signal and the depth sensitivity of the ellipticity signal. The Kerr rotation signal is more sensitive to surface magnetism, whereas the ellipticity signal is more sensitive to the Co/Cu interface. They observed that the ellipticity decay and recovery occur earlier than the Kerr rotation, as depicted in Fig. 26. Their calculations, based on the spin-dependent electron diffusion model, indicate that the spin-polarized current flowing from Co to Cu causes the demagnetization of Co before electron thermalization. As the electronic system begins to thermalize, the spin current diminishes, and spin flipping becomes the dominant mechanism of demagnetization. However, this model necessitates the determination of the film's thickness sensitivity function, signifying the high requirement for correlational analysis regarding thickness dependency. This finding seems to contradict the earlier conclusions of Schellekens et al. [336]. Schellekens et al. investigated the demagnetization of Ni films deposited on insulating sapphire substrates, where spin transport may be suppressed. They found no significant difference in the demagnetization dynamics when the pump light was incident from the front and back of the sample. Even when a wedge-shaped conductive aluminum buffer layer was introduced between the substrate and the nickel film to amplify the ultrafast demagnetization resulting from spin transport, the demagnetization of the Ni/Al film remained unaccelerated compared to a single-layer nickel film. Therefore, they concluded that the contribution of spin transport was negligible.

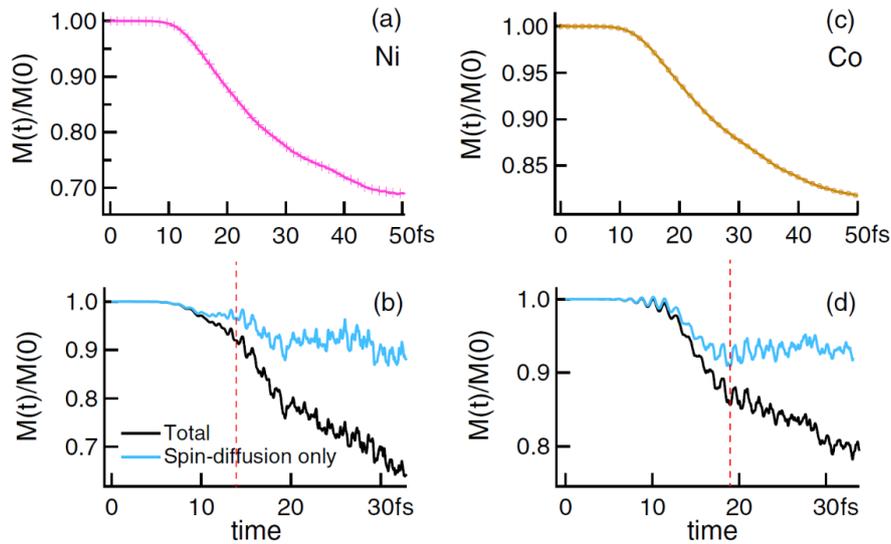

Fig. 27. The complete demagnetization process of (a) Ni and (c) Co films, along with the calculated spin transport-induced demagnetization (blue) and overall demagnetization (black) for (b) Ni and (d) Co films [249]. Reprinted with permission from V. Shokeen *et al.*, Phys. Rev. Lett., **119**, 107203 (2017). Copyright (2013) by the American Physical Society.

In 2017, Shokeen *et al.* [249] compared the influence of spin-flipping and spin transport on ultrafast demagnetization by detecting the TR-MOKE signal from the front and back sides of Ni and Co films of varying thicknesses. They also performed *ab initio* calculations on the ultrafast spin dynamics of Ni and Co films using time-dependent density functional theory. This approach allowed them to differentiate between the total demagnetization signal and the spin transport-induced demagnetization signal. The results are presented in Fig. 27. For Ni films, spin-flipping was found to be the dominant mechanism in the demagnetization process. In the case of Co thin films, the demagnetization within the first 20 fs was primarily driven by spin transport, before transitioning to the spin-flipping mechanism after 20 fs.

Although these studies present seemingly contradictory results, they provide valuable insights into the contribution of spin transfer in the demagnetization process.

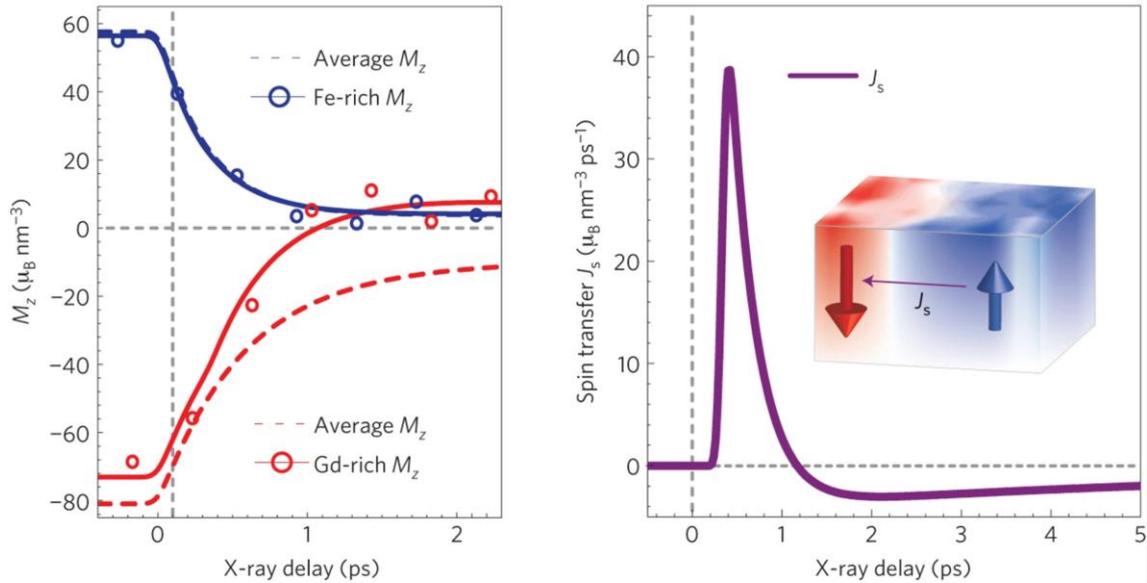

Fig. 28. (Left) Time evolution of $M_z$ in the Gd-rich and Fe-rich regions compared with the sample average. (Right) Time-resolved angular momentum flow, $J_s$, into the Gd-rich regions [192]. Reprinted with permission from C. E. Graves *et al.*, Nat. Mater., **12**, 293–298 (2013) Copyright © 2013, Springer Nature Limited.

Furthermore, researchers have extensively investigated the role of spin transport in the demagnetization dynamics of RE-TM alloys [128, 192, 337]. In 2013, Graves *et al.* [192] examined the demagnetization behaviors of a compositionally inhomogeneous GdFeCo sample. As depicted in Fig. 28, both Fe and Gd exhibited demagnetization in the Fe-rich regions of the film. However, in the Gd-rich regions, a magnetization reversal of Gd was observed at 1 ps, while the magnetization direction of Fe remained unchanged. The change in magnetic susceptibility of Gd was significantly greater than that of Fe. They attributed this phenomenon to the Gd-Fe exchange interaction and spin-torque scattering induced by non-collinear spins and abrupt chemical variations on the electron scattering length scale at the interface of the Gd-rich regions. This interaction caused angular momentum to flow from the Fe-rich regions to the Gd-rich regions, reaching its maximum at 1 ps. In 2020, Hennes *et al.* [128] observed that the demagnetization time of $Co_{88}Tb_{12}$ samples with stripe domains was shorter than previously reported values [338]. They supposed that superdiffusive spin transport might contribute to this phenomenon. However, the magnetic domains did not exhibit the anticipated ultrafast broadening. Since the study was conducted at low laser power, the local temperature of the film might have been too low to induce a spin-polarized current.

### 4.2.2 Electron magnon scattering-induced spin transfer

The above studies are based on the superdiffusion model of non-thermal electrons proposed by Battiato *et al.* [53, 95]. In 2015, Tveten *et al.* [339] introduced a non-equilibrium theory of ultrafast spin dynamics in magnetic heterostructures based on the *s-d* model of ferromagnetism. Under the influence of *s-d* interactions, itinerant *s* electrons exhibit a finite spin density in equilibrium. Following laser excitation of the sample, an out-of-equilibrium spin accumulation ($\mu_s \equiv \delta\mu_\uparrow - \delta\mu_\downarrow$) arises within the material. This spin accumulation subsequently leads to electron-magnon scattering. Tveten *et al.* demonstrated that electron-magnon scattering alters the magnon temperature and spin density. During this process, the splitting of the chemical potential of the itinerant *s* electrons system induces spin currents, which further enhance the demagnetization dynamics [328, 339-342].

In 2020, Beens *et al.* [341] investigated the interaction between local and non-local spin dynamics in a collinear magnetic heterostructure using a simplified *s-d* model. Their calculation results demonstrated that electron-magnon scattering induces short-term spin accumulation at the interface, which can partially counteract demagnetization. The resulting spin current (interface spin transport) generated by this spin accumulation can subsequently enhance the demagnetization rate. Soon after, Beens *et al.* [328] further calculated the spin transport induced by magnons and spin-polarized electrons in the Ni(5 nm)/Pt(3 nm) bilayer. Figure 29(a) presents the calculated results of the interface spin current (the spin current carried by conduction electrons and the interfacial magnon current, blue solid line) and the time derivative of the magnetization (red solid line) following laser excitation. These results indicate that the spin current injected into the non-magnetic layer is closely related to the rate of magnetization variation. Additionally, they calculated the different contributions to the spin current in the heterojunction at $t = 50$ fs [Fig. 29(b)]. The findings reveal that the magnon current ($j_m$) and spin-polarized electron transport ($j_{s,e}$) contribute equally to the total spin current within the ferromagnet. Specifically, the electron spin transport is primarily driven by bulk electron-magnon scattering, and the resulting negatively polarized spins (spins polarized in the opposite direction to the majority spin direction) are transferred to the receiving layer through spin diffusion. The negative $j_m$ indicates that electron-magnon scattering at the interface produces thermal magnons, which propagate in the -*x* direction. The magnon current at the interface ($x = 0$) is mainly dependent on the temperature difference between the magnons and the electrons in the non-magnetic layer. Beens *et al.* also conducted a numerical analysis of the roles of interfacial electron-magnon scattering and magnon transport. The results showed that the contributions of interfacial electron-magnon scattering and the resulting magnon transport to demagnetization are significant and cannot be ignored.

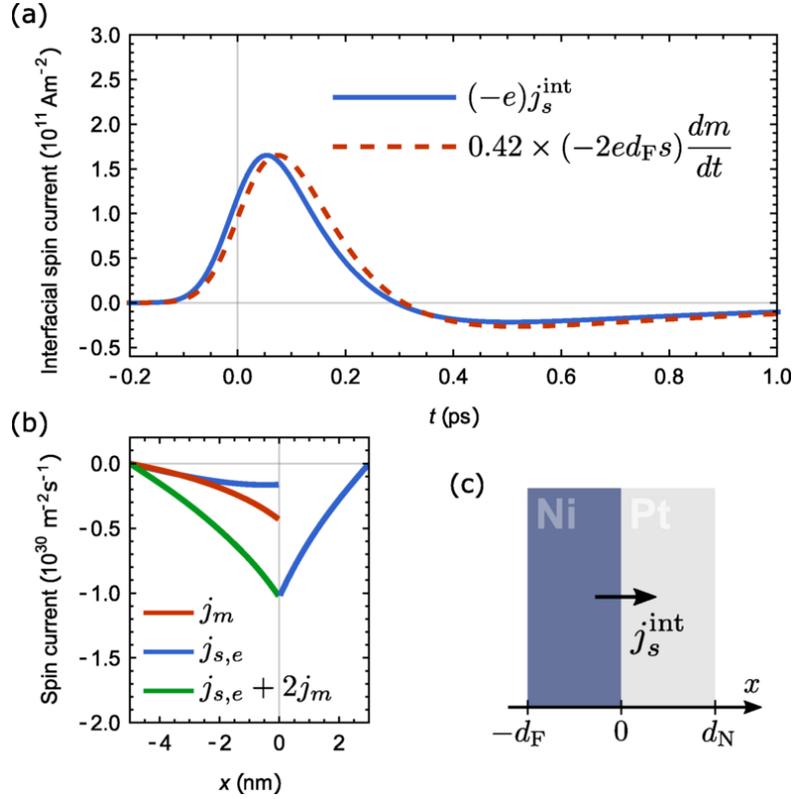

Fig. 29. Laser-induced spin transport in a Ni/Pt bilayer with insulating boundary conditions. (a) The interfacial spin current (blue) as a function of time $t$ following laser-pulse excitation at $t = 0$. The dashed red line represents the temporal derivative of the magnetization. (b) Different spin current contributions as a function of spatial coordinate $x$, evaluated at $t = 0.05$ ps. The blue line denotes the electronic contribution, the red line denotes the magnonic contribution, and the green line indicates the total contribution. (c) Schematic overview of the system [328]. Reprinted with permission from M. Beens *et al.*, Phys. Rev. B, **105**, 13 (2022) Copyright (2022) by the American Physical Society.

To more accurately describe the effect of spin current generated by electron-magnon scattering in heterostructures on ultrafast demagnetization, researchers have been dedicated to enhancing the *s-d* model [165, 342, 343]. In recent years, Remy *et al.* [343] extended the *s-d* model and applied it to Co/Pt multilayers. Their model considers a dynamic exchange splitting of the *s* electrons, energy transfer from *d* to *s* electrons, and the spin current reflection mechanism. This model also successfully reproduces the magnetization reversal of the free layer in a ferromagnetic spin valve, which is impossible to achieve in previous models. However, this model is currently limited to providing qualitative results. We anticipate the development of models capable of making quantitative predictions and being applicable to a broader range of material systems in the future.

**4.2.3 Optically induced intersite spin transfer (OISTR)**

In 2017, Hofherr *et al.* [344] observed that demagnetization in Ni/Au (42 ± 8 fs) was three times faster than in Ni/insulating substrate (120 ± 14 fs) [39, 77]. By using approximate calculations, assuming that the loss of the magnetic moment in Ni on Au is entirely driven by optically generated spin currents, and by comparing the demagnetization dynamics of Ni on different substrates, they demonstrated that this ultrafast demagnetization is predominantly caused by these optically generated spin currents. The following year, Dewhurst *et al.* [96] proposed another mechanism for purely optically excited spin transfer, termed optically induced intersite spin transfer (OISTR). Under the influence of laser irradiation, a spin-selective charge flow is generated. This charge flow moves from the majority spin channel to the minority spin channel, leading to the loss of local magnetic moments in each layer, and ultimately altering the global magnetic order. Their calculations suggest that OISTR can trigger magnetization flipping in Co/Mn multilayer films during the initial phase of demagnetization (< 40 fs). At this stage, the influence of SOC and the contribution of spin-flip scattering are insignificant. This finding also suggests that in experimental studies of OISTR, the laser pulse width should be minimized to approximately 20 fs. The proposal of this effect provides researchers with a deeper understanding of ultrafast non-local spin dynamics [136, 137, 172, 345].

In 2019, Yamamoto *et al.* [345] measured the demagnetization dynamics of Pt in single-crystalline $L1_0$ FePt films at the Pt $L_3$ edge using TR-XMCD and compared it with the demagnetization dynamics measured using TR-MOKE under the same laser parameters. They found that Pt exhibited a slower demagnetization rate but a higher degree of demagnetization. They stated that since the majority of the DOS of the 3*d* bands of Fe are below the Fermi level ($E_F$), and even exceed the DOS of the 5*d* bands of Pt near $E_F$, during the initial stage of demagnetization, the 3*d* electrons in Fe can be easily excited directly by the laser to the Pt states via OISTR. Alternatively, these electrons can first be excited to the unoccupied states of Fe and then transition to the Pt states through superdiffusive transport.

Additionally, Willems *et al.* [97] analyzed the ultrafast demagnetization dynamics measured at the Co $M_{2,3}$ edge of Co films and CoPt alloys. They found that the demagnetization time of Co in the CoPt alloy (86 ± 3 fs) is significantly shorter than that of the Co film (124 ± 4 fs), which contradicts previous experimental and theoretical reports [39, 346]. As shown in Fig. 30, their results indicate that after laser excitation, the increase in minority electrons in the Co film is equal to the loss of majority electrons. For

81 / 122

the CoPt alloy, the increase in minority electrons in Co exceeds the loss of majority electrons, and minority electrons in Pt partially transfer to the Co 3*d* states. Their experimental results are highly consistent with those obtained by time-dependent density functional theory (TD-DFT). This suggests that the demagnetization of CoPt is driven not only by the increased local spin-flip rate but also partly by the OISTR. In the same year, Hofherr *et al.* [136] observed an optically induced spin transfer effect between the Ni and Fe magnetic subsystems in Fe$_{50}$Ni$_{50}$ alloy using time-resolved EUV T-MOKE. They demonstrated that this effect is responsible for the difference in initial demagnetization times between Fe and Ni.

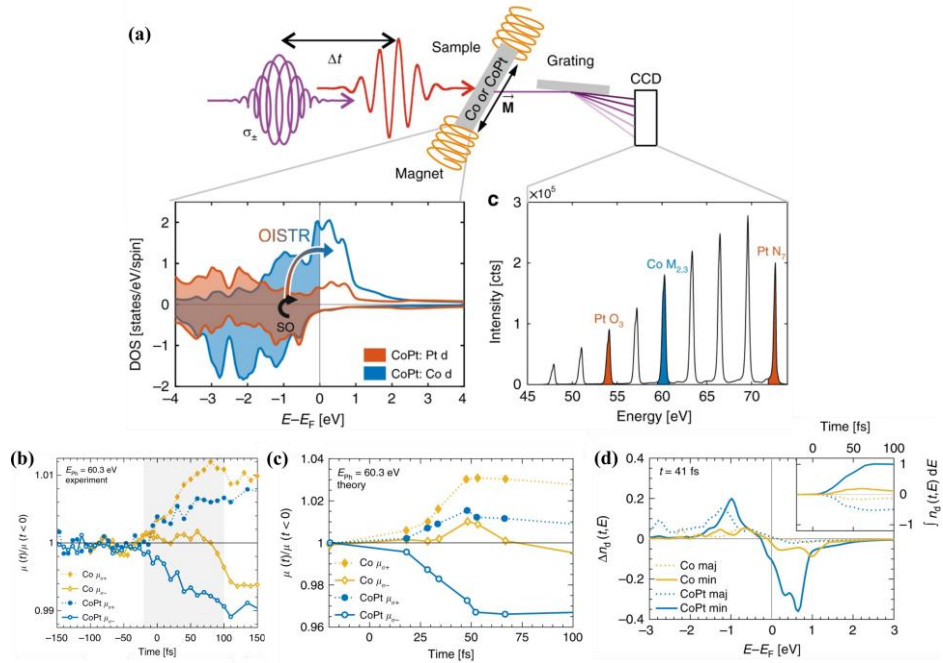

Fig. 30. (a) Experimental schematic diagram and microscopic mechanism of OISTR in CoPt. (b) Experiment and (c) calculated helicity-dependent transient absorption. (d) Theoretically calculated difference in occupied minority and majority *d*-states for Co and CoPt [97]. Reprinted with permission from F. Willems et al., Nat. Commun., 11, 871 (2020) CC BY 4.0.

Since the commonly used technique of T-MOKE in the EUV range may have issues with unwanted crosstalk between different elemental signals and energy-dependent non-linear response, the conclusions of Hofherr *et al.* [136] have been scrutinized. Recently, Jana *et al.* (2022) [98] compared results obtained from EUV T-MOKE and XMCD in the Fe$_{50}$Ni$_{50}$, and confirmed the delayed demagnetization of Ni in Fe$_{50}$Ni$_{50}$ alloy through both experimental techniques. Their findings suggest that the delayed demagnetization of Ni in Fe$_{50}$Ni$_{50}$ is an intrinsic effect, rather than a result of T-MOKE nonlinearity or crosstalk issues. However, they could not verify the contribution of OISTR to the demagnetization process. Recently, Probst *et al.* (2024) [139] further questioned the validity of EUV T-MOKE experiments. They observed differing transient dynamics even under nearly identical experimental conditions, demonstrating that direct comparisons between time-resolved T-MOKE data and magneto-optical data, in general, with time-resolved calculations of the spin-resolved density of states are not always feasible. They concluded that the phenomena observed in EUV TR-MOKE by Hofherr *et al*. [136] were attributable to optical artifacts. In previous studies, a delayed onset of the Ni demagnetization, as well as an asymmetry increase for photons tuned to energies below the Ni edge, was considered evidence supporting the existence of the OISTR effect. However, von Korff Schmising *et al*. (2024) [347] recently questioned these two key observational indicators. By employing XUV T-MOKE measurements, they compared the ultrafast magnetic responses of $Fe_{50}Ni_{50}$ after direct and indirect photoexcitation. Their experimental results demonstrated that both fingerprint observables were present in both excitation scenarios. Although von Korff Schmising *et al*. [347] did not entirely rule out the possibility of coherent light-wave magnetization dynamics, they raised critical questions for the research community: Does OISTR genuinely exist? Can it be experimentally confirmed? Furthermore, how does the experimentally observed increase in magnetic asymmetry validate the transfer of spins from Ni to Fe in the alloy?

In the same year, Möller *et al*. (2024) [138] employed EUV T-MOKE to investigate spin transfer in $Fe_{50}Ni_{50}$, $Fe_{19}Ni_{81}$, and Ni. They observed an increase in magnetic asymmetry in pure Ni as well. Through extended analysis [139], they attributed this rise in magnetic asymmetry to a transient rotation of the off-diagonal element of the dielectric tensor in the complex plane. By comparing the transient evolution of this rotation with TDDFT theory, they were able to identify the spin transfer effect. Consequently, Möller *et al*. [138] concluded that while the observed increase in the magneto-optical signal is insufficient to conclusively prove the existence of the OISTR effect, its presence cannot be still ruled out.

Furthermore, OISTR has been employed to clarify ultrafast demagnetization in RE metal systems. In 2023, Liu *et al*. [337] observed that, following laser excitation of a Gd/Fe bilayer, the spin polarization of the spin-mixed surface state of Gd rapidly decreased by 20% within the first ~100 fs. The electron temperature of the Gd/Fe system then returned to its initial value, suggesting that the laser-excited carriers were transferred from the Gd layer to the Fe layer. This spin transport mechanism was further supported by an observed increase in spin polarization within the Fe layer in a separate Fe/Gd bilayer. Consequently, it was proposed that ultrafast spin transport governs the demagnetization process in the Gd/Fe bilayer. Fe acts as a spin filter, facilitating the transfer of majority spins from Gd to Fe, which are then converted into minority spins within the Fe layer.

These studies demonstrate that spin transport is sufficient to induce significant ultrafast magnetization dynamics and that its contribution to demagnetization is substantial and cannot be overlooked.

**4.3 Non-thermal electronic distribution**

When a femtosecond laser is directed onto the surface of the magnetic film, it modifies the distribution of electrons near the Fermi level. These electrons become excited into a significantly non-equilibrium state, referred to as non-thermal electrons. Their distribution deviates from the Fermi-Dirac distribution. According to previous reports, direct energy transfer between non-thermal electrons and spins [64], as well as the non-thermal electronic distribution [55], result in spin-flip scattering events in the higher binding-energy region. Subsequently, the non-thermal electronic distribution rapidly undergo thermalization, reverting to a Fermi-Dirac distribution through interactions with other thermal electrons or subsystems. Although the temperature of non-thermal electrons is difficult to determine and their thermalization time is extremely short, studies have shown that the non-thermal electronic distribution promotes demagnetization [48, 55, 64, 76, 99, 100].

Initially, the 2TM, 3TM, 4TM, M3TM, and EM3TM models considered only thermal electrons, assuming a uniform electron or phonon temperature. These models were later expanded to include non-thermal electrons and non-thermal phonons. For example, the E3TM extends the conventional 3TM model by incorporating the effects of non-thermal electrons, which are crucial for accurately describing the ultrafast dynamics following laser excitation. The inclusion of non-thermal electron distributions in the E3TM allows for a more detailed analysis of the initial energy transfer processes between electrons, spins, and the lattice. Specifically, the E3TM can capture the rapid energy redistribution among these subsystems, which is driven by the non-thermal electron population immediately after laser excitation.

In the early stages of investigation, some researchers focused on the ultrafast dynamics induced by non-thermal electronic distribution in nonmagnetic metal materials. For instance, Sun *et al.* (1994) [257] conducted experiments to measure transient transmittance and reflectance in thin Au films to investigate electron thermalization dynamics. They employed the multi-wavelength pump-probe technique and utilized the extended 2TM to simulate electron distribution dynamics. While this model simplifies electron distribution into thermalized and non-thermalized components, it provides a qualitative description of electron behavior. The results offer fresh insights into electron dynamics in metals and provide a valuable foundation for future research.

In 2006, Carpene *et al.* [240] proposed another extended version of the 2TM by considering non-thermal electronic distribution. This model not only accounts for the correlation between electron temperature and photon energy but also enables the electron-lattice system to distinguish between different heating terms. This advancement provides a more effective strategy for characterizing ultrafast demagnetization behavior. As discussed in Section 3.2.1, despite the limitations of the 2TM, these studies underscore the significance of non-thermal electronic distribution in the ultrafast demagnetization process.

For magnetic materials, Kim *et al.* (2009) [64] analyzed the step-like demagnetization in amorphous TbFe alloys occurring near the electronic thermalization timescale, using E3TM. Their investigation unveiled that this phenomenon can be ascribed to the energy conversion between non-thermal electrons and spin systems.

In 2011, Carva *et al.* [76] calculated the contribution to demagnetization induced by E-Y scattering in thin Ni films, employing first principles calculations. They observed that the demagnetization caused by non-thermal electronic distribution is more significant than that arising from thermal electron spin-flipping due to E-Y scattering. In a subsequent study, Carva *et al.* [55] conducted computational simulations examining demagnetization behaviors resulting from thermal and non-thermal electronic distribution in transition metals, specifically Fe, Co, and Ni. Their findings revealed that the demagnetization effect induced by non-thermal electronic distribution is more pronounced than that attributable to thermal electron distribution. Consequently, they concluded that non-thermal electronic distribution plays a vital role in the demagnetization process. These studies introduce a novel mechanism into the study of ultrafast demagnetization, posing new challenges to prevailing explanations that solely consider spin-flipping and spin transport mechanisms during the ultrafast demagnetization process.

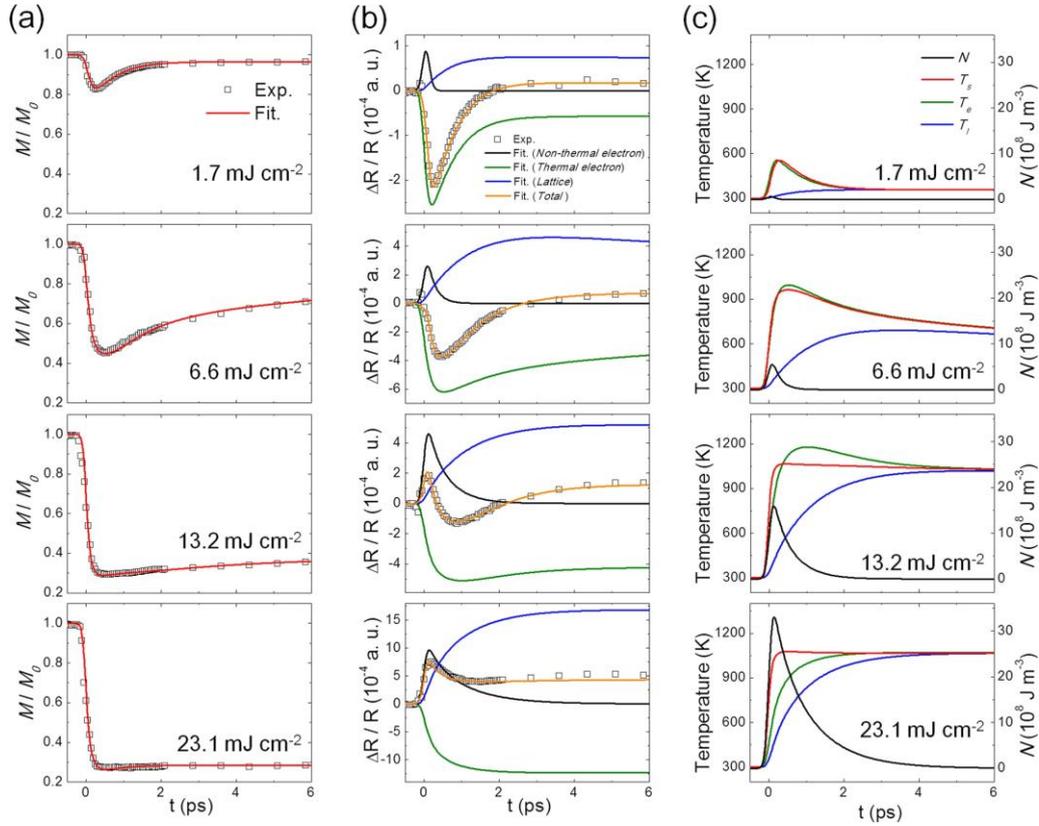

Fig. 31. E3TM fitting results of (a) TR-MOKE and (b) time-resolved reflectivity (TR-R) experimental data at different laser fluences. The contributions of non-thermal electrons (black), thermal electrons (green), lattice (blue), and the total sum of these components (orange) are plotted in (b). (c) Evolution of extracted non-thermal electron density ($N$), spin temperature ($T_s$), thermal electron temperature ($T_e$), and lattice temperature ($T_l$) [48]. Reproduced Reprinted with permission from J. H. Shim *et al.*, Sci. Rep., **10**, 6355 (2020). CC BY 4.0.

In 2020, Shim *et al.* [48] reported that the traditional 3TM could not adequately fit the time-resolved reflectivity (TR-R) and TR-MOKE experimental results they observed in Co/Pt multilayer films. To gain a better understanding of the energy conversion from non-thermal electrons to other subsystems, they employed the E3TM, which accounts for the distribution of non-thermal electrons, to quantitatively analyze their experimental data. In this study, the energy exchange coefficient, $G_{ee}$, between non-thermal and thermal electrons significantly influenced the TR-R fitting, but had minimal impact on the TR-MOKE fitting, indicating its insensitivity to the contribution from the non-thermal electron distribution. Therefore, combining TR-R and TR-MOKE measurements is necessary for a comprehensive analysis. As shown in Fig. 31, the fitting results closely match the experimental data. Their research demonstrated that in scenarios of low laser fluence, the influence of non-thermal electronic distribution is weak. Consequently, magnetization swiftly recovers within a few picoseconds after demagnetization. In cases of high laser fluence, the contribution of non-thermal electronic distribution significantly increases. The excitation of non-thermal electronic distribution significantly enhances the TR-R signal of the Co/Pt multilayer film initially. During this brief period, the contribution of non-thermal electronic distribution rivals that of thermal electrons. Under the influence of non-thermal electronic distributions, the degree of demagnetization in the sample increases, consistent with previous reports [55, 76, 100]. Subsequently, non-thermal electrons transfer energy to other subsystems, leading to a slower recovery of the magnetization following demagnetization.

Subsequently, Chekhov *et al.* [125] investigated the ultrafast demagnetization of a Fe film when excited by optical and terahertz pump pulses. Their findings indicated that the optical pump induces a highly non-thermal electronic distribution, whereas the terahertz pulse results in a Fermi-Dirac electron distribution at elevated temperatures. Despite these differences, the demagnetization dynamics induced by both the terahertz and optical pumps are nearly identical. Consequently, they concluded that the rate of electron-phonon energy transfer is unaffected by the configuration of the non-thermal electronic distribution. Previous research has also suggested that the impact of non-thermal electronic distribution on ultrafast demagnetization is inconsequential [39, 272, 285].

However, the use of TR-MOKE alone is insufficient for measuring the non-thermal electronic distribution. Previously, Eich (2017) [306] and Gort (2018) [309] obtained non-thermal electronic distribution through photoemission spectra by using ARPES. In 2024, Pierantozzi *et al.* [262] utilized ARPES to monitor the changes in the electronic energy reservoir within the first Brillouin zone of an Fe film. They also analyzed the spin polarization of photoelectrons through a Mott-scattering experiment to detect the evolution of the magnetic state. The TR-ARPES experimental results indicated that non-thermal electronic distribution is present only when the Fermi vector $k_F$=1.65Å$^{-1}$. Subsequently, Pierantozzi *et al.* incorporated non-thermal contributions into the differential equations of electrons and phonons based on the M3TM model and the previous report by Carpene *et al.* [240]. They used these modifications to analyze the changes in electron temperature measured by TR-ARPES and the variations in relative magnetic moment measured by the time-resolved spin-polarization (TR-SP) experiment. Their results suggest that at low energy flux densities, transient spin changes are driven by thermal fluctuations. The increase in electron temperature during pump excitation leads to spin changes once non-thermal electrons transfer their excess energy to the entire electron bath.

Consequently, further verification is required to ascertain the genuine contribution of non-thermal electronic distribution.

## 4.4 Laser-induced lattice strain effect

In recent years, researchers have proposed that pulsed lasers can alter crystal field anisotropy on a picosecond timescale, with the induced ultrafast lattice response leading to magnetization precession [27, 52, 102, 103, 206, 254]. Conversely, changes in magnetization will, in turn, affect the lattice stress. We refer to this lattice response as the laser-induced lattice strain effect. Since this effect generally occurs on a picosecond timescale, its impact on ultrafast magnetization dynamics is likely to be primarily observed in the process of magnetization relaxation occurring within this timescale. This stands out as one of the most conspicuous distinctions between the laser-induced lattice strain effect and other demagnetization origins, such as spin-flipping, spin transport, and non-thermal electronic distribution.

In 2018, Reid *et al.* [206] utilized ultrafast electron diffraction to directly measure the lattice response of FePt nanoparticles on a sub-picosecond timescale. They proposed that there are three types of stress contributions to the demagnetization process: electronic stress, phononic stress, and magnetic stress. As shown in Fig. 32, within the first picosecond following laser excitation, the lattice experiences rapidly evolving magnetic and electronic stresses. This results in an increase in the tetragonal distortion of the unit cell, as depicted. On the timescale of a few picoseconds, phonon and macroscopic stresses become significant, driving the lattice toward its equilibrium state. In their work, both electronic and phononic stresses exhibit anisotropy and are positive along the crystallographic axis. The magnetic stress is proportional to the square of the change in magnetization, implying that it can be induced by ultrafast demagnetization. They employed this model to fit the experimental data, and the results indicated the presence of non-zero magnetoelastic stress in FePt nanoparticles on the timescale of ultrafast demagnetization. They also observed an anisotropic lattice displacement, moving inward from the sample boundary in the form of strain waves. These displacements are primarily caused by the magnetic stress, which forms on a sub-picosecond timescale, matching the characteristic time of ultrafast demagnetization.

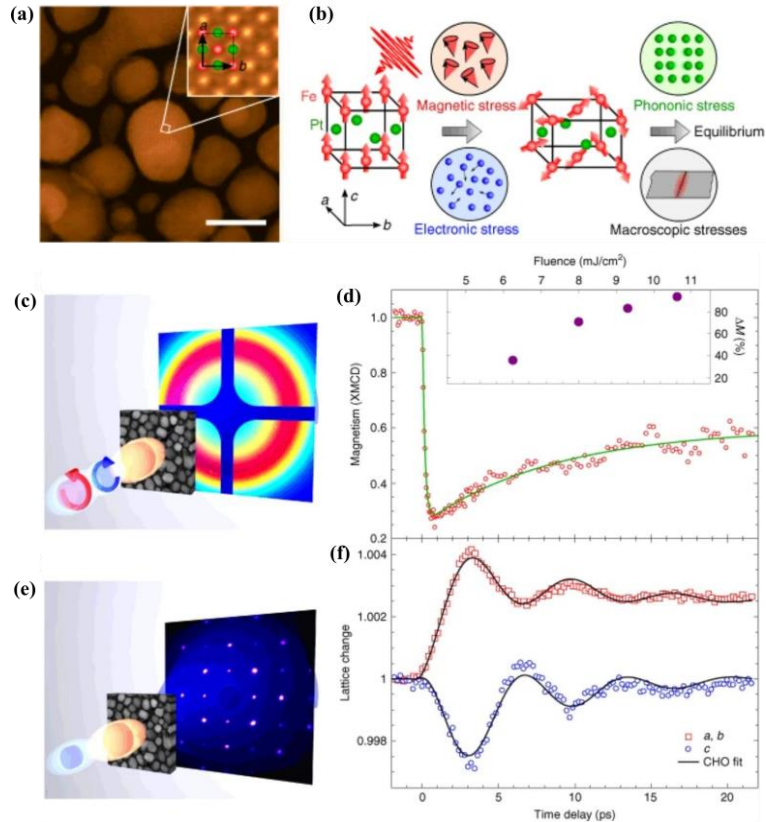

Fig. 32. (a) High-angle annular dark-field scanning transmission electron microscopy (HAADF-STEM) images of the morphology and *a*-, *b*-plane atomic arrangement (inset) for FePt nanoparticles embedded in an amorphous carbon matrix. (b) The crystal structure, magnetic arrangement, and the sequence of stress contributions in FePt. Ultrafast spin (c-d) and lattice (e-f) dynamics of the FePt nanoparticles, with (c-d) measured by time-resolved soft X-ray scattering and (e-f) measured by ultrafast electron diffraction. Reproduced Reprinted with permission from A. H. Reid et al.[206], Nat. Commun., 9, 388 (2018). CC BY 4.0.

Subsequently, Reppert *et al.* [101] compared transient stresses in granular and continuous FePt thin films and found that both samples exhibited out-of-plane expansion. However, the granular FePt thin film also showed lattice contraction along the short out-of-plane *c*-axis of the tetragonal unit cell. Later, Reppert *et al*. [102] further analyzed the stresses on granular and continuous FePt thin films during laser heating. They observed that when the spin system is in a disordered state, the lattice contraction under high laser flux reaches saturation. They attributed this contraction to spin entropy. When the pulse laser energy is low, this lattice contraction driven by spin stress occurs after 3 ps, the process at this time being dominated by expansive lattice stresses. Additionally, their double-pulse excitation scenarios reveal a direct connection between spin disorder and contractive stress. The first strong excitation pulse saturates the spin excitations, while the second, weaker pulse induces subsequent dynamic behavior with a delay time $\Delta t$. If the second pulse arrives approximately 100 ps later, partial recovery of spin order occurs, and the second pulse induces lattice contraction. The timescale for the recovery of contractive stress is determined by thermal transport and corresponds to the remagnetization timescale observed in TR-MOKE measurements. In another work by Reppert *et al*. [103], the strain response of the Dy layer in a metallic heterostructure under femtosecond laser excitation was investigated using ultrafast electron diffraction. They were able to distinguish between lattice expansion caused by strain pulses and thermal diffusion. Their findings revealed the presence of contraction stress in the sample, which increased with a picosecond time constant. This timescale was consistent with subpicosecond electron-spin coupling and subsequent phonon-spin coupling, aligning with previous reports on demagnetization [254, 301, 348]. The results also indicated that the stress was dependent on the excitation density. At higher excitation densities, the spin stress increased over a longer timescale. This is because deeper regions of the sample received less light excitation, resulting in reduced energy in the phonon system, causing energy transfer from the demagnetized surface area to the deeper parts of the sample.

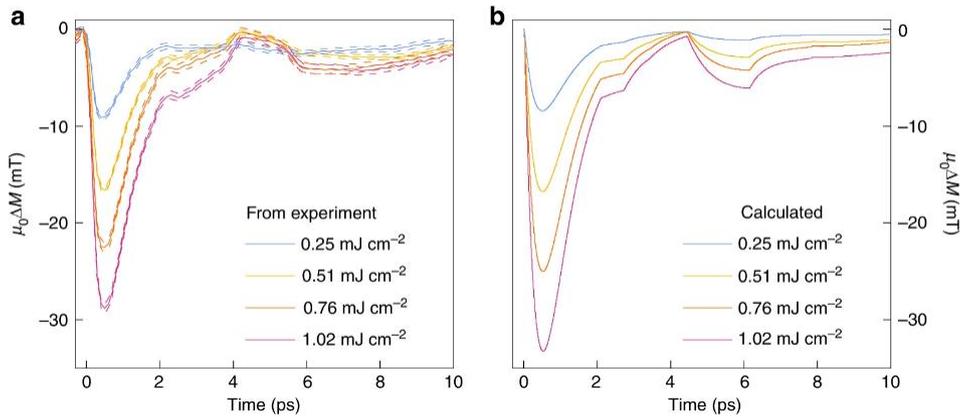

Fig. 33. (a) Measured and (b) simulated magnetization dynamics $M(t)$ of MgO/Fe/MgO film under different laser fluences. The demagnetization driven by the strain wave effect occurs on a timescale of 2~7 ps [57]. Reproduced Reprinted with permission from W. Zhang *et al.*, Nat. Commun., **11**, 4247 (2020). CC BY 4.0.

The mechanical rotation and associated strain waves resulting from the Einstein-de Haas effect provide valuable insights into the coupling between magnetic and mechanical properties in materials. In 2019, Dornes *et al.* [52] used time-resolved X-ray diffraction to observe an ultrafast demagnetization process induced by a strain wave in Fe. This effect transfers angular momentum to the lattice, contributing to 80% of the demagnetization process. Subsequently, Tauchert *et al.* (2022) [79] employed ultrafast electron diffraction with terahertz-compressed electron pulses to investigate the lattice dynamics of Ni. Their experimental and simulation results revealed that, in addition to inducing isotropic atomic displacements, the laser also triggered anisotropic lattice motion, with the oscillation direction oriented perpendicular to the initial magnetization. They further reported that, preceding the Einstein-de Haas rotation, a significant number of circularly polarized and chiral phonons were generated on the femtosecond timescale, which subsequently facilitated phonon-based or phonon-assisted spin transport. Consequently, they proposed that the direct spin-phonon coupling observed by Dornes *et al.* [52] arises from the preferential coupling between spins and circularly polarized phonons near the center of the Brillouin zone.

In 2020, Zhang *et al.* [57] simultaneously observed two ultrafast demagnetization phenomena in a magnetic system. The first phenomenon involves magnon excitation on the femtosecond timescale, while the second relates to the strain wave effect over a period of 2 to 7 picoseconds. For lattice strain, they considered three mechanisms: i) the force induced by optically excited electrons; ii) the force induced by ultrafast demagnetization, where a new electronic surface potential is generated following magnon emission; iii) the thermal strain due to heating of the lattice. The first and second mechanisms originate from the deformation potential mechanism and occur on a timescale below 100 fs [349]. In these processes, the strain caused by acoustic phonons depends on the deformation potential and the absorbed laser energy. The third mechanism is driven by the changing temperature distribution in a thermoelastic model, establishing a linear relationship between the stress pulse and the varying temperature distribution. These results are depicted in Fig. 33. The researchers elucidated that these lattice strain dynamics stem from the deformation potential mechanism and the temperature distribution within the system. However, their demagnetization model in this case does not take into account the spin transport process and heat transfer. Consequently, it is unable to provide a comprehensive understanding of the true microscopic origins of ultrafast demagnetization on the femtosecond scale. Furthermore, Shim *et al.* [48] also observed phonon oscillations within the 5 to 10 picosecond range in Co/Pt multilayer films. The oscillation frequency exhibited a slight decrease with increasing laser fluences. As their primary focus was on the contribution of non-thermal electronic distribution to the demagnetization process, they did not consider the influence of the strain wave effect in their study.

In addition to the dynamic strain effect mentioned above, Shin *et al.* (2022) [350] discovered that quasi-static strain (QSS) induced by lattice thermal expansion following ultrafast photoexcitation has a significant impact on ultrafast demagnetization. The ultrafast spin dynamics driven by QSS and thermal effects are challenging to differentiate temporally. However, QSS primarily affects plasmonic [351] or electronic bands [352], whereas thermal effects alter electron populations. Consequently, they can be distinguished using ultrafast Sagnac interferometry and MOKE measurements. The former can confirm the presence of QSS by directly measuring lattice expansion dynamics. Their findings indicate that QSS governs ultrafast spin dynamics from the first picosecond to nanosecond timescales in ferromagnetic films, even surpassing thermal effects. Recently, Mattern *et al.* [27] reported that QSS and demagnetization, as driving forces of magnetization precession, counteract each other.

However, there is still limited research on the strain effects produced on the ultrafast scale. Its contribution to demagnetization dynamics appears to be primarily reflected in the magnetization relaxation process. The significant impact on the demagnetization process at the sub-picosecond scale remains a subject for further investigation.

### 4.5 Summary of ultrafast demagnetization origins

Since the initial discovery of ultrafast laser-induced demagnetization of nickel films in 1996, the microscopic origins of ultrafast demagnetization have remained a contentious topic. In this chapter, we mainly introduce four microscopic demagnetization origins, including spin-flipping, spin transport, non-thermal electronic distribution, and the laser-induced lattice strain effect. The E-Y scattering (including electron-phonon scattering [74, 75, 118] and electron-electron Coulomb scattering [76, 77, 254]), electron-magnon interaction [79, 81, 82], and the photon-spin interaction [83, 255] are strongly related to spin-flipping processes. Superdiffusion transport [53, 95], electron magnon scattering-induced spin transfer [328, 341], and OISTR [97, 345] are closely connected to spin transport processes. Subsequently, we further elaborated on the research progress concerning various origins of demagnetization.

When a laser pulse irradiates the surface of a magnetic material for an ultrashort duration, the opto-magnetic field generated by the laser can directly interact with the spins in the material, leading to magnetization oscillations. These effects include photon-spin interactions [322] and OISTR [138]. However, compared to other demagnetization mechanisms, photon-spin interactions do not result in significant demagnetization. Simultaneously, the laser pulse rapidly excites electrons to high-energy states, generating non-thermal electrons that do not follow the Fermi-Dirac distribution. As the laser fluence increases, the contribution of this non-thermal electron distribution grows, leading to a more pronounced degree of demagnetization [48]. Consequently, photon-spin interactions in spin-flipping, OISTR in spin transport, and non-thermal electronic distributions play critical roles in the early stages of ultrafast demagnetization. These processes typically occur within 100 femtoseconds.

Subsequently, the non-thermal electrons rapidly thermalize within a short period, transferring the energy absorbed from photons to the thermal electron, spin, and lattice systems. During this phase, the excited electrons move randomly in all directions, generating spin currents [53, 95]. Simultaneously, these electrons undergo scattering with other electrons, magnons, and phonons [84, 307, 339], further contributing to the demagnetization process [350]. As a result, superdiffusive transport, electron-magnon scattering-induced spin transfer, E-Y spin-flip scattering, and electron-magnon spin-flip scattering dominate this stage. These processes typically occur on a sub-picosecond timescale. In addition, on the picosecond timescale, the influence of laser-induced strain waves becomes significant. The lattice could generate anisotropic strain wave effects, during which angular momentum is transferred from the spin system to the lattice. This leads to magnetization precession occurring on the picosecond timescale [52].

For most of the research findings, demagnetization primarily driven by the spin-flipping mechanism typically occurs in single-domain and single-layer systems, while demagnetization dominated by the spin transport mechanism generally takes place in multi-domain (stripe domain or labyrinth domain) and multilayer systems. The results suggest that there are slight differences in demagnetization speed induced by spin-flipping and spin transport mechanisms. For demagnetization driven by the spin-flipping mechanism, the duration is generally greater than 200 fs, such as 275~350 fs for FeRh film [129], 260 fs for Co film, and 280~400 fs for Fe/NM [353] single-domain films. In contrast, demagnetization driven by the spin transport mechanism occurs slightly faster, with demagnetization times of less than 200 fs, such as 75~200 fs for Co/Pd [54, 126, 326, 329] and Co/Pt [127, 197, 329] multilayers with a stripe domain. Specifically, in 2014, Moisan *et al*. [114] studied the demagnetization processes of CoPd and CoPt alloys. Their experimental results indicated that the demagnetization speed of CoPt and CoPd alloys with multi-domain structures (not reaching magnetic saturation, CoPt ~180 fs and CoPd~162 fs) is slightly faster than that of single-domain structures (reaching magnetic saturation, CoPt ~200 fs and CoPd~180 fs)). Recently (in 2023), Chen *et al* [354] utilized TR-MOKE measurements in conjunction with micromagnetic simulations to investigate the ultrafast demagnetization phenomena in a wedge-shaped $Ni_{80}Fe_{20}$ film. Their findings also indicated that the demagnetization rates were more rapid in the thicker regions of the film (characterized by a stripe domain structure, <120 fs) compared to the thinner regions (characterized by an in-plane single domain structure, >120 fs). Although these demagnetization time differences are small, current research generally suggests that spin transport can introduce additional demagnetization channels, thereby accelerating the demagnetization process.

In many research reports, with the exception that the laser-induced lattice strain effect can be readily distinguished from the other origins in terms of timescales, the contributions of spin-flipping, spin transport, and non-thermal electronic distribution often overlap. Although researchers have used time-resolved reflectivity [48] and ARPES [262] to detect non-thermal electronic distribution, its contribution to demagnetization is not as significant as that of spin-flipping and spin transport. To explore the impact of non-thermal electronic distribution on ultrafast demagnetization, relying solely on TR-MOKE is insufficient. As a result, there remains ongoing debate about the influence of non-thermal electronic distribution on the demagnetization process [39, 272, 285].

## 5. Conclusion and outlook

When a femtosecond laser pulse is directed onto the surface of a magnetic material, the irradiated region within the material exists in an extremely non-equilibrium state on a sub-picosecond scale. Subsequently, magnetization undergoes rapid reversal in a short time. This phenomenon has opened up a new avenue for dynamically manipulating magnetic moments, and it has also piqued the keen interest of many researchers. A substantial body of theoretical and experimental studies has been conducted to investigate the methods of energy and angular momentum transfer in laser-induced ultrafast demagnetization. Various phenomenological models and distinct microscopic mechanisms have been proposed.

Different models for ultrafast demagnetization share similarities but also exhibit variations. It is challenging to distinguish the demagnetization contributions resulting from various microscopic origins. Consequently, numerous researchers predominantly engage in discussions surrounding the following questions: Which model can more accurately elucidate the phenomenon of ultrafast demagnetization? Is the ultrafast demagnetization phenomenon in different systems governed by a single origin or multiple origins? How can one ascertain the existence of this origin in the process?

In recent years, many reports have unquestionably enhanced our comprehension of the interactions between different subsystems in magnetic materials. Researchers have put forward more comprehensive models to unravel the enigma of the microscopic mechanisms involved in ultrafast demagnetization. This review aims to provide a systematic introduction to the research progress of the ultrafast demagnetization phenomenon since its discovery. It also describes the various ultrafast demagnetization models and their microscopic mechanisms proposed by researchers. We primarily present seven different phenomenological models (2TM, 3TM, E3TM, MMTM, 4TM, M3TM, and EM3TM) and discuss their applicability and limitations. Among these models, five of them (2TM, 3TM, E3TM, MMTM, and 4TM) do not account for angular momentum transfer, potentially leading to significant discrepancies between the established ultrafast demagnetization reconstruction process and real-world scenarios. These five models lack a predefined mechanism for the demagnetization process, making it challenging to precisely analyze the demagnetization contributions from various microscopic mechanisms.

In contrast to the preceding five models, both M3TM and EM3TM incorporate angular momentum dissipation processes, with parameters obtained through *ab initio* calculations. Previous reports have confirmed spin-lattice angular momentum transfer through time-resolved X-ray diffraction measurements [52, 79]. Although the M3TM model is widely used to reconstruct ultrafast demagnetization phenomena, it only considers angular momentum transfer between electrons and lattices, which often leads to an overestimation of electron-lattice coupling. To address this, Beens *et al.* (2019) [275, 276] introduced the EM3TM, incorporating the spin-lattice angular momentum transfer channel into the M3TM framework. In this model, the phonon system is assumed to be internally thermalized, and phonons are treated within the Debye model. However, this approach also has limitations that make it challenging to accurately predict spin-lattice interactions. Consequently, there is an urgent need for a detailed theoretical description of spin-lattice angular momentum transfer.

Furthermore, in comparison to M3TM, EM3TM introduces a lattice cooling process that facilitates energy exchange between the phonon system and the external environment, thereby promoting a magnetic moment relaxation process that aligns more closely with real-world conditions.

It is noteworthy that traditional 2TM, 3TM, and 4TM models do not fully account for non-thermal electrons and phonons. This oversight leads to significant errors in reconstructing the demagnetization process, such as underestimating the demagnetization effects caused by non-thermal electron transport, the electron-phonon coupling constant, and the magnetization recovery time. Although the M3TM model assumes that most electrons and phonons are in local thermal equilibrium with a few in a non-thermal state, it still lacks accurate calculations. To address this shortcoming, researchers have modified the 2TM and 3TM models to incorporate non-thermal electronic distribution. Afterward, unlike these extended versions (modified 2TM, E3TM, and M3TM), the MMTM model introduces a distribution for non-thermal phonons. This model provides a comprehensive description of the non-equilibrium processes involved in ultrafast demagnetization, contributing to a deeper understanding of the phenomenon. However, the current MMTM can only approximate the effective electron-magnon coupling by thermalizing the spin system.

We later introduce four widely discussed microscopic origins in ultrafast demagnetization, namely spin-flipping, spin transport, non-thermal electronic distribution, and the laser-induced lattice strain effect, respectively. The spin-flipping mechanism can be induced by Elliott-Yafet scattering (electron-phonon scattering and electron-electron Coulomb scattering), electron-magnon interaction, and the photon-spin interaction. In the discussion of spin transport mechanism, in addition to superdiffusion transport, we also introduced electron magnon scattering-induced spin transfer and OISTR. For nearly two decades, researchers have debated the roles of the spin-flipping and spin transport mechanisms in demagnetization. Among these results, most studies indicate that ultrafast demagnetization induced by the spin-transport mechanism occurs more rapidly than the spin-flipping mechanism. They have analyzed the contributions of the spin-flip mechanism and the spin-transport mechanism with respect to the speed at which the demagnetization process occurs. While these conclusions are somewhat consistent, certain contradictions cannot be overlooked. Most of the results suggest that the spin-flipping-dominated mechanism generally occurs in single-domain and single-layer systems, while the spin-transport-dominated mechanism typically occurs in multi-domain (stripe domain or labyrinth domain) and multilayer systems. However, some studies have also shown that there is no significant difference in the demagnetization speed between single-domain and multi-domain systems for the same material. Therefore, a quantitative understanding of the microscopic mechanism's contribution to the demagnetization process and the true contributions of different origins to ultrafast demagnetization necessitate further investigation.

When the laser irradiates the sample surface, photon-spin interactions occur within the sample. Simultaneously, due to the OISTR, the laser directly excites a spin-polarized current. Additionally, the sample absorbs the photon energy, rapidly generating a non-thermal electron distribution within an extremely short time. Early studies lacked consideration of non-thermal electronic distribution, which has subsequently been proven to be a non-negligible part of the demagnetization contribution by different techniques (time-resolved reflectivity and TR-ARPES). The contribution of non-thermal electronic distribution is comparable to that of thermal electrons, particularly at high laser fluences. Subsequently, the non-thermal electrons transfer energy to other subsystems, and the angular momentum of the electron and spin systems is transferred to the lattice, resulting in a significant reduction in the sample's magnetization. At this stage, the primary mechanisms driving demagnetization include spin-flip, spin-transport mechanisms, and laser-induced QSS. The influence of spin-flip and spin-transport mechanisms persists for approximately 100 fs, while the QSS effect can extend to the nanosecond timescale. Although the sample's magnetization largely recovers on the picosecond timescale, the laser-induced lattice strain wave effect begins to take effect, inducing magnetization precession.

In comparison to the non-thermal electronic distributions, spin-flipping, and spin-transport mechanisms, the lattice strain-induced demagnetization has received less attention because it typically occurs on the picosecond timescale, and the amplitude of the magnetization oscillation is also quite weak. While existing models such as the 3TM, E3TM, and others have successfully described many experimental observations, there remain challenges in fully capturing the intricate details of ultrafast demagnetization. Future research should aim to enhance these models by incorporating additional physical mechanisms, such as opto-electronic response, non-equilibrium spin dynamics and electron-electron correlations, to provide a more comprehensive understanding. Rather than simply adding complexity, these enhancements should be guided by experimental insights to address specific unresolved phenomena. Specifically, although extensive research has been conducted on laser-induced spin and orbital momentum transfer (i.e., OISTR) [96, 97, 136, 138, 347], definitive experimental evidence for the existence of OISTR requires linking the instantaneous optoelectronic response with the spin system's response. This highlights the urgent need for the development of new tools with attosecond time resolution to further explore the emerging field of attosecond magnetism.

The advent of attosecond laser pulses presents a promising frontier for ultrafast demagnetization studies. These pulses can now be generated in the EUV and soft X-ray regions, and they offer the ability to probe electronic dynamics on unprecedented timescales. This enables deeper insights into the initial stages of laser-induced demagnetization, as well as the complex interplay between electronic excitation and spin dynamics, which occur within femtoseconds or even attoseconds. Pioneering works using attosecond pulses have already begun to shed light on these ultrafast processes, and further studies could reveal new aspects of the interaction between light and magnetic materials that are not accessible with longer pulses.

In 2019, Siegrist *et al*. [355] combined attosecond transient XUV absorption spectroscopy detection (attosecond XAS) with attosecond time-resolved MCD. They observed that the MCD contrast of the Ni/Pt system rapidly decreased by 40% of its initial value within the first ~10 fs during and after electronic excitation, while no significant change was observed in pure Ni. Using TD-DFT, which accounts for the coupled dynamics of charge and spin, they elucidated the coherent electronic processes in the presence of an oscillating light field. The results closely matched both theoretical calculations and experimental observations. They concluded that OISTR dominated the demagnetization process in the Ni/Pt system during the first 20 fs after photoexcitation, while from 20 to 50 fs, the SOC-mediated spin-flip mechanism became the dominant process. This work by Siegrist *et al*. marked a significant progress in attosecond magnetism. Subsequently, in 2020, Hofherr *et al*. [356] combined TR-EUV-MOKE with TD-DFT to study the laser-induced coherent spin transfer process in $Fe_{50}Ni_{50}$. By analyzing MOKE asymmetry data, they revealed that spin polarization in Ni and Fe is energy-dependent, demonstrating that the $Fe_{50}Ni_{50}$ alloy enables laser-induced coherent spin transfer (i.e., OISTR) from the Ni to the Fe spin system. In the future, attosecond techniques can be applied to a wider range of material systems, such as antiferromagnetic materials, enabling the exploration of ultrafast magnetization dynamics on even shorter timescales.

Since antiferromagnetic (AFM) materials lack stray fields and possess zero net magnetic moments, they remain impervious to external magnetic influences. The dense arrangement of AFM materials lends itself to enhancing storage density and performance in memory devices. Consequently, AFM materials exhibit promising applications in future high-frequency spintronics. In recent years, researchers have begun to divert their attention towards the laser-induced ultrafast spin dynamics of AFM materials [168, 193, 348, 357-370]. Nevertheless, due to the absence of a net magnetic moment in AFM materials, it proves challenging for researchers to employ MOKE to yield a linear variation of the Kerr signal. This complication heightens the difficulty of studying ultrafast spin dynamics in AFM materials. Various research methods have been proposed to investigate the ultrafast spin dynamics of AFM materials, including:

i) The indirect reflection of ultrafast spin dynamics in the AFM layer via the MOKE signal obtained in the FM layer [357, 358].

ii) The direct detection of the ultrafast spin dynamics of AFM using the magneto-optical Faraday effect [359-362] or magneto-optical Voigt effect [363-365].

iii) The utilization of time-resolved resonance X-ray diffraction (TR-RXD) to monitor the ultrafast spin dynamic behavior of AFM materials [193, 348, 366], etc.

Nonetheless, the exploration of ultrafast spin dynamics in AFM materials is still in its infancy. In the future, researchers must strive to discover simpler and more efficient approaches for detecting the spin dynamics of AFM. Additionally, they should delve into more suitable models and possible demagnetization mechanisms pertaining to ultrafast spin dynamics in AFM materials. This endeavor will chart a new course for the study of manipulating spin on ultrafast timescales.

## Acknowledgments

The work at Qingdao University is supported by the Natural Science Foundation of Shandong Province (Grants No. ZR2022MA053), the National Natural Science Foundation of China (Grants No. 11704211, No. 11847233, and No. 12205093), the Fundamental Research Funds for the Central Universities (Grants No. lzujbky-2022-kb01), China and Germany Postdoctoral Exchange Program (Helmholtz-OCPC), China Postdoctoral Science Foundation (Grants No. 2018M632608), Applied basic research project of Qingdao (Grants No. 18-2-2-16-jcb). The work at the Research Center Jülich was performed within JuSPARC (Jülich Short-pulse Particle Acceleration and Radiation Center), a strategy project funded by the BMBF.

Competing interests: The authors declare no competing interests.

Magnetically brightened dark electron-phonon bound states in a van der Waals antiferromagnet, Nat. Commun., 13 (2022) 98.